%% file: manuscript.tex
\documentclass[useAMS,usenatbib]{mn2e}

\usepackage{aas_macros,amsmath,graphicx,longtable,amssymb,supertabular}

\newcommand{\rearth}{$R_{\oplus}$}

\newcommand{\kep}{{\it Kepler}}
\newcommand{\logg}{$log~g$ }

\newcommand{\teff}{$T_{\rm eff}$}

\newcommand{\feh}{[Fe/H]}
\newcommand{\etaearth}{$\eta_{\oplus}$}

\title[Small Worlds After All]{They are Small Worlds After All: Revised Properties of \kep{} M Dwarf Stars and Their Planets}

\author[Gaidos et al.]{E. Gaidos,$^{1,2,3,4,5}$\thanks{E-mail: gaidos@hawaii.edu} A.~W. Mann,$^{6,7,8}$ A. L. Kraus,$^{6}$ and M. Ireland$^{9}$\\
$^{1}$Department of Geology \& Geophysics, University of Hawaii at M\={a}noa, Honolulu, Hawaii 96822 USA\\
$^{2}$Visiting Scientist, Max Planck Institute f\"{u}r Astronomie, Heidelberg, Germany\\
$^{3}$Visiting Scientist, Kavli Institute for Theoretical Physics, Santa Barbara, California USA\\
$^{4}$Visiting Scientist, Center for Astrophysics, Harvard University, Cambridge, Massachusetts USA\\
$^{5}$Visiting Scientist, Observatoire de Gen\`{e}ve, Sauverny Switzerland\\
$^{6}$Department of Astronomy, University of Texas at Austin, Austin, Texas 78712 USA\\
$^{7}$Visiting Researcher, Department of Astronomy, Boston University, MA USA\\
$^{8}$Harlan J. Smith Postdoctoral Fellow\\
$^{9}$Research School for Astronomy \& Astrophysics, Australia National University, Canberra ACT 2611, Australia
}
\begin{document}
\date{Accepted to MNRAS}
\pagerange{\pageref{firstpage}--\pageref{LastPage}} \pubyear{2015}

\maketitle

\label{firstpage}

\begin{abstract}
We classified the reddest ($r-J>2.2$) stars observed by the NASA \kep{} mission into main sequence dwarf or evolved giant stars and determined the properties of 4216 M dwarfs based on a comparison of available photometry with that of nearby calibrator stars, as well as available proper motions and spectra.  We revised the properties of candidate transiting planets using the stellar parameters, high-resolution imaging to identify companion stars, and, in the case of binaries, fitting light curves to identify the likely planet host.  In 49 of 54 systems we validated the primary as the host star.  We inferred the intrinsic distribution of M dwarf planets using the method of iterative Monte Carlo simulation.  We compared several models of planet orbital geometry and clustering  and found that one where planets are exponentially distributed and almost precisely coplanar best describes the distribution of multi-planet systems.  We determined that \kep{} M dwarfs host an average of $2.2 \pm 0.3$ planets with radii of $1-4$\rearth{} and orbital periods of $1.5-180$d.  The radius distribution peaks at $\sim 1.2$\rearth{} and is essentially zero at 4\rearth{}, although we identify three giant planet candidates other than the previously confirmed Kepler-45b.  There is suggestive but not significant evidence that the radius distribution varies with orbital period.  The distribution with logarithmic orbital period is flat except for a decline for orbits less than a few days.   Twelve candidate planets, including two Jupiter-size objects, experience an irradiance below the threshold level for a runaway greenhouse on an Earth-like planet and are thus in a ``habitable zone".  
\end{abstract}


\input{intro}
\input{stars}
\input{planets}
\input{occurrence}
\input{discussion}

\section*{Acknowledgments}

Jonathan Swift kindly provided the posterior MCMC chains from \citet{Swift2015}.  This research was supported by NASA grants NNX10AQ36G and NNX11AC33G to EG.  EG was also supported by a International Visitor grant from the Swiss National Science Foundation.  ALK acknowledges support from a NASA Keck PI Data Award administered by the NASA Exoplanet Science Institute, and from NASA XRP grant 14-XRP14\_2-0106. SNIFS on the UH 2.2-m telescope is part of the Nearby Supernova Factory project, a scientific collaboration among the Centre de Recherche Astronomique de Lyon, Institut de Physique Nuclaire de Lyon, Laboratoire de Physique Nuclaire et des Hautes Energies, Lawrence Berkeley National Laboratory, Yale University, University of Bonn, Max Planck Institute for Astrophysics, Tsinghua Center for Astrophysics, and the Centre de Physique des Particules de Marseille. Based on data from the Infrared Telescope Facility, which is operated by the University of Hawaii under Cooperative Agreement no. NNX-08AE38A with the National Aeronautics and Space Administration, Science Mission Directorate, Planetary Astronomy Program. Some/all of the data presented in this paper were obtained from the Mikulski Archive for Space Telescopes (MAST). STScI is operated by the Association of Universities for Research in Astronomy, Inc., under NASA contract NAS5-26555. Support for MAST for non-HST data is provided by the NASA Office of Space Science via grant NNX09AF08G and by other grants and contracts.  This research has made use of the NASA/IPAC Infrared Science Archive, which is operated by the Jet Propulsion Laboratory, California Institute of Technology, under contract with the National Aeronautics and Space Administration.  This research made use of the SIMBAD and VIZIER Astronomical Databases, operated at CDS, Strasbourg, France (http://cdsweb.u-strasbg.fr/), and of NASA’s Astrophysics Data System, of the Jean-Marie Mariotti Center SearchCal service (http://www.jmmc.fr/searchcal), co-developed by FIZEAU and LAOG/IPAG.

\include{dwarf_table}
\include{binary_table}
\include{table_tap}
\include{planet_table}

\input{appendix_a}
\input{appendix_b}
\input{appendix_c}

\end{document}

%% file: intro.tex
\section{Introduction} \label{sec.intro}

In the twenty years since the discovery of 51 Pegasi b \citep{Mayor1995} the catalog of known planets around other main-sequence stars has grown by leaps and bounds.  One such leap was the advent of large and precise surveys of the brightest solar-type stars using the Doppler radial velocity method to systematically search for giant (Saturn- to Jupiter-size) planets \citep{Udry2006,Marcy2008}.  Another bound was the launch of the NASA \kep{} spacecraft in 2009, which monitored about 200,000 stars for up to four years in search of transiting planets \citep{Borucki2010}. Several thousand candidate or confirmed planets have been discovered in data from \kep{} \citep{Mullaly2015}, permitting the robust statistical analyses of much smaller (Earth- to Neptune-size) planets. Many analyses have reconstructed the planet population around the solar-type stars observed by \kep{} \citep{Youdin2011,Howard2012,Petigura2013,Foreman-Mackey2014,Silburt2015,Burke2015}.

\kep{} also observed many dwarf stars cooler than 3900~K and having M spectral types.  The mission's target catalog is mostly magnitude-limited and these intrinsically faint stars comprise only about 2\% of all targets, nevertheless these have yielded a disproportionate number of the smallest candidate planets \citep{Muirhead2012b}, multi-planet systems \citep{Muirhead2015}, and those planets irradiated at levels similar to Earth and thus potentially ``habitable'' \citep{Mann2013b, Quintana2014}.  The planets around \kep{} M dwarfs represent a distinctive population that can be compared with those around solar-type stars to test models of planet formation \citep[e.g.,][]{Gaidos2014b}.

There have been several recent works estimating the occurrence of planets around M dwarfs, the distribution with radii, and the number of planets within a ``habitable zone" (HZ) bounded by minimum and maximum irradiance levels.  Both \citet{Dressing2013} (hereafter, DC 13) and \citet{Gaidos2013} (hereafter, G13) carried out Bayesian estimations of properties of host stars of candidate planets with priors based on a stellar population model and likelihoods based on comparing the available photometry to predictions of stellar evolution and atmosphere models.  They estimated \etaearth{}, the occurrence of Earth-size planets in the HZ.  DC13 focused exclusively on the M dwarfs while G13 considered all host stars, although the less-luminous late K and M dwarfs dominate any estimation of \etaearth{}.  

\citet{Gaidos2014} used similar calculations for M dwarfs in the \kep{} target catalog and applied a method of iterative Monte Carlo (MIMC) simulation to estimate the occurrence per star, radius distribution and period distribution of planets.  Unlike previous approaches, MIMC analyzes planets as discrete values or distributions of values that can include non-trivial priors on parameters.  \citet{Silburt2015} pointed out that because of detection bias, errors in radii of small planets will not be symmetric: observed values are more likely to be underestimates of a larger value than overestimates of a smaller value.  Importantly, MIMC does not require binning, which may blur important features in a distribution.

\citet{Gaidos2014} used MIMC to estimate an occurrence $f$ of $2.0 \pm 0.3$ for 1-4\rearth{} planets with orbital periods $P < 180$~d, a radius distribution that peaks at $\sim 1$\rearth{} and a period distribution that was flat in logarithmic $P$.  \citet{Morton2014} also found $f = 2.0 \pm 0.45$ for $P < 150$~d and a similar radius distribution using a weighted density kernel estimation.  \citet[][hereafter DC15]{Dressing2015} performed an analysis of the \kep{} detection efficiency independent of the \kep{} project pipeline.  They identified candidate transit signals and estimated the detection completeness by injecting and recovering artificial transit signals. They incorporated available spectroscopic values for stellar parameters and and estimated $f \approx 2.5 \pm 0.2$.  \citet{Swift2015} performed a uniform re-analysis of \kep{} M dwarf light curves with candidate planets, accounting for the effect of transit timing variation where necessary.

There are several reasons to re-visit the properties of M dwarfs observed by \kep{} and the population of planets they host since these publications.  First, the latest (DR24) release of \kep{} ``Objects of Interest" (KOIs, candidate planets or astrophysical false positives) is the first uniform analysis of the entire \kep{} dataset using the project pipeline and includes numerous newly-detected candidate planets.  Second, the properties of \kep{} M dwarfs can be more accurately characterized by comparison with new, model-independent estimates of the properties of nearby calibrator stars. \citet[][hereafter M15]{Mann2015b} published the properties of 183 M dwarfs with effective temperatures (\teff{}), radii ($R_*$), and luminosities ($L_*$) determined by parallaxes and either interferometric or spectroscopic methods of angular diameters. Flux-calibrated visible-wavelength and near-infrared spectra were obtained for all these stars from which photometry was synthesized.  Third, many of the stars hosting candidate planets have been observed with high-resolution imaging techniques, i.e. adaptive optics and aperture-masking image synthesis \citep{Kraus2015}.  About a quarter of M dwarf systems are multiples and the \kep{} data alone often cannot disambiguate which component is hosting the candidate planets, information necessary to accurately estimate the planet's radius and probability of transit \citep{Barclay2015}.  Moreover, high-resolution imaging can help identify background eclipsing binaries masquerading as planetary signals and not identified by the \kep{} data validation process.  Finally, we wished to use the statistical methods described in \citet{Gaidos2014} to calculate the planet population.

%% file: stars.tex
\section{The Stars}
\label{sec.stars}


\subsection{Stellar Sample \label{sec.sample}}

Most \kep{} target stars lack spectra, and classification and parameter estimation of these stars were performed by comparing Sloan $griz$ photometry from the Kepler Input Catalog (KIC) \citep{Brown2011} and $JHK_S$ photometry from the Two Micron All Sky Survey (2MASS) with equivalent photometry for nearby M dwarf reference stars.  We supplemented (and superseded) KIC photometry with $gri$ photometry from the Kepler Input Survey (KIS) Data Release 2 \citep{Greiss2012a,Greiss2012b}.  The KIS was especially important to replace missing $g$ and $r$ photometry from KIC.  We retrieved or computed equivalent photometry for a set of nearby M dwarf calibrator stars with precisely determined properties (M15) as well as giants selected by asteroseismic signal. Corrections were applied to these data to bring them into a common photometric system and remove systematic errors (Sec. \ref{sec.photometry}).

For our catalog of M stars (dwarfs and giants) we retained the 8732 stars with $r-J > 2.2$.  This criterion was based on an established relationship between $r-J$ and \teff{} for M dwarfs that is only weakly metallicity-dependent \citep[][M15]{Gaidos2014}, and the high fraction of \kep{} host stars with $r-J > 2.2$ which were already observed by \citet{Kraus2015} with high-resolution imaging methods (Section \ref{sec.binaries}).

\subsection{Correction and Verification of Photometry \label{sec.photometry}}

We first converted Kepler Input Catalog photometry into the Sloan system using the relations of \citet{Pinsonneault2012}.  We corrected the AB system of the KIS photometry to the Pinsonneault-corrected KIC system using polynomial best-fits to the matches between the two catalogs, i.e. 
\begin{equation}
g - g_{\rm KIS} = 0.154 + 0.0462 \left(g_{\rm KIS} - r_{\rm KIS}\right),
\end{equation}
\begin{equation}
r - r_{\rm KIS} = 0.130 + 0.0213 \left(r_{\rm KIS} - i_{\rm KIS}\right), 
\end{equation}
and 
\begin{equation}
i - i_{\rm KIS} = 0.352 + 0.0694 \left(r_{\rm KIS} - i_{\rm KIS}\right).
\end{equation}

\begin{figure}
\includegraphics[width=84mm]{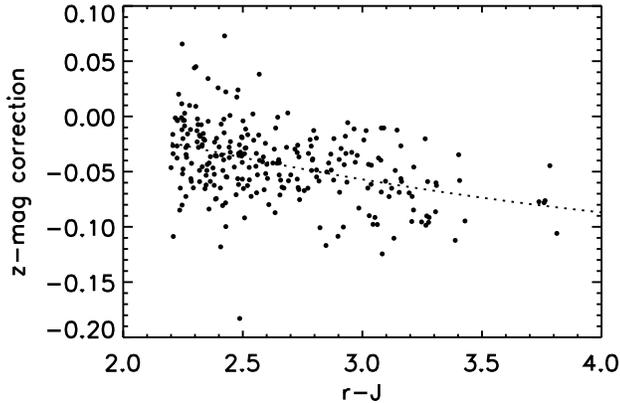}
\caption{Difference between $z$-band SDSS and KIC magnitudes corrected to SDSS using the relations of \citet{Pinsonneault2012} vs $r-J$ color.  The best-fit quadratic is plotted as a dotted line.}
 \label{fig.sdss}
\end{figure}

The color corrections derived by \citet{Pinsonneault2012} were based on a sample that included few red stars and thus may be erroneous when applied to M dwarfs.  We checked this by matching KIC stars with sources from Data Release 9 of the Sloan Digital Sky Survey \citep[SDSS,][]{Ahn2012} with less than 3\arcsec{} separation.  We regressed SDSS-corrected KIC magnitudes against $r-J$ color and tested the significance of corrections of different order against no correction using an F-test of the ratio of variances.  For the $g$, $r$, and $i$ passbands no significant correction was found, but a significant correction ($p = 1.2 \times 10^{-5}$) was found for $z$-band.  The $z$-band differences are plotted in Fig. \ref{fig.sdss}.  The negative sign and trend of the correction $\Delta z$ suggest that KIC $z$ magnitudes are affected by an uncorrected red leak which is larger for redder stars.  We use the best-fit quadratic with $2.5\sigma$ rejection (solid line in Fig. \ref{fig.sdss} and Eqn. \ref{eqn.leak}) to correct the KIC photometry; a constant correction of -0.0823 was applied to stars redder than $r-J > 3.813$ where there are no data for deriving a correction.
\begin{equation}
\label{eqn.leak}
\Delta z = -0.024 - 0.047\left(r-J-2.2\right) + 0.064\left(r-J-2.2\right)^2
\end{equation}

Because they are bright, most M dwarf calibrators in the catalog of \citet{Mann2015b} lack SDSS-band $griz$ photometry and some lack 2MASS $JHK_s$ photometry, therefore we synthesized photometry from flux-calibrated spectra using the corrected response functions of \citet{Cohen2003} and \citet{Mann2015a}.  For a small number of stars, we validated the synthetic $griz$ photometry with new photometry that we obtained as described in Appendix A.  Figure \ref{fig.synvsobs} compares the synthetic vs. observed $g-J$, $r-J$, $i-J$, $J-H$, and $H-K$ colors vs. synthetic $r-J$.  We performed F-tests, comparing the $\chi^2$ with no offset to those after removal of a mean offset or a weighted linear regression.  In no cases was there a significant decrease in $\chi^2$ compared to that of a mean offset.  Only in the cases of the $g-J$ and $i-J$ colors were there (marginally) significant improvements to $\chi^2$ from linear regressions, with $p$ values of 0.08 and 0.10, respectively.  Thus, no corrections were made.

\begin{figure}
\includegraphics[width=84mm]{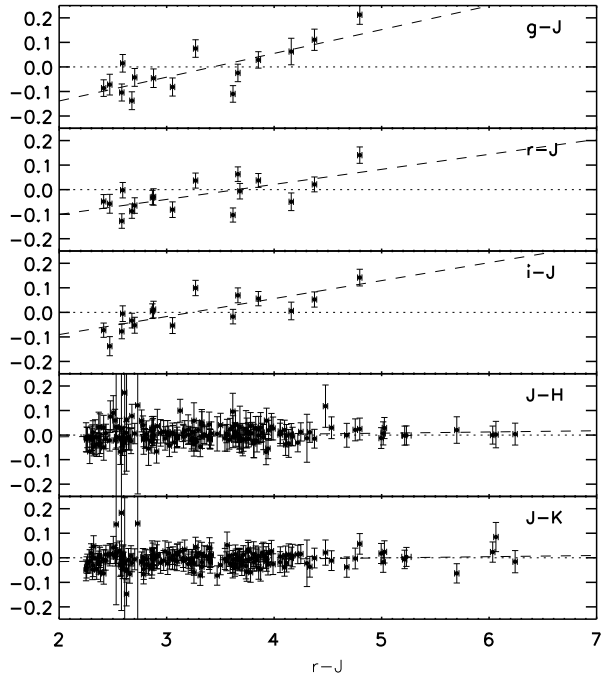}
\caption{Difference between synthetic and actual SDSS-tied photometry for calibrator stars for the $g-J$, $r-J$, $i-J$, $J-H$, and $J-K$ colors.  No data was obtained for $z$-band.  Dashed lines are weighted linear fits.   Only for $g-J$ and $i-J$ are non-zero trends marginally significant (F-test $p$ values of 0.08 and 0.10, respectively).  }
 \label{fig.synvsobs}
\end{figure}

Finally, we identified 8 KIC stars for which we have both SNIFS and SpeX spectra and thus can derive synthetic photometry.  Since we do not have photometry independent of the KIC to flux-calibrate the spectra, we can only calculate colors, and only the $i-z$ color which does not span gaps in the spectra, i.e. between the two SNIFS channels or SNIFS and SpeX.  After correction of the KIC to the SDSS system using \citet{Pinsonneault2012} and Eqn. \ref{eqn.leak}, we found a mean difference of only 0.007 mags with the synthetic $i-z$ colors, and a standard deviation of 0.009 mags, consistent with the photometric errors of the KIC.  The offset is only twice the error in the mean, and an F-test of the ratio of variances has a $p$ value of 0.26 so the offset is not significant and we did not remove it.

\subsection{Classification \label{sec.classification}}

The vast majority of M stars observed by \kep{} lack spectra or parallaxes to decisively assign a luminosity class, but they have photometry in varying numbers of bands and sometimes proper motion ($\mu$).   The original KIC classification \citep{Brown2011} and subsequent analyses by DC13 and G13 were model-dependent, i.e. to infer \teff{} and \logg{} they compared observed colors to grids of predicted values from stellar atmosphere models.  An empirical classifier based on a comparison with the colors of established M dwarfs and giants must work with discrete samples that are distributed non-uniformly through multi-dimensional color space, rather than grids of arbitrary resolution.  The classifier must also be flexible in terms of the availability and weight of different data; for very red, late-type M stars, two colors are often sufficient to  separate giants from dwarfs \citep[e.g.,][hereafter H14]{Huber2014}, but this is not the case for hotter M stars.  

\begin{figure}
\includegraphics[width=84mm]{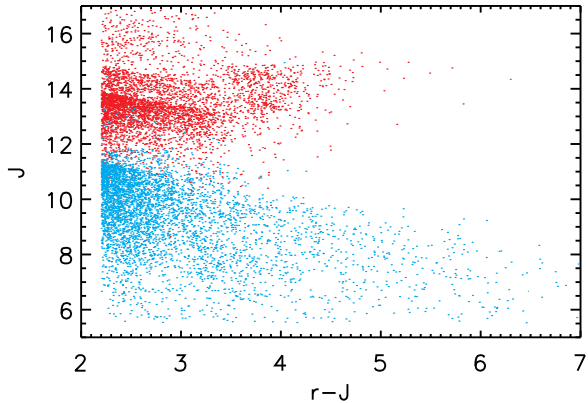}
\caption{Distribution of 8732 red \kep{} ($r-J>2.2$) stars classified as M dwarfs (red) or M giants (blue) with $r-J$ color and apparent $J$-magnitude.  The diagonal boundaries are artifacts as a result of \kep{} target selection to a limiting \kep{} magnitude (similar to $r$-band for red stars).}
 \label{fig.colormag}
\end{figure}

We developed a Bayesian, Gaussian process-based procedure to perform a uniform classification on the 8732 M stars selected in Sec. \ref{sec.sample}, incorporating all measurements for each star.  Details of the classification algorithm are provided in Appendix B and only a summary is given here.    We calculated the probability that a star is a dwarf as
\begin{equation}
P({\rm dwarf}) = \frac{p_d}{p_d + p_g},
\end{equation}
where $p$ is the Bayesian posterior probability that the star is a dwarf (d) or giant (g).  These are in turn given by
\begin{equation}
\label{eqn.prob}
p_{d,g} = \tilde{p}(J;r-J) p(H_J){\displaystyle \prod_{n}} p_n(c_n;r-J) 
\end{equation}
where $\tilde{p}$ is a prior function calculated from the expected color-magnitude distribution of dwarf or giant stars based on a stellar population model of the Galaxy, $p(H_J)$ is a likelihood based on the reduced proper motion $H_J$ (a proxy for absolute magnitude based on proper motion $\mu$ rather than parallax), and $p_n$ are likelihoods calculated by comparing the colors $c_n$ (e.g., $i-J$ or $J-K$) of a star with those of established M dwarfs and giants.  We used $r-J$ as an independent parameter that serves as a proxy for \teff{}.  We calculated $p(H_J$) using the distributions for asteroseismic giants and dwarfs with high $\mu$.  We calculated color-dependent likelihoods using the method of binary Gaussian process (GP) classification and training sets of dwarf and giant stars selected independently of color.  Where available, we also used measurements in the gravity-sensitive Mg~Ib resonant line at 5100\AA{} (recorded as a ``D51" magnitude in the KIC) to better discriminate between high-surface gravity dwarfs and low-surface gravity giants.  We classified 4218 stars with $r-J > 2.2$ as M dwarfs, and 4513 as giants (classification failed for one star, KIC~9897227).  Figure \ref{fig.colormag} plots apparent $J$-magnitude and $r-J$ color for the classified stars, showing the dichotomy between the two classifications at $J \approx 11.5$.

We compared our classifications to those of DC13, who fit KIC photometry of 3897 cool stars to a grid of atmospheric and evolutionary model predictions \citep{Dotter2008}, and to those of H14 who fit KIC parameters, where available, to stellar evolution model predictions, but also incorporated spectroscopic data to characterize Q1-16 target stars. H14 also assigned parameters to formerly unclassified stars using the Bayesian priors of their model.  Of our 4218 dwarf stars, 1436 are not in the \citet{Dressing2013} catalog; 810 of these were observed for a single quarter, mostly in Q17, and 1117 of them were observed for fewer than 5 quarters.  Most of the 1436 missing stars are at the cooler ($\sim 3000$K) or hotter ($\sim 4000$K) ends of the \teff{} range and may have been excluded because of insufficient color information or the difficulty of discriminating between giants and dwarfs, respectively.  We identified 1109 stars from DC13 that were not in our dwarf sample.  Two-thirds (735) of these are hotter than 3900K, the nominal limit of our sample, and nearly all have \teff{} $>3700$K.  Six stars from  DC13 were classified as giants, but classifications for three are ambiguous.  Two of the other stars KIC 9330545 and 10731839, have reduced proper motions favoring a giant classification, and a third, KIC 9330545, has colors indicating it is a giant.

There are 764 dwarfs and 1910 giants stars not listed in H14, almost all were observed only during the Q0 or Q17 quarters that are excluded from the H14 catalog.  The missing dwarfs were mostly (551) observed in Q17, but another 170 were observed in Q6.  The giants were almost entirely (1890) observed in Q0 and subsequently removed from the target list.  Forty-four of our dwarfs are classified by H14 as giants (i.e., \logg{} $ < 3.5$).  Ten of these are ambiguous cases and have posterior probabilities that only marginally favor the dwarf classification because they are fainter ($J > 11.5$) and/or have small but significant proper motion (PM) in the PPMXL catalog.  The majority have classification probabilities of $>0.95$ because of their colors and high PM, many from the high-quality SUPERBLINK catalog.  208 of our giants are classified by H14 as dwarfs (\logg $>3.5$).  Only one of these, KIC~7341653, has colors and a SUPERBLINK PM that favor a dwarf classification: this star was spectroscopically confirmed by \citet{Riaz2006} and is also an X-ray source.

We also inspected 13 stars that have $H_J$ with one magnitude of the dwarf locus but were classified as giants.  All of the PMs are from PPMXL and none exhibit the expected motion of a few arc-sec in a comparison between POSS 1 and POSS 2 images.  Instead, in each case there is another star within a few arc seconds and confusion may have produced erroneously high PMs in PPMXL.

\subsection{Stellar Properties \label{sec.properties}}

We estimated the temperature, metallicity, radius, luminosity, and mass of all but two of the 4218 stars classified as M dwarfs (Sec. \ref{sec.classification}).  The procedure is described in detail in Appendix C and only briefly described here.  We related \teff{} and \feh{} to Sloan colors using the 172 calibrator M dwarfs with $r-J > 2.2$ in M15 and the method of Gaussian process regression.  Each of the two regressions used a stationary squared exponential function with 7 hyperparameters: one for amplitude and one for each of the six color terms in the exponential.  Hyperparameter values were optimized to minimize $\chi^2$ when comparing estimated \teff{} and \feh{} to the known values of the calibrator stars.  We applied these relations to the KIC stars to estimate \teff{} and \feh{}.  We then in turn applied the relations described in M15 to estimate radius $R_*$, luminosity $L_*$, and mass $M_*$.  For 101 stars, i.e. most host stars of candidate or confirmed transiting planets, spectroscopically-determined \teff{} and \feh{} are available and these were substituted.  

\begin{figure}
\includegraphics[width=84mm]{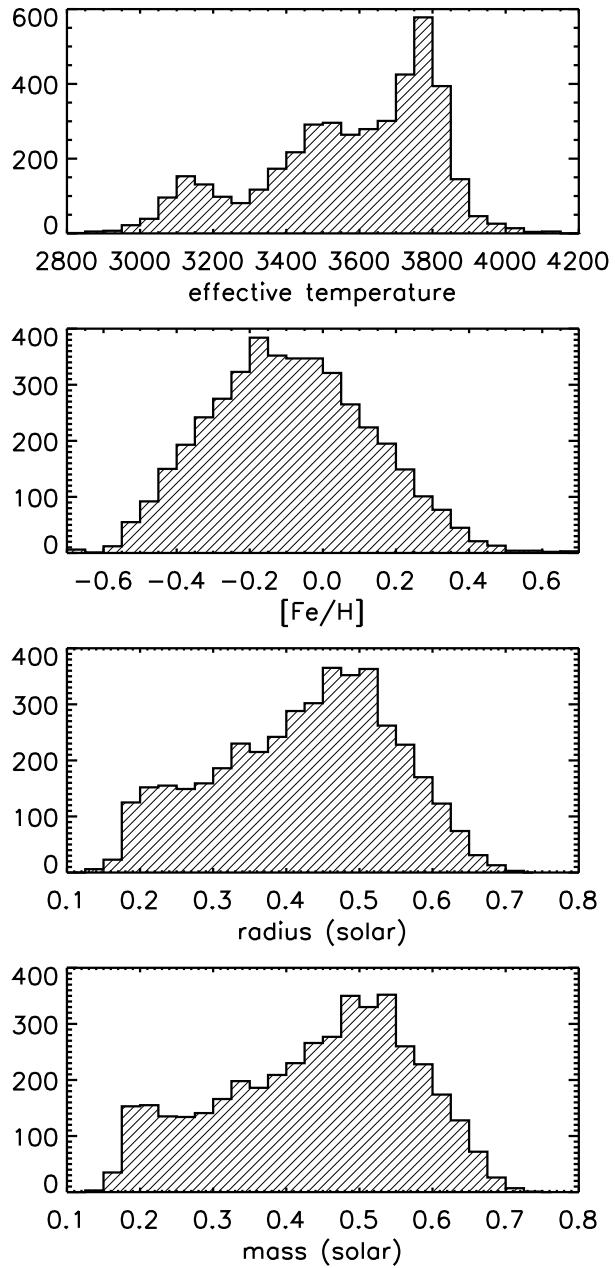}
\caption{Distributions of estimated \teff{}, [Fe/H], radius, and mass of \kep{} M dwarfs.}
 \label{fig.dwarfs}
\end{figure}

The estimated properties of the M dwarfs are reported in Table \ref{tab.dwarfs}.  Figure \ref{fig.dwarfs} plots the distributions of the sample with \teff{} and  \feh{} assigned to the stars.  The distribution of \teff{} peaks at 3800-3850~K and falls sharply with cooler or hotter temperature.  The former is a manifestation of the Malmquist bias produced by the flux-limited nature of the \kep{} target list.  The latter is produced by our $r-J > 2.2$ cutoff.  The sharp color cut does not produce an equally sharp cutoff in \teff{} because the $r-J$ colors of M dwarfs are also weakly dependent on [Fe/H].   There are ten very red ($r-J > 5$) dwarfs, all of which have assigned \teff{} $<2900$~K.  Two stars (KIC~4076974 and 9834655) have $r-J > 7$, much redder than our M dwarf calibration sample, and \teff{} assignment failed.  It is likely that the colors of these two stars have been affected by variability or photometric errors.  The first of these two is the high proper-motion M dwarf LSPM J1945+3911 \citep{Lepine2005}.  The other 8 of the 10 stars have not been previously identified.  The distribution of [Fe/H] has a mean value of -0.09 and a standard deviation of 0.22 dex.  The dispersion is indistinguishable from that found among nearby M dwarfs as well as the \kep{} field \citep{Mann2013b,Gaidos2014}, in part due to a small adjustment in the $J$-band magnitudes (see Appendix C), and the intrinsic scatter is 0.16\,dex after subtracting measurement error.  Distributions of radius and mass calculated according to the relations in M15 are also shown in Fig \ref{fig.dwarfs}.

\begin{figure}
\includegraphics[width=84mm]{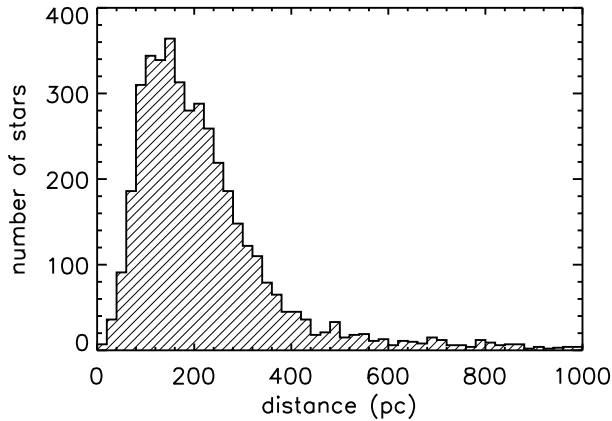}
\caption{Distribution of distances of \kep{} M dwarfs based on apparent $K$-band magnitude and estimated luminosities.}
 \label{fig.dist}
\end{figure}

A photometric distance was calculated as 
\begin{equation}
d_{\rm phot} = 10^{(K-M_K)/5 + 1},
\end{equation}
and is plotted in Fig. \ref{fig.dist}.  The median distance is $\approx190$~pc.  The are three stars with reliable ($\pm 2-3$pc) photometric distances within 25~pc.  The closest of these is KIC~3629762 (J19051335+3845050) at $d \approx 15.2$~pc.  This high proper motion star was previously selected by \citet{Reid2007} as likely to be within 20~pc of the Sun.   KIC~10453314 is assigned a photometric distance of $\approx 13$~pc but this estimate is based on a \teff{} and hence $M_K$ which are much more uncertain for this star.  This star was identified as a nearby ($\lesssim 20$~pc) M3 dwarf by \citet{Reid2004} and has a high proper motion \citep[LSPM J1854+4736,][]{Lepine2005}.

\begin{figure}
\includegraphics[width=84mm]{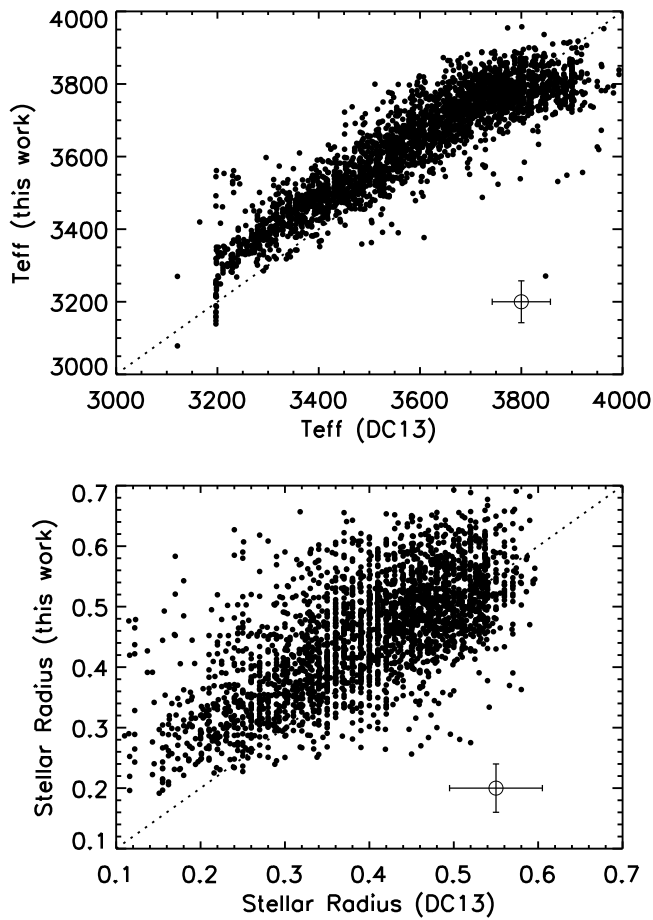}
\caption{Comparison of stellar \teff{} (top) and radius (bottom) estimated in this work with the model-based values of \citet{Dressing2013} (DC13).  The dotted lines represent equality and the points in the lower right-hand corner illustrate median uncertainties.  The \teff{} range of the model grid used by DC13 did not go below 3200K although they adopted some cooler spectroscopic values.}
 \label{fig.dc_compare}
\end{figure}

Figure \ref{fig.dc_compare} compares our estimates of \teff{} and $R_*$ to that of DC13.  Our values are systematically larger: the median differences are 51~K and 14\%, respectively.  The difference in \teff{} presumably reflects the use of observed colors vs. those predicted by the PHOENIX stellar atmosphere model used by DC13.  The difference in $R_*$ is produced both by the difference in \teff{} and the use of empirical radii vs. those predicted by the Dartmouth stellar evolution model, which is known to systematically under-predict empirical M dwarf radii \citep[e.g.,][M15]{Boyajian2012}.  The model grid used by DC13 also did not extend below 3200K, although they included some cooler spectroscopic values.

\subsection{Multiple Systems \label{sec.binaries}}

Multiplicity is an important stellar property in its own right.  It also affects estimates of transiting planet properties and statistics via the properties of the possible host stars and the dilution of the transit signal.  To identify possible multiple systems, adaptive optics (AO) and non-redundant aperture masking (NRM) data was taken from \citet{Kraus2015}, which contains significantly more detail on the observations, reduction, and derivation of detection limits. To briefly summarize, observations were taken with the Keck~2 telescope using natural guide star or laser guide star AO (depending on target brightness and observing conditions) between 2012 May and 2014 August. All observations were conducted with the NIRC2 detector and the facility AO imager in the $K$-prime ($K'$) pass-band, as well as a 9-hole aperture mask for NRM.  All observations used the narrow-field camera and smallest pixel scale (9.952 mas pixel$^{-1}$). Observations of each object consisted of four components; an acquisition exposure, at least one shallow image (10-20s), two deep images ($>20$s each), and 6-8 20-sec exposures with a 9-hole mask in place (NRM interferometry exposures).  Modest adjustments were made to the number of exposures and exposure times to compensate for poor weather conditions. 

\begin{figure}
\includegraphics[width=84mm]{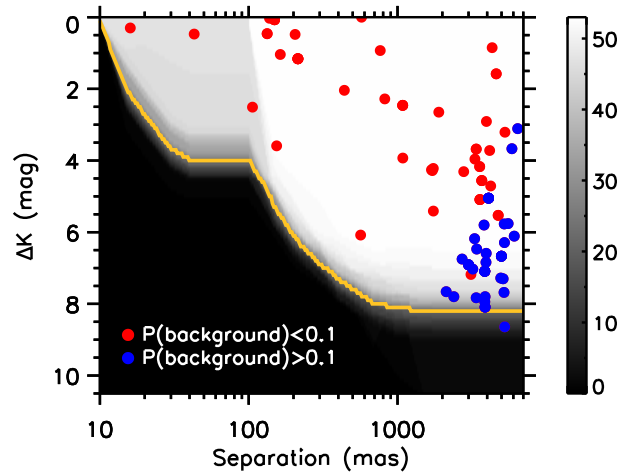}
\caption{Detection limits for AO data in this survey as a function of $\Delta K'$ and angular separation (in milli arc seconds). Detections are shown with red circles for objects that are likely to be bound systems and blue circles for objects are likely to be background stars (See Section~\ref{sec.binaries} for details). Detection limits are shown as a shading gradient from black (no detections possible) to white (a detection is possible around all 60 targets). The yellow solid line shows the median limit for the survey.}
 \label{fig.ao}
\end{figure}

For each AO image we corrected for geometric distortion using the NIRC2 distortion solution from \citet{Yelda2010}, flagged dead and hot pixels, and removed cosmic rays. We fit each detected source with a template point-spread function (PSF) built from sources in images from the same night. Separation and position angles were calculated between the centroids of the PSF and $\Delta K$' values were estimated between the primary and all other sources in the frame, as presented in \citet{Kraus2015}.  NRM data were analyzed following the method of \citet{Kraus2008} and \citet{Kraus2011b}. This includes flat-fielding, bad pixel removal, dark subtraction, extraction and calibration of squared visibility and closure phase, and binary model fitting.  This survey included 54 of the 73 host stars in our sample.  Detections and detection limits for all observations are summarized in Fig.~\ref{fig.ao}.

\begin{figure}
\includegraphics[width=84mm]{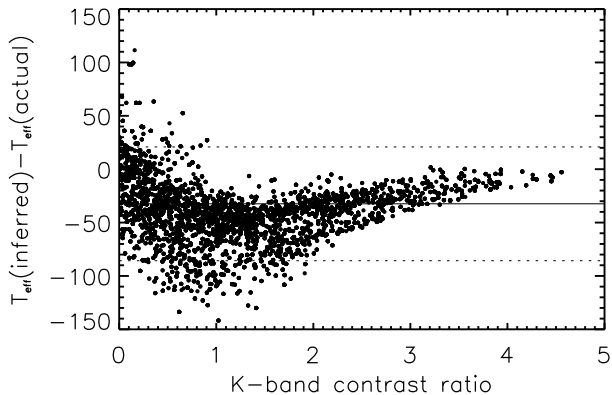}
\caption{Error in estimates of the primary \teff{} produced by an unresolved companion vs. the primary$-$companion contrast in $K'$-band.  For comparison the solid line is the mean offset between the color-predicted and actual \teff{} values (-33~K) among the M dwarf calibration stars, and the dashed lines are plus and minus one standard deviation (53~K).}
 \label{fig.bincolor}
\end{figure}

Systems resolved into multiples required additional analysis to ascertain the properties of the individual components.  We first investigated the effect of multiplicity on the primary parameters estimated from colors (see Appendix B).  This was done by adding the fluxes of randomly-selected pairs of stars from our M dwarf training set, recomputing colors, estimating \teff{} from the combined fluxes, and comparing it the estimated \teff{} of the primary.  Physical pairs of stars originate from the same molecular cloud and should have very similar metallicities.  Thus only pairs with metallicity values within 0.07 dex of each other were included; this is the limiting resolution of our metallicity estimates and stars with less difference in [Fe/H] were considered indistinguishable.  Figure \ref{fig.bincolor} plots the difference between estimated and actual primary \teff{} due to the presence of a companion with a certain $K'$-band contrast ratio ($\Delta K'$).  The effect on \teff{} is small for equal-contrast pairs which have similar \teff{} and [Fe/H] and hence colors, reaches about -60~K for contrasts of $\sim 1$~mag (as expected, companions make primaries appear cooler), and approaches zero again for large $\Delta K'$ because the secondary contributes negligible flux to the total.  Because we assume that the metallicities of the components are very similar, it follows that the effect of the companion on the estimated radius of the primary is small.  For all cases the error is comparable to the systematic offset and uncertainty in estimating \teff{} from colors (solid and dashed lines in Fig. \ref{fig.bincolor}) and was neglected in further analysis.

For each additional source found in our high-resolution imaging we calculated the probability that this is a background star based on a star count calculation with the TRILEGAL version 1.6 stellar population model at the centers of each of the 84 \kep{} CCD fields (see Appendix B).  Every simulated star from every field contributes to this calculation with a weight $w$ that is a Gaussian function of the angular separation between the KOI and the center of the field of the simulated star\footnote{TRILEGAL synthesizes populations at a single celestial coordinate, thus all synthetic stars in a CCD field share the same coordinate, i.e. the field center.}, with a smoothing parameter of 1 degree.   We calculated the probability that a putative companion was a background star as the likelihood that a brighter star (in the $K$-band) would fall closer to the primary, i.e. 
\begin{equation}
p_{\rm back} = 1 - \exp \left[-\frac{\pi \theta^2}{\Omega}\frac{\sum\limits_n^{\rm fields} w_n N_{n,K<K_0}}{\sum\limits_n^{\rm fields} w_n}\right],
\end{equation}
where $\theta$ is the angular separation between the primary and putative companion, $\Omega$ is the field of view of an individual CCD, and $N_{n,K<K_0}$ is the number of stars in the $n$th field with $K$ magnitudes brighter than that of the detected source ($K_0$).  We also calculated a false-positive probability that if the planet candidate were around the fainter companion/background star it would be larger than Jupiter and thus probably not a planet.   These probabilities are reported in Table \ref{tab.binaries}.     

We then estimated the properties of each possible companion by computing the absolute magnitude in the $K'$ pass-band, using the contrast ratio and assuming the stars are at the same distance.  We then estimated the companion star \teff{} that reproduce that absolute $K'$ magnitude using a GP regression of the synthetic photometry against the established properties of our calibrator M dwarfs, and assuming the primary and companion metallicities are equal.  Observed and estimated companion properties are given in Table \ref{tab.binaries}.  In some cases, often when the source is probably a background star, the regression failed or is an upper limit ($\sim 2500$K) because such a companion would have to be cooler than any of our calibrator stars.  Synthetic $K'$ photometry was synthesized from calibrated infrared spectra following the procedures in \citet{Mann2015a}.  A range of \teff{} values was tried, while [Fe/H] was set to the primary value (estimated from photometry or obtained from spectroscopy), and each value was weighted by a Gaussian with estimated $\Delta K$ centered on the correct value of $\Delta K'$ and with a standard deviation equal to the formal observational error of $\Delta K'$.  The \teff{} of the potential companion was taken to be the weighted mean of all \teff{} values.  A contrast ratio for the visible-wavelength \kep{} bandpass $K_p$ was also calculated by GP regression in a similar fashion.  \teff{} and [Fe/H] values were then used to estimate the other properties of the potential companion as described in Appendix C.

%% file: planets.tex
\section{The Planets}
\label{sec.planets}

\subsection{KOI Selection}
\label{sec.koi}

We drew our planet candidates from the DR24 TCE release of the \kep{} project pipeline.  We selected 4261 KOIs flagged as candidates plus 385 flagged as potential false positives (FPs) by the automated classification procedure, as of 24 October 2015.  In the latter case we selected exclusively those KOIs which only have the flag set for significant offset or motion of the stellar image centroid.  The Data Validation (DV) reports three behaviors of the centroid; correlation of centroid motion with variation in flux, offset of the centroid of the transit signal image (the difference between in- and out-of-transit images) with respect to the out-of-transit image, and offset of the transit signal centroid with respect to the nominal KIC position.  The centroid FP flag is set if any of these show a $>3\sigma$ effect.  Of these KOIs, there are 115 candidates and 5 potential FPs around the M dwarfs selected in Sec. \ref{sec.classification}.

Eleven other KOIs are affiliated with stars that we classified as M giants.  Three of these are in the KOI-314 (\kep{}-42) system, discussed below.  KOI~977.01 was confirmed as a false positive by \citet{Hirano2015}.  KOI~4948.01 was classified as a false positive based on significant centroid motion and the transit-based stellar density ($1.8 \pm 0.8$~g~cm$^{-3}$) is inconsistent with an M dwarf.  A spectrum of the host star of KOI~5129.01 suggests it is a carbon (AGB) star (D. Latham, comment in the Community Follow-up Observing Program website) and the estimated ``planet'' radius equals that of the Sun.  KOI~5346.01 is flagged as a false positive due to the $\approx 11$" offset of the transit signal from both the out-of-transit signal and KIC coordinates.  The stellar density for the host of KOI~5346.01 based on the transit fit is consistent with a dwarf, but with very large error bars.  KOI~5752.01 is a similar case, but the offset of the transit signal is only marginally significant.  KOI~6165.02 is affiliated with an eclipsing binary \citep{Slawson2011}.  An $R \approx 60,000$ visible spectrum of KOI~6391 (KIC~4160669) suggests a low log~$g$ (M. Endl, private communication) The signal of KOI~6391.01 is actually offset by $\approx 7.5$" south of the target star, where there is a hotter, fainter ($\Delta J \approx 5.2$ magnitudes) source (2MASS J19291781+3917504).  KOI~6475.01 is an ambiguous system (P(giant)$ = 0.62$) consisting of a pair of stars separated by 2", possibly both rotational variables with periods $\approx 4$~d \citep{Reinhold2013}.

We inspected the \kep{} light curves and DV reports of our sample and identified 13 that are known or likely FPs (a rate of 11\%):  KOI~256.01 which is a known post-common envelope binary with a white dwarf \citep{Muirhead2013}:  KOI~950.01 is also listed as a EB in \citet{Slawson2011} and has significant flux-weight centroid motion statistic.  KOI~1225.01 also appears in \citet{Slawson2011} and has a very large transit depth inconsistent with a planet-sized object.  KOI~1654.01 was identified as an EB by \citet{Slawson2011} and the planet transit fit parameters are inconsistent with those of a dwarf star.  KOI~1902.01 has an asymmetric lightcurve inconsistent with a planetary transit, a duration that is too short (1.7~hr) for its orbital period (137~d), and is assigned a 93\% FP probability by \citet{Swift2015}.  KOI~2015.01 has a variable light curve and significant centroid-flux correlation suggesting that the KIC star is not the source of the transit.  The transit signal of KIC~2570.01 is offset 2.7" to the NE of the KIC star, where there is a faint source in a UKIRT $J$-band image.  Likewise, KOI~2688.01 is significantly offset to the SW.  KOI~3749.01 exhibits a distinctly non-planetary V-shaped light curve and was assigned a FP probability of 86\% by \citet{Swift2015}.  KOI~3773.01 has a significant transit offset but also a V-shaped lightcurve and ellipsoidal variation suggestive of an EB.  KOI~3810.01 also has a V-shaped transit lightcurve, an estimated radius of 9.8\rearth{}, near the planetary limit, and a best-fit impact parameter $b \sim 0.9$.  KOI~4808.01 is offset 1.7" from the host star where there is a fainter star in the UKIRT image.  The transit signal of KOI~7137.01 is about 7" from the KIC and unlikely to be related to the star.  Finally, \citet{Swift2015} assigned KOI~3263.01 a FP probability of 71\%, however this is a particularly interesting but ambiguous object which we retain to motivate follow-up observations and discuss in Section \ref{sec.special}.

Three notable M dwarf systems do not appear in our catalog: KOI~314 (Kepler-138), 961 (Kepler-42), and 2704 (Kepler-445).  The first is mis-classified as a giant because of its low metallicity and unusual colors, bright $J$-band magnitude, and lack of a $z$-magnitude.  The other two lack CCD $r$-band magnitudes for constructing an $r-J$ color.  These are among the very coolest \kep{} stars, which has the unfortunate side-effect of limited reliable photometry in the KIC.

\subsection{Light Curve Fitting}\label{sec.transitfit}

For stars with no AO data or no detected companions within 2~\arcsec, we used the available transit fit posteriors from \citet{Swift2015}. They fit long- and short-cadence \kep{} data for 165 of the KOIs orbiting M dwarfs using an MCMC framework including improved handling of systems with significant transit timing variations. The fits from \citet{Swift2015} are generally consistent with those produced by the \kep{} data validation fits \citep{Rowe2014,Rowe2015}, with a median difference in $R_P/R_*$ of just 0.2\%, and 76\% of $R_P/R_*$ values agreeing within 1$\sigma$.

We refit the light curves of 20 KOIs missing from \citet{Swift2015}, mostly KOIs identified after publication.  We downloaded Q1-17 light curves for the remaining KOIs from the Barbara A. Mikulski Archive for Space Telescopes (MAST). For each KOI we extracted the Pre-search Data Conditioning \citep[PDCSAP,][]{Stumpe2012} light curve data during transit along with a 3 hour interval on each side.  Data with obvious non-transit artifacts (e.g., stellar flares) and data covering less than half of a transit were identified by eye and removed. We fit the out of transit light curve with a third order polynomial to remove trends in the light curve unrelated to the transit (mostly from stellar variability), and then stacked the data into a single light curve for each planet candidate.

We fit the stacked curves using a modified version of the Transit Analysis Package \citep[TAP,][]{Gazak2012}, which uses a \citet{MandelAgol2002} model with a quadratic limb-darkening law. We modified TAP to calculate the host star's density using the equations from \citet{Seager:2003lr} during each MCMC step. TAP compares this value to the density from the spectroscopic/GP stellar parameters derived previously, and applies an additional penalty to the likelihood. Similar methods have been used to derive orbital eccentricity \citep{Dawson:2012fk} and identify FPs \citep{Sliski2014}. Here we assumed that the planets are real, all orbit the same star, and have low eccentricities as found for small planets \citep{Van-Eylen2015}, and instead we used the density data to better constrain the transit parameters, and in the case of binaries, help identify the host star (Section~\ref{sec.hoststar}) 

We applied a prior drawn from the model-derived limb-darkening coefficients from \citet{Claret2011}. For each star we interpolated our stellar parameters (log~$g$, \teff, [Fe/H]) onto the \citet{Claret2011} grid of limb-darkening coefficients from the PHOENIX models, accounting for errors from the finite grid spacing, errors in stellar parameters, and variations from the method used to derive the coefficients (Least-Square or Flux Conservation). By stacking the transits, we effectively erased information about the planet period, so we used a prior from the \kep{} fits. Comparison to fits done by \citet{Swift2015}, suggested this has a negligible effect on the results provided there are no significant transit timing variations. Eccentricities were forced to be small ($<0.1$), and red noise was set to zero. Other fitted parameters (impact parameter, transit duration, $R_P/R_*$, argument of periastron, white noise) were fit without constraint (other than physical limits) and uniform priors. 

For the 20 single stars we refit our results are generally consistent with those from the \kep{} data validation fits. Excluding the two most discrepant fits (3090.02 and 6635.01), the median difference between our $R_P/R_*$ values and those from \citep{Rowe2015} fits is 0.1\%. Further, 88\% of our  $R_P/R_*$ values are within 1$\sigma$ of those from \kep, and all are within 3$\sigma$ (again excluding 3090.02 and 6635.01). The \kep{} fit for 3090.02 assigned an impact parameter $>1$, yielding an unphysical $R_P/R_*$ of 92$\pm$8. For 6635.01 the \kep{} fit has an $R_P/R_*$ twice our own (0.033 versus 0.015), although the exact reason for this large disagreement is unclear. 

\begin{figure}
\includegraphics[width=84mm]{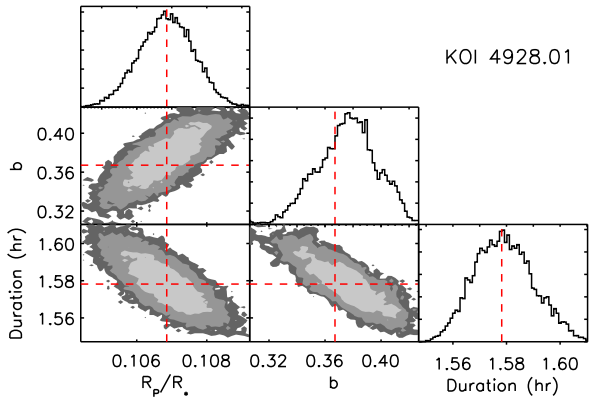}
\caption{Posteriors of fits for transit duration, impact parameter ($b$), and planet-to-star radius ratio ($R_P/R_*$) for the light curve planet candidate KOI~4928.01. Two-dimensional distributions are shaded to include 68\%, 95\%, and 99.7\% of the MCMC steps. The locations of maximum likelihood are marked with dashed red lines. The posteriors for these three parameters resemble Gaussians, but show significant correlations with each other. }
 \label{fig.transitfit}
\end{figure}

We report the results of our transit fits in Table~\ref{tab.tap} along with the 1$\sigma$ errors. The fit posteriors are typically near-Gaussian, but also show strong correlations between parameters (Figure~\ref{fig.transitfit}), so such errors should be used with caution.

\subsection{Host Star Assignment and Transit Dilution\label{sec.hoststar}}

{\it Kepler} pixels are 3.98\arcsec\ across, and photometric data is extracted typically using apertures 2-4 pixels on a side (depending on the target and crowding). As a result, a given light curve can contain flux from multiple stars which dilutes the transit signal, causing the estimated planet-to-star radius ratio to be erroneously low.   {\it Kepler} Pre-search Data Conditioning \citep[PDCSAP,][]{Stumpe2012} removes contaminating light from nearby stars using their estimated $K_P$ magnitudes from the KIC, but this is ineffective if the stars are not resolved in the KIC ($\lesssim2\arcsec$). For KOIs with such companions, transit depths from \citet{Swift2015}, which are based on PDCSAP data, will be underestimates.  The identity of the host star also matters to planet parameters if the stars within a system are not identical.  Moreover, if planet occurrence is calculated per primary star rather than per system then planets found to orbit companions must be excluded.

For each KOI with one or more additional stars within the {\it Kepler} aperture that are identified in the KIC and are bright enough to affect the light curve more than statistical errors ($\Delta K_P\lesssim3$) we refit the \kep{} light curves using TAP. Dilution was included as an additional parameter following the method from \citet{Johnson2011}. Dilution in the \kep{} bandpass was determined (with errors), along with companion stellar parameters from the primary stellar parameters and $\Delta K'$ values using the GP analysis described in Section~\ref{sec.binaries}, which was incorporated as a prior in the TAP fit. This derivation of $\Delta K_P$ assumes that the nearby star is a true companion, however, based on our background star calculations (Section~\ref{sec.binaries}) this is likely to be case for all or nearly all systems we refit. Background stars are much more likely to be $>2\arcsec$ away from, and much fainter than the primary, and hence are generally either identified in the KIC (and removed in PDCSAP data) or have a negligible effect on the transit depth. 

Potentially a more significant problem than transit dilution is identification of the host star. If two or more stars land within the \kep{} aperture the planet may potentially orbit any of them. We used two methods to identity the host star: inspection of the \kep{} difference images \citep{Bryson2013}, and comparison of the stellar densities derived from the transit light curve \citep{Seager:2003lr} to those determined from the GP analysis (Section~\ref{sec.binaries}).

We ignored nearby stars that are too faint to reproduce the light curve unless the transiting object is much larger than a planet ($>3R_J$). While it possible that the transit signal is from a background eclipsing binary, such systems typically exhibit characteristic signs in their light curves (e.g. V-shaped transit, secondary eclipse), which can be detected by the \kep{} vetting process. Comparison with the astrophysical FP estimates \citep{Morton2012,Swift2015} suggests this accounts for less than 3\% of systems. 

Difference images were constructed by subtracting stacked in-transit images from stacked out-of-transit images, resulting in an image of the transit \citep[see][for more information]{Bryson2013}. We compared the centroid of the difference image to the KIC coordinates after correcting for proper motions (where available) between the \kep{} observations (2009-2012) and the KIC epoch \citep[taken from 2MASS][]{Skrutskie:2006lr}. This was sufficient to identify the source of the transit signal from systems with separations $\ge 2$" (and in some cases smaller separations).  In all cases the signal was associated with the primary star. This left 16 tight systems with ambiguous parent star identifications. 

We refit each of these 16 systems twice, each time assuming a different parent star (KOI~2626 is a triple system and hence fit three times). For each fit we adjusted the dilution and limb-darkening priors based on the assumed host star. We adopted a uniform prior on stellar density, and forced eccentricity to 0, but otherwise the fits are unchanged from the method outlined in Section~\ref{sec.transitfit}. As with the efforts to remove dilution this assumes the nearby stars are true companions, but as before this is likely always the case as background stars are generally too faint to produce the transit signal or separated enough from the primary to be ruled out by the difference images.

\begin{figure*}
\includegraphics[width=84mm]{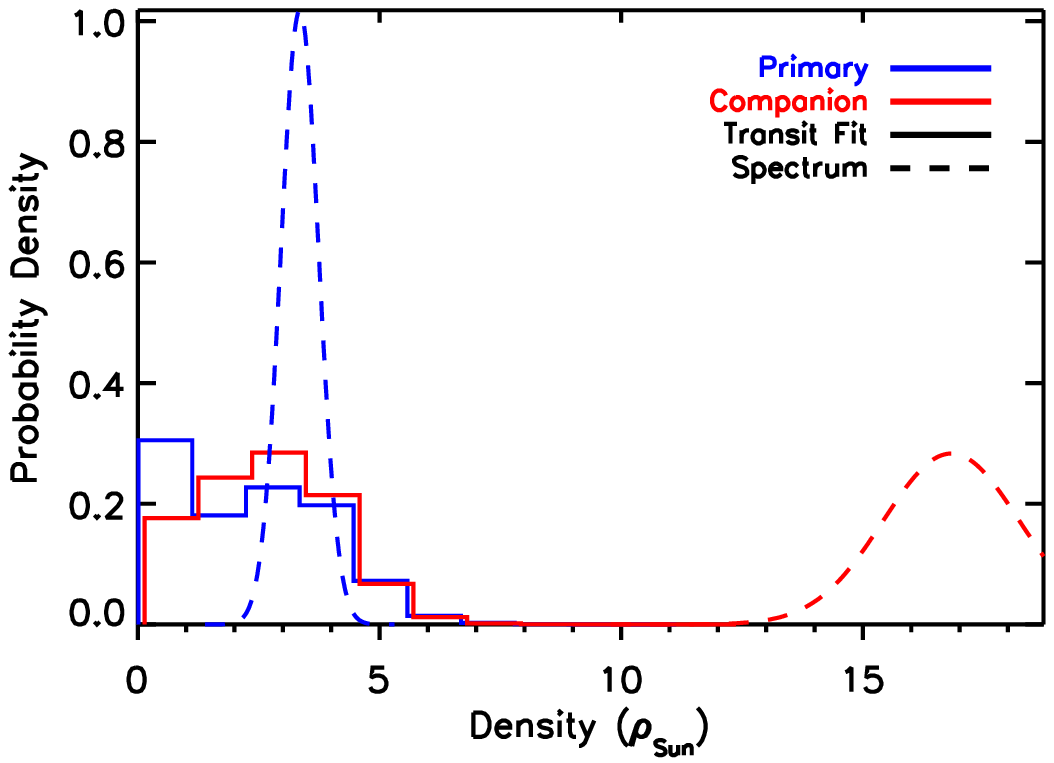}
\includegraphics[width=84mm]{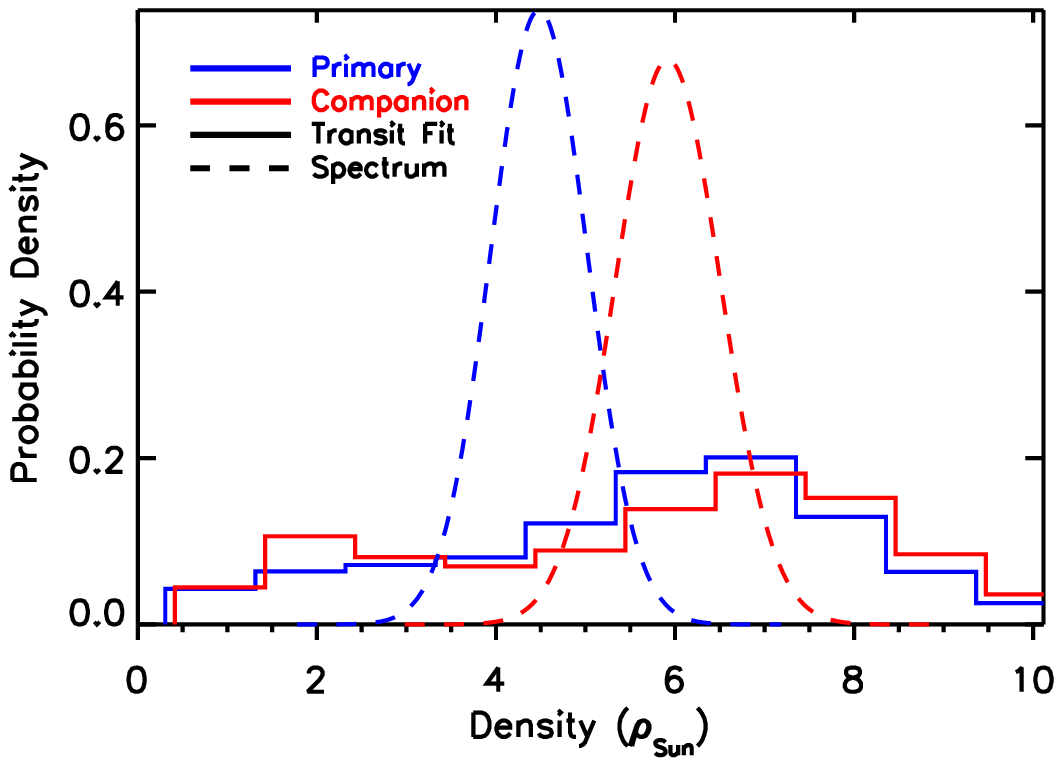}
\caption{Posterior probability density as functions of stellar density ($\rho$) for KOI 3284 (left) and KOI 2179 (right). The dashed lines are the posteriors derived from analysis of the spectroscopy and photometry (Section~\ref{sec.binaries}, while the solid lines are derived from the transit light curve fit. In the case of KOI~3284 the planet is around the primary, while for KOI~2179 the host star is ambiguous.}
 \label{fig.binarypos}
\end{figure*}

We took the posterior output chains from each TAP fit and computed the stellar density posterior (Fig. \ref{fig.binarypos}) following the formula from \citet{Seager:2003lr}. Because our fits assumed $e=0$, we multipled these densities by an eccentricity correction assuming a Rayleigh distrubtion with mean $e=0.05$ \citep[consistent with the Solar system and small explanets,][]{Van-Eylen2015}. For both the primary and companion(s) we then computed the probability that the value of the stellar density at each link in the MCMC is consistent with the spectroscopic/GP density with the formula:
\begin{equation}
S_i = \frac{1}{\sigma_{\rho_{GP}}\sqrt{2\pi}}\rm{exp}\left(\frac{-(\rho_{Tr,i}-\rho_{GP})^2}{2\sigma_{\rho_{GP}}^2}\right), \\
\end{equation}
where $\rho_{GP}$ is the density of the star from the spectrum or Gaussian process with associated error $\sigma_{\rho_{GP}}$, and $\rho_{Tr,i}$ is the stellar density derived from the transit for link $i$. $S_i$ is computed for the primary and companion separately.  Assuming that the transit signal can only arise from one of the stars considered (so the sum of the probabilities must be one) then the probability that a planet orbits the primary ($P_{prim}$) is
\begin{equation}
P_{prim} = {\displaystyle \frac{\sum_iS_i,prim}{\sum_iS_i,prim+\sum_iS_i,comp}}.
\end{equation}
The same calculation can be done for the companion ($P_{comp}$, or any star) by swapping the $S_i$ for the primary star with that of the companion. This can also be generalized to more than two components by adding the relevant term to the sum in the denominator.  For our sample, two stars were sufficient for all but KOI~2626.  We report the derived transit parameters (period, impact parameter, transit duration, and $R_P/R_*$) for each of the stars separately in the bottom of Table~\ref{tab.tap}, along with the probability that the planet orbits a given component in the binary. 

These parent star assignments are only statistical estimates; it may be that an individual system is unusually eccentric and thus deviates significantly from the assumed Rayleigh distribution. Our analysis also assumes that all candidate planets in a detected system orbit the same star. In at least one system (KOI 3444) this does not appear to be the case (see Section~\ref{sec.special}).

%% file: occurrence.tex
\section{Planet Population Inference \label{sec.inference}}

We calculated the occurrence {\it per primary star}, i.e. we excluded any planets determined to orbit a putative companion star, as determined in Section \ref{sec.hoststar} (we are obviously excluding the companion stars from our statistic as well).   Since the determination of the host star is probabilistic, we weighted each planet by the probability that it orbits this primary.  This turned out to be a relatively minor correction: the probabilities that the planets orbit a companion sum to $\sim 4$\% of the total planet tally.  We used the planet properties determined in Section \ref{sec.planets}, calculated by combining the revised stellar parameters from Section \ref{sec.stars} with the revised transit parameters from Section \ref{sec.planets}, accounting for known stellar multiplicity (Section \ref{sec.hoststar}).  

\subsection{Method of Iterative Monte Carlo \label{sec.imc}}

The intrinsic distribution of planets with radius and orbital period was calculated using the method of sequential Monte Carlo with bootstrap filtering \citep{Cappe2007}, which we call the Method of Iterative Monte Carlo (MIMC).  This method treats planets as discrete objects, does not rely on binning of data \citep[e.g.,][]{Howard2012}, and readily permits the incorporation of probabilistic (prior) descriptions of planet or stellar properties.  It was previously employed in calculating the distribution of planets around solar-type stars \citep{Silburt2015} and \kep{} M dwarfs \citep{Gaidos2014}.  Briefly, the process of inference begins with a large trial population of simulated planets on isotropically oriented orbits, parameters distributed over the range of possible values, and placed uniformly among stars.  A simulation of \kep{} detections that accounts for both the geometric probability of a transiting orbit and the signal-to-noise of any transits is performed on this population.  Systems that are ``detected" are then replaced with randomly selected actual detections.  The revised trial population is shuffled among the host stars and the process is repeated until simulated detections are statistically indistinguishable from the actual set of detections.  After convergence, the trial population represents the intrinsic population of planets, to the degree that detection by \kep{} is accurately modeled.

\kep{} pipeline detection efficiency was modeled as a function of the Multiple Event Statistic (MES), which describes the statistical significance with which the data can be described by a box-shaped model transit signal, and is approximated by
\begin{equation}
\label{eqn.snr}
{\rm MES} \approx 8.4 \times 10^{-5} \left(\frac{R_p}{R_*}\right)^2 \sqrt{\frac{90~d}{P}\sum\limits_n \frac{1}{\rm {CDPP}(\tau)_n^2}},
\end{equation}
where the CDPP$(\tau)$ is the value of the Combined Differential Photometric Precision in the $n$th \kep{} observing quarter interpolated or extrapolated from 3, 6, and 12~hr intervals to the transit duration via a linear fit vs. inverse square root of the duration.  The summation is over all quarters in which a star was observed.

\citet{Christiansen2015} performed injection-and-recovery tests with artificial transit signals in the lightcurves of 10,000 \kep{} stars and found that the detection efficiency was about half the ideal value of 50\% at MES = $7.1\sigma$, and could be described by a sigmoidal function of MES with parameters that depend on stellar type.  There are 96 injection-and-recovery experiments on the M dwarfs in our sample, and we found the parameters of a logistic function $1/(1 + \exp(-({\rm MES}-{\rm MES}_{\rm 50\%})/\sigma))$ that maximized the likelihood:  MES$_{\rm 50\%} = 8.98$ and $\sigma = 1.43$.  This best-fit logistic curve lies between those for FGK dwarfs and M stars (mostly giants) in Fig. 3 of \citet{Christiansen2015}.  A second set of injection-and-recovery experiments was performed by the \kep{} project to account for changes between the Q1-16 and the DR24 pipelines\footnote{http://exoplanetarchive.ipac.caltech.edu/docs/DR24-Pipeline-Detection-Efficiency-Table.txt}.  We identified 2272 M dwarfs in this experiment and found best-fit values MES$_{\rm 50\%} = 10.01$ and $\sigma = 1.46$.  We found a negligible difference between the best-fit values for experiments with $P < 40$d and with $40 < P < 180$d orbits.  These detection efficiencies were incorporated into MIMC as follows: A Monte Carlo planet was detected if a random number drawn from a uniform distribution was less than the best-fit model detection efficiency for the estimated MES. 

To accurately model the uncertainties in planet radii, the nominal radius of each planet in the pool of actual detections was replaced by up to 1000 values with a dispersion representing the uncertainty in radius, specifically the uncertainties in $R_*$ and $R_p/R_*$, taken to be independent and added in quadrature.  The values of $R_p/R_*$ were randomly drawn from the posteriors generated during transit light curve fits (Sec. \ref{sec.transitfit}), either from our own fits, or provided by J. Swift (private communication).  Values of $R_*$ were drawn from a Gaussian distribution.

As described in \citet{Silburt2015}, the radius of a planet near the detection limit of \kep{} is a biased value.  Given a finite uncertainty in radius and a detection bias towards larger planets, the value is more likely to be an underestimate of a larger value than an underestimate of a smaller value because the latter is less likely to be detected and appear in the catalog.  This effect can be approximately accounted for by taking each of the values from the distribution described above and calculating the fraction of all target stars around which the planet with that radius (and orbital period) would have been detected.  The distribution of possible radii of a small planet near the detection threshold becomes a truncated, strongly asymmetric distribution.  The radius distribution of a large, easily-detected planet is unaffected.

\begin{figure}
\includegraphics[width=84mm]{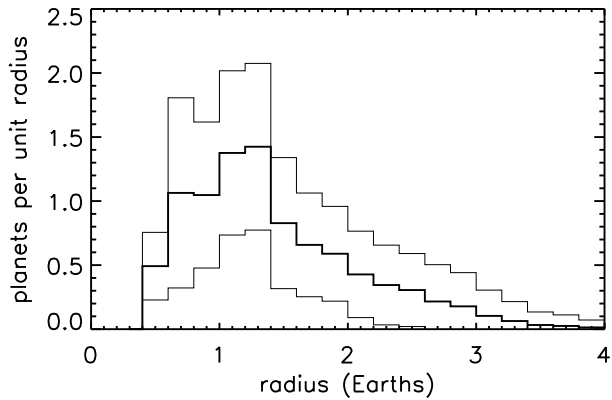}
\caption{Intrinsic radius distribution of \kep{} M dwarf planets.  The grey lines are $1\sigma$ intervals as established by Monte Carlo bootstrap calculations.  There are no detections in the smallest radius bin.}
 \label{fig.radius}
\end{figure}

\begin{figure}
\includegraphics[width=84mm]{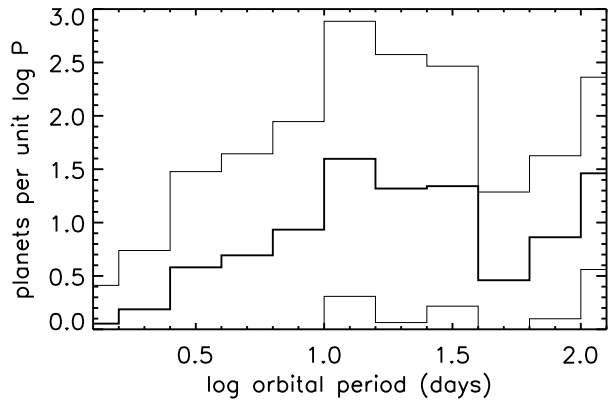}
\caption{Intrinsic period distribution of \kep{} M dwarf planets.  The grey lines are $1\sigma$ intervals as established by Monte Carlo bootstrap calculations.}
 \label{fig.period}
\end{figure}

Errors in the inferred distributions were calculated by the Monte Carlo bootstrap method.  For each bootstrap iteration, the total size of the simulated sample of detections was drawn from a Poisson distribution with mean equal to the actual size; the simulated catalog was then drawn by sampling with replacement.  The inferred distribution was calculated as described above, and standard errors were constructed by computing the variance of the number of planets at a given radius over all simulated samples.  Figures \ref{fig.radius} and \ref{fig.period} plot the inferred radius and period distributions of planets on with $P = 1.2 - 180$~d.  There are no planets smaller than 0.2\rearth{} in the sample so the smallest radius bin has no significance.  The total occurrence for 1-4\rearth{} planets using the \citet{Christiansen2015}-based detection efficiency model is $f = 1.9 \pm 0.3$.  Using the detection efficiency model based on the DR24 experiments we found $f = 2.2 \pm 0.3$.  The difference is only $1\sigma$ but we adopted the latter value as more accurate.  

\subsection{Planet Clustering and Coplanarity \label{sec.coplanar}}

\begin{figure}
\includegraphics[width=84mm]{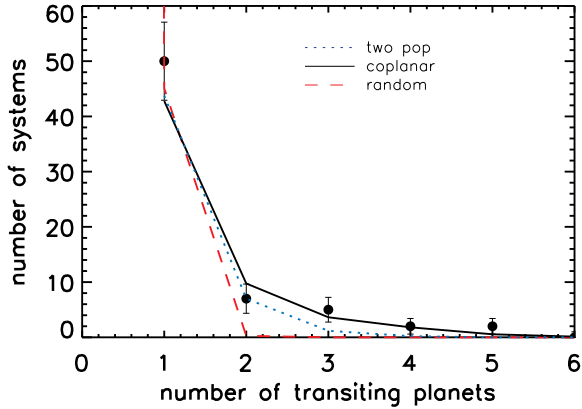}
\caption{Distribution of detected \kep{} M dwarf systems among single- and multi-planet systems.  The red dashed line is a model which matches the overall occurrence and assumes the planets are Poisson-distributed among stars and on isotropically-oriented orbits.  The blue dotted line assumes a combination of single planet systems and multi-planet, near coplanar systems with a constant occurrence of planets; the solid black line is a single population model with exponentially-distributed numbers of planets on nearly coplanar orbits.}
 \label{fig.distribution}
\end{figure}

A full description of the occurrence of planets should include a description of their non-Poisson correlated distribution  among stars \citep{Swift2013,Muirhead2015,Ballard2015} and the coplanarity of their orbits \citep{Lissauer2011,Fang2012}.  Orbital coplanarity will reduce the number of detected systems compared to the isotropic case, but the number of transiting planets per system will be higher.  Figure \ref{fig.distribution} shows the distribution of the 66 single and multi-planet systems comprising 97 planets with orbital periods between 1.2 and 180~d.    Our sample includes 50 singletons, 7 systems with two detected planets, 5 with three planets, and two each with four and five planets.  The errors are based on Poisson statistics and thus are overestimates if planets are clustered.  

We compared the observed distribution with calculated distributions based on assumptions about the clustering and coplanarity of planets.  We assumed that the period and radius distribution of planets is independent and separable from the distributions of occurrence and mutual inclinations, as well as the properties of the host star.  This was in part a practical requirement because the small size of the sample of detected planets prohibits too many divisions.  This assumption allowed us to calculate the distribution of transiting planets without regard for detection completeness corrections because these can be factored out of the equations describing the distribution of planets in multi-planet systems.  These calculations used information on the densities of the host stars and the distribution with orbital period, as these affect the transit probability.  The predicted absolute planet occurrence - one of the parameters of the models - is thus incorrect because detection efficiency is not accounted for.  However, the distribution among single- and multi-planet systems is still accurate.

In each of these calculations, we simulated $10^5$ systems with the numbers of planets drawn from a specified distribution and placed randomly around the stars in our entire \kep{} dwarf sample.  Orbital periods are drawn from the distribution inferred in Sec. \ref{sec.imc}.  For coplanar systems, the inclination of the normal of a  common (mean) plane to the line of sight $\theta$ was selected from an isotropic distribution.  Orbital inclinations $\delta$  relative to the common plane were selected from a Rayleigh distribution\footnote{In the limit of small $\delta$, the random walk of an orbit normal with respect to the normal of the common plane becomes a two-dimensional walk with $\delta$ as the distance from the origin and described by a Rayleigh distribution.} with dispersion $\sigma$ and an ascending node $\omega$ drawn from a uniform unit distribution over [0,$2\pi$].   For small mutual inclinations, and $\theta$ near $\pi/2$ (the condition for transits to occur), the minimum angular separation between star and planet as seen by the observer is
\begin{equation}
b = \delta \cos \omega + \pi/2 - \theta.
\end{equation}
The condition for a transiting orbit is 
\begin{equation}
b < 0.238 \rho_*^{-1/3} P^{-2/3},
\end{equation}
where the stellar density $\rho_*$ is in solar units and $P$ is in days.  

We considered several combinations of planet distribution and orbital geometry.  The first and simplest scenario has a uniform number $N$ of planets per star on isotropically-oriented orbits.  Setting $N = 1$ minimizes $\chi^2$ to be 63.3 ($\nu = 6$) but this model can be rejected out of hand because it does not explain multi-planet systems.  In the second scenario, planets are Poisson-distributed among stars and placed on isotropically-oriented orbits and there is a single fitting parameter, mean occurrence, chosen to reproduce the total number of detections.  This scenario provides a better fit ($\chi^2= 15.3$, $\nu=6$, red-dashed line in Fig. \ref{fig.distribution}), but significantly underestimates the number of multi-planet systems, as found by \citet{Ballard2015,Muirhead2015}.  The best-fit occurrence is 0.50, an under-estimate because the detection efficiency of \kep{} for the smallest, numerous planets is $\ll 1$, even around M dwarfs.  We also considered two other scenarios where the planets are nearly co-planar.  The best-fit co-planar model with a Poisson distribution of number of planets per star gives $\chi^2 = 13.2$ ($\nu = 5$).  However, an exponential distribution of planets per star yields a far superior fit ($\chi^2 = 3.2$; solid line in Fig. \ref{fig.distribution}).\footnote{An exponential distribution differs from the Poisson distribution in that the variance is equal to the square of the mean, rather than equal to the mean.}  The dispersion in the mutual inclinations is only 0.25 deg.

Previous analyses of solar-type \kep{} stars have likewise found that the best-fit single-population models for planets around solar-type stars significantly under-predict the number of single-transit systems \citep{Lissauer2011,Hansen2013}.  A distinct population of systems with single planets or large mutual inclinations was proposed; \citet{Ballard2015} estimated that about half of all M dwarf systems are members of this population. The dotted blue curve in Fig. \ref{fig.distribution} shows the best-fit model ($\chi^2 = 6.0$, $\nu =4$)  is a mixture of 69\% coplanar (mutual inclinations $\sim 0.2$ deg), and 31\% planets with isotropic orbits.  However, as we have shown, two-population model ares neither unique nor necessarily the best explanation for the observed distribution.  

We tested the effect of clustering and orbital coplanarity on occurrence by determining the {\it ratio} of the mean occurrence of planets for the chosen best-fit model to that of the Poisson/isotropic model.  The ratio of the mean occurrences is independent of the \kep{} detection efficiency, but only if the period/radius distribution of planets are independent of the coplanarity and clustering.  In the case of the coplanar/exponential planet distribution model this ratio was found to be indistinguishable from unity: coplanar systems are less likely to intersect the line of sight than a collection of independent isotropically-oriented orbits, but will produce more transiting planets if they do.

%% file: discussion.tex
\section{Analysis and Discussion} 
\label{sec.discussion}

\subsection{The M Dwarf Planet Population \label{sec.planetpop}}

\begin{figure}
\includegraphics[width=84mm]{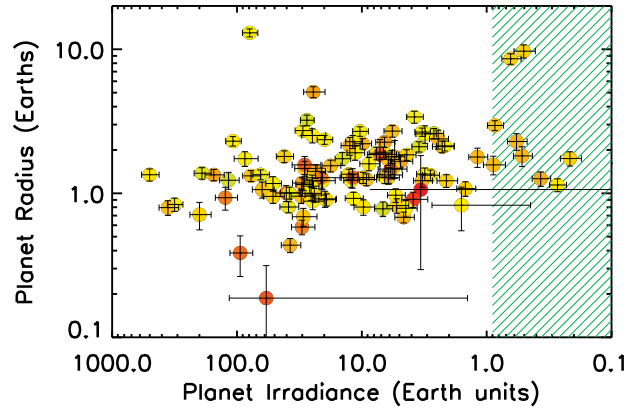}
\caption{Radius vs. irradiance (in terrestrial units) of candidate and confirmed planets orbiting the \kep{} M dwarf sample.  Points are color-coded by the \teff{} of the host star, with yellow representing 4000~K and deep red 3000~K.  The green hatched area is a nominal "habitable zone" with an inner edge defined by the conservative runaway greenhouse condition of \citep{Kopparapu2013}.}
 \label{fig.radirrad}
\end{figure}

Estimated planet radius vs. orbit-averaged irradiance for all 106 candidate planets around M dwarfs are reported in Table \ref{tab.planets} and plotted in Fig. \ref{fig.radirrad}.  The small effect of non-zero orbital eccentricity was neglected in calculations of irradiance.  The vast majority of planets experience stellar irradiance well in excess of the terrestrial value, but 12 objects (KOIs 571.05, 854.01, 1422.04, 2418.01, 2626.01, 3263.01, 3444.02, 4290.01, 4427.01, 4463.01, 5416.01, 7592.01) have values no more than $1\sigma$ above 90\% of the terrestrial value (green hashed zone), a nominal runaway greenhouse limit for Earth-like planets orbiting M dwarfs \citep{Kopparapu2013,Mann2013}.  

We find that, on average, each M dwarf hosts $2.2 \pm 0.3$ planets with radii of 1-4\rearth{} on orbits with $P < 180$~d.  Our estimate within one standard deviation of previous estimates by DC13, \citet{Gaidos2014}, \citet{Morton2014}, and DC15.   The major differences between our analysis and that of DC15 is our use of empirical rather than model-based stellar parameters and our employment of a simple function with signal-to-noise ratio (SNR, or multiple event statist, MES) to describe the detection efficiency calculated by numerical transit injection and recovery experiments, rather than the experiments themselves.  However, \citet{Silburt2015} found that a single SNR criterion approximately reproduced the injection-and-recovery-based estimates of \citet{Petigura2013} over a wide range of $P$ and $R_p$.  (The exception was at $P \sim 1$~yr comparable to the three-transit limit, which is not relevant to the range we consider, and only for the very smallest planets near the detection threshold.)  Our figure for the uncertainty in the occurrence includes the effect of random errors in the stellar and planet parameters, but does not include systematic errors in the detection efficiency of the \kep{} mission.

We find that the distribution of M dwarf planets can be matched if planets orbit in the same plane, with negligible mutual inclinations ($\sim 0.2^{\circ}$), and if planets are clustered, i.e. the system-to-system variance is larger than Poisson.  An exponential probability distribution works well, but this description may not be the only adequate one.  One possible explanation for an exponential distribution is that planet formation is a sequential process, e.g. see \citet{Chatterjee2014}.  The dynamical "coolness" of M dwarf planetary systems might be due to the paucity of Neptune- and Jupiter-size planets, at least on close-in orbits \citep{Johnson2010,Gaidos2014b}, that would tend to dynamically excite the system. 

Multiple studies of \kep{} M dwarfs, including this one, have found an elevated number of Earth- to Neptune-size planets relative to the numbers around solar-type stars \citep[e.g.,][]{Howard2012,Mulders2015}.  This trend of increasing small-planet occurrence with decreasing \teff{} runs counter to a possible trend of {\it decreasing} giant planet occurrence \citep{Johnson2010,Gaidos2013b,Gaidos2014b} but since the occurrence of the latter is much lower than the former, the elevated numbers of small planets cannot be the product of a population of ``failed Jupiters" \citep[e.g.,][]{Adams2004,Mann2010}.  One possible explanation is that the protoplanetary disks around young M dwarfs are more compact than their counterparts around solar-type stars, spawning more planets at shorter orbital periods.  Another possible explanation is that planet migration is somehow more efficient in M dwarf disks; the existence of multi-planet systems, particularly those with planet pairs near mean-motion resonances, suggests that migration is important \citep{Muirhead2015,Mulders2015}.

Our inferred intrinsic planet radius distribution peaks at $\approx 1.2$\rearth{} and declines at smaller and larger radii.  The radius distribution is essentially zero by 4\rearth{} and larger planets appear to be exceptionally rare close to M dwarfs.  A notable exception is the hot Jupiter Kepler-45b \citep[KOI~254.01,][]{Johnson2012}.  We estimated that the host star has a radius of 0.63$R_\odot$ and thus the planet is slightly larger than previously estimated: 1.16$R_J$, increasing the likelihood that it is inflated relative to predictions by evolution models.  In addition to Kepler-45b, we identified two other candidate Saturn-size planets; KOI~3263.01 and 5416.01, as well as a 5\rearth{} "super-Neptune" (KOI~4928.01).  KOI~3263.01 is discussed further in Sec. \ref{sec.special}.  We caution that it is possible that the paucity of giant planets around \kep{} M dwarfs is in part due to their very large transit depths causing mis-identification as eclipsing binaries, as was the case of KOI-254 \citep{Slawson2011}.

A significant difference in our radius distribution compared to that of DC15 is a decrease at smaller radius, whereas DC15 found a plateau like that found by similar analyses of solar-type stars \citep[e.g.,][]{Petigura2013}.  We attribute this to our de-biasing of the radius distribution for individual detections, i.e. we account for the likelihood that radii are more likely to be underestimates of larger values than overestimates of smaller values.  The behavior of the distribution at small radii also depends on the model of the \kep{} detection efficiency.  We experimented with different descriptions of the detection efficiency, including a step function at MES=7.1, and the gamma distributions of \citet{Christiansen2015} for M stars and FGK stars.  We find that the exact shape of the radius distribution depends on the assumed detection efficiency, but the peak and the turnover at smaller radii are robust.

The distribution with $\log P$ is approximately flat except for a decline at orbital periods less than a few days, and possibly at $\sim 50$~d.  The former is consistent with all previous analyses of both solar-type stars and M dwarfs \citep[e.g.,][]{Howard2012,Petigura2013,Silburt2015} and presumably reflects some inefficiency of planet formation or migration very close to host stars.   

\begin{figure}
\includegraphics[width=84mm]{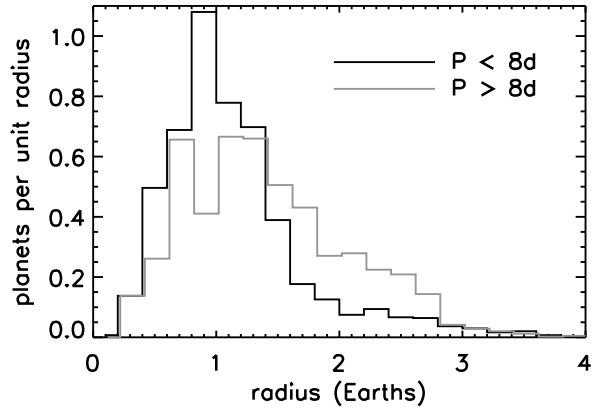}
\caption{Normalized distributions with radius of planets on orbits with $P< 8$~d (black) and $P>8$~d (grey).  There are no detections in the smallest radius bin.  The two histograms are slightly offset for clarity.}
 \label{fig.radper}
\end{figure}

Our method of inferring the intrinsic planet population assumed that the distributions of planet radius and orbital period are independent and separable.  However, there are theoretical expectations that planets on wider orbits will retain envelopes of low molecular weight volatiles and have larger radii \citep[e.g.,][]{Owen2013,Lopez2015}.  To evaluate the possible dependence of radius distribution with orbital period, we split the KOI sample into two equal-sized samples for short ($< 8$~d) and long ($> 8$~d) periods.  We calculated the intrinsic radius distributions from both subsamples and the normalized versions are compared in Fig. \ref{fig.radper}.  The corresponding values of occurrence are 0.34 and 1.59 but the detection efficiency of short-period planets is much higher.  The distributions suggest that there are comparatively more super-Earths on $P>8$~d orbits than at shorter periods, but each of these distributions has large uncertainties (not shown for clarity but see Fig. \ref{fig.radius}).  
To evaluate whether the distributions are significantly different relative to these uncertainties, we calculated a two-sided Kolmogorov-Smirnov (K-S) $D$ value of 0.23 between the short-period and long-period distribution.   We then repeated this with 1000 pairs each of short-period only, and long-period only distributions selected from the bootstrap Monte-Carlo calculations.  For each value of $D$ from the short-long K-S comparison we calculated the fraction of short-short or long-long $D$ values that were above this value; this is the probability that the short- and long-period distributions are actually both drawn from the same distribution and have a $D$ value at least as large as 0.23 simply due to random fluctuations.  These probabilities are 0.07 and 0.28, respectively, and thus the short-period and long-period distributions are only marginally distinguishable.  Our assumption that the radius distribution is period-independent for $P<180$~d is, to first order, correct, but a larger sample is required for a more definitive answer.   

While the relationship between the metallicity of dwarf stars and the occurrence of close-in giant planets, including around M dwarfs is clearly established \citep{Gaidos2014b}, any connection between metallicity and smaller planets is much more controversial. \citet{Sousa2008} found no relationship for Neptune-size planets around stars in a large high-precision Doppler survey. \citet{Mann2013b} found no significant difference between spectroscopy-based metallicities for \kep{} M dwarf planet hosts vs. the background population. In contrast, \citet{Buchhave2014} used a sliding Kolmogorov-Smirnof (K-S) statistic to establish three planet radius ranges with different metallicity distributions for solar-type stars. \citet{Schlaufman2015} criticized \citet{Buchhave2014} on statistical grounds and concluded that the data support a continuous variation of the metallicity-dependence of planet occurrence with radius, rather than distinct ``regimes''. \citet{Wang2015} adjusted photometry-based metallicities using spectroscopic values and claimed that the occurrence of Earth- to Neptune-size planets around metal-rich stars is a factor of two higher than average.        

\begin{figure}
\includegraphics[width=84mm]{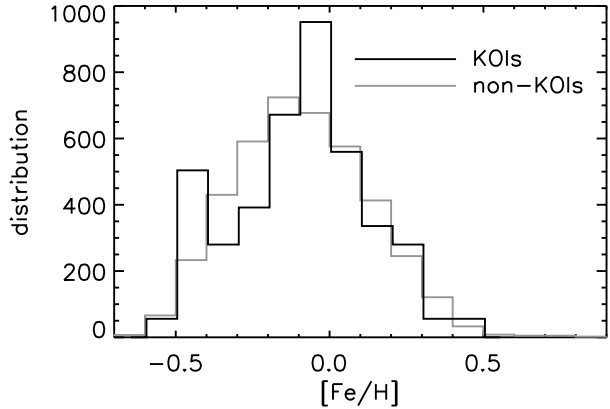}
\caption{Distributions of [Fe/H] for \kep{} M dwarfs with (black) and without (grey) detected transiting planets.  The two histograms are slightly offset for clarity.}
 \label{fig.planetmetal}
\end{figure}

Our M dwarf sample includes a preponderance of systems with small planets suitable for such a comparison, although it may be hampered by the large uncertainties associated with photometry-based metallicities.  Figure \ref{fig.planetmetal} plots the distribution of estimated [Fe/H] for stars with (solid line) and without (dashed line) detected exoplanet hosts, after adjusting for the offset found between photometric and spectroscopic values (Appendix C).  The distribution of the host-star sample is scaled to the size of the non-host star sample for ease of comparison.  A K-S comparison of the two distributions suggests they are the same ($D = 0.083$, $p = 0.68$).  For comparison we calculated 10,000 values of $D$ by constructing Monte Carlo versions of [Fe/H] values adjusted by adding errors according to normal distributions, and comparing host-star-sized subsamples to the non-host star sample.  The $p$ value given by the fraction of K-S $D$ values above 0.072 is similar (0.66).  These finding agree with those of \citet{Mann2013b}, i.e. the metallicity distribution of M dwarf planet hosts (almost exclusively small planets) is indistinguishable from that of the overall population.  We rule out any mean differences between the two sets of stars $>0.064$ dex with a confidence of 99\% (K-S probability $<0.01$).

\begin{figure}
\includegraphics[width=84mm]{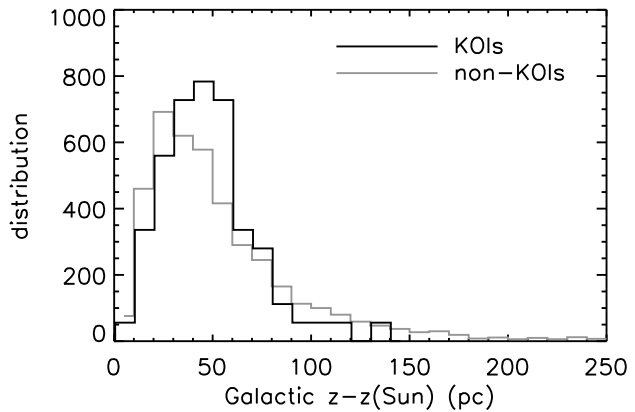}
\caption{Distributions of height above the Galactic plane (relative to the Sun) for \kep{} M dwarfs with (black) and without (grey) detected transiting planets.  The two histograms are slightly offset for clarity.}
 \label{fig.planetglat}
\end{figure}

Stellar age may also be an underlying determinant of planet occurrence.  (Age and metallicity are weakly related except for the oldest stars).  As stars become older their space motions increase and they spend more time away from the natal Galactic plane.  \citet{Ballard2015} reported that multi-transiting planet systems among \kep{} M dwarfs tend to be closer to the mid-plane of the Galaxy compared to single-planet systems.  Figure \ref{fig.planetglat} plots the distribution of distance above the Galactic plane relative to the Sun for M dwarfs with (solid line) and without (dashed line) detected transiting planets (the former is scaled up).  Stars with detected planets tend to appear closer to the plane than those without.  A K-S comparison of the two samples has a $D$ parameter of 0.18 and $p = 0.015$ that the two samples are drawn from the same population).  The significant issue here is that of detection bias; the \kep{} field limits the range of Galactic latitude observed and stars closer to the mid-plane are likely to be closer, brighter and thus more amenable to planet detection.  We controlled for this by controlling for apparent $r$-band magnitude (similar to the \kep{} magnitude for M dwarfs).  We constructed a new non-host sampling by selected with replacement from the original, retaining ony those stars with magnitudes within 0.1 magnitude of a randomly chosen host star.  The $D$ value for the comparison between this sample and the host sample is only 0.11 and $p=0.28$.  Thus, after proper control of bias, there is no significant evidence from our catalog that M dwarf host stars of detected transiting planets are preferentially closer to the Galactic plane.  

\subsection{Systems of Special Interest \label{sec.special}}


{\it KOI~2626:} is the only planetary system in our sample in a known triple star system \citep[see also][]{Cartier2015}. Because all three stars are close on the sky (separation$\lesssim2\arcsec$) and have similar brightness ($\Delta K \lesssim1$), the probability that any are background stars is essentially zero. Unfortunately, the stellar densities based on the analysis of photometric colors are similar for all three stars, so we were unable to unambiguously determine around which star the planet orbits, although the primary is favored.

{\it KOI~2705:} is one of the few systems that, based on the spectroscopic and transit-inferred density, does not orbit the primary ($p > 0.99$), although this is based on the assumption of low eccentricity. The centroid of the transit signal from the in-out difference images \kep{} is inconsistent with both the primary and companion. However, once we accounted for the high proper motion of this star \citep[0.14" yr$^{-1}$,][]{Deacon2015} and the difference in epochs between the KIC (i.e., 2MASS) and \kep{} observations ($>10$ yr), the location of the transit signal is consistent with both, but favors the companion (2.0$\sigma$). To reach consistency between the spectroscopic and transit-fit density the planet would need an orbital eccentricity $\gg0.2$. The planet has a tidal circularization timescale of $\ll1$\,Gyr, while the lack of H$\alpha$ or significant activity in the light curve or spectrum suggests the star is much older and hence probably orbiting the companion. However, this ignores the influence of the second star and undetected planets. We conclude that this is either a mini-Neptune sized (2.3$\pm$0.6\rearth) planet around a mid M-type dwarf (\teff{} = 2938$\pm$70\,K) or a super-Earth sized ($1.4\pm0.2$\rearth) eccentric planet orbiting an early M-type dwarf.

{\it KOI~3263:} Taken at face value, this object is a Jupiter-size planet on a $\sim 77$~d orbit in the habitable zone of a mid M-type dwarf.  However, \citet{Swift2015} assigned the single candidate planet a FP probability of 71\% based on the shape of the transit light curve and the ratio of the prior probability of the transiting planet scenario vs. various astrophysical FP scenarios.  The light-curve is V-shaped, but this could be the product of a grazing (high impact parameter) transit rather than an eclipsing binary, although a significantly non-zero eccentricity is required to keep the transit duration and orbital period consistent.  KOI~3263 was resolved into a binary system ($\theta = 0.82$", $\Delta K' = 2.28$).  If the transiting objects is around the 3000~K secondary then the transiting object must be significantly larger than Jupiter and on a highly eccentric orbit and thus likely a third star or brown dwarf.  Given the scarcity of transiting giant planets around M dwarfs, especially giant planets with cool equilibrium temperatures, Doppler observations of this object should be undertaken. The expected Doppler amplitude is about 180 m~s$^{-1}$ for a Jupiter mass and near-circular orbit, but the faintness ($K_P\simeq15.9$) makes it a challenging target.    

{\it KOI~3444:} There were three other sources detected in high-resolution imaging of the primary, but the two faintest have appreciable probabilities of being background sources \citep[see also][]{LilloBox2014}.  Our initial fit of the light curves from the four planets in this system yielded stellar densities more consistent with the possible companion than the primary. However, this would require that two of the four KOIs may orbit too close ($P$ of 12.67 and 14.15 d) to be dynamically stable over the age of the star. If the planets orbit the companion their larger sizes and hence masses should be sufficient to induce transit timing variation (TTV) visible in the light curves, but none are seen. We therefore refit the four KOIs {\it without} assuming that they all orbit the same star. The results favor a scenario where three objects orbit the primary, and a fourth orbits the companion. The large size of the last also suggests that it is a binary star, rather than planet, consistent with its identification as a FP in the DR24 catalog.  

{\it KOI~6705:} is the smallest candidate planet in our sample, with a radius of only 0.22$^{+0.11}_{-0.14}$\rearth{}. It was only detectable because of its short period (0.995 days). If confirmed this would replace Kepler-37b \citep[0.30\rearth{},][]{Barclay2013} as the smallest known (exo)planet.  This system was investigated in detail by \citet{Gaidos2015b} who validated the transit signal but found it to vary with time and likely be a false positive produced by charge transfer inefficiency in a \kep{} detector.  

\subsection{Summary and Future Outlook}

We have classified the reddest (M-type) stars observed by the \kep{} prime mission into dwarfs and giants and estimated the properties of the dwarfs using empirical, model-independent relations with photometric colors derived from a set of nearby calibrator stars with established parallaxes (and sometimes angular radii).  We also investigated the multiplicity of 54 of the 73 M dwarfs hosting candidate planets using adaptive optics imaging, and estimated the properties of candidate stellar companions based on the contrast ratio in the near-infrared $K$-band, assuming they are at the same distance as the primary.  We combined the revised properties of the stars with fits of the \kep{} lightcurve transits to re-estimate the properties of the planets.  For systems with candidate stellar companions we calculated the probability for each component that it hosts the planet(s).  In all but 5 cases we validated the planet as orbiting the primary; in one system the planet orbits the companion and in four other cases a definitive assignment cannot be made.  Using these revised parameters, we calculated the occurrence and distributions with planet radius and orbital period: the overall occurrence is $2.2 \pm 0.3$ for $P < 180$~d,  the radius distribution peaks at $\sim 1.2$\rearth{}, and the period distribution is approximately flat with $\log P$.  The distribution of planets among systems can be explained by a {\it single} population of exponentially-distributed planets on coplanar orbits.  There is suggestive but not conclusive evidence that there are fewer larger planets close ($P<8$d) to the star, but we find no evidence that occurrence depends on metallicity or height above the Galactic plane (a proxy for age).  

Even more accurate parameters for \kep{} M dwarfs should become available in the next few years.  The ESA {\it Gaia} spacecraft is obtaining precise spectrophotometry as well as parallaxes with precisions as good as 10 $\mu$as \citep{deBruijne2012} of all \kep{} target stars, including the M dwarfs.  These can combined to generate significantly more accurate determinations of stellar radii, limited only by precise determination of bolometric fluxes and \teff{}.  The former can be generated by fitting appropriate ``template" spectra to the {\it Gaia} and KIC, KIS, and 2MASS photometry, as well as low-resolution spectra from large-scale spectroscopic surveys such as LAMOST \citep{DeCat2015}. \citet{Mann2015b} showed that {\it Gaia} BP-RP colors can be used to generate a comparatively metallicity-insensitive estimate of \teff{}, but of course this calibration should be re-performed once actual colors of standard stars are available.  Metallicities can be estimated from the high-resolution spectra obtained by the {\it Gaia} Radial Velocity Spectrometer in the vicinity of the Ca~II infrared triplet, provided the confounding effect of stellar activity can be accounted for.  

Planet radius alone is much less interesting than the combination of mass and radius.  This yields the mean density of a planet, probably the single most informative piece of information about a planet's composition. Near-infrared spectrographs will observe M dwarfs at wavelengths near the peak in their emission spectrum and can be used to detect and measure the masses of their Earth- to Neptune-size planets by the Doppler radial velocity method \citep{Artigau2014,Kotani2014,Mahadevan2014,Quirrenbach2014}.   Most \kep{} M dwarfs are too faint to be practical targets for Doppler observations, but mass measurements of planets around a separate set of nearby, brighter M dwarfs could be combined with \kep{} radii to construct a statistical mass-radius relation, assuming both samples of systems derive from the same population \citep{Gaidos2012}.

The photometric sensitivity of \kep{} has enabled us to probe the Earth-size and sub-Earth regimes of planet parameter space, and to quantitatively re-construct the overall occurrence and distributions with radius and orbital period, especially around M dwarfs.  These statistics will be discriminating tests of planet population synthesis models and hence the planet formation theory on which those models are based \citep{Benz2014}.  Of particular importance will be the models ability to reproduce the peak in the radius distribution at $\approx 1.2$\rearth{} (Fig. \ref{fig.radius}) and their consistency with the relationship between radius distribution and orbital period (or lack thereof, Fig. \ref{fig.radper}) as it relates to evaporation of H/He envelopes \citep[e.g.,][]{Sheng2014}.  

Failure of a second reaction wheel in May 2013 terminated the primary \kep{} mission, but this was succeeded by the ``K2" program of two wheel operation \citep{Howell2014}.  K2 targets include many M dwarfs; the extra noise produced by larger spacecraft pointing error in two-wheel mode is less significant for a planet of a given radius around a smaller star (at a given distance), since the transit depth but also the photon noise is larger.  Several candidate planets orbiting M dwarfs have already been reported \citep{Crossfield2015,Montet2015,Mann2015c}.  Some of these stars are brighter than \kep{} M dwarfs and may lend themselves to direct inquiry into the mass-radius relation of Earth- to Neptune-size planets, although the much shorter observation time baseline (about 81 days instead of 4 years) and the period distribution of M dwarf planets (Fig. \ref{fig.period}) suggests that the yield per star will be much lower than that of \kep{}.  The methods of M dwarf characterization described in this work may lend themselves to determination of the properties of the K2 M dwarf target catalog, at least until {\it Gaia} data become available.

%% file: dwarf_table.tex
\begin{table}
\centering
\begin{minipage}{140mm}
\caption{\kep{} M Dwarfs$^{a}$ \label{tab.dwarfs}}
\begin{tabular}{@{}rr|rr|rr|r|r|r|r@{}}
\hline
\multicolumn{1}{c}{KIC ID} & \multicolumn{1}{c}{$p_{\rm dwarf}$} & \multicolumn{2}{c}{\teff{}} & \multicolumn{2}{c}{[Fe/H]} & \multicolumn{1}{c}{$R_*/R_{\odot}$} & \multicolumn{1}{c}{$M_*/M_{\odot}$} & \multicolumn{1}{c}{$M_K$} & \multicolumn{1}{c}{$d$ (pc)} \\
\hline
  892376 & 0.993 & 3623$\pm$ 51 & PHOT & -0.26$\pm$0.09 & PHOT & 0.42$\pm$0.03 & 0.44$\pm$0.04 & 6.17$\pm$0.22 &   81 \\
 1025859 & 1.000 & 3549$\pm$ 57 & PHOT &  0.05$\pm$0.16 & PHOT & 0.45$\pm$0.04 & 0.47$\pm$0.05 & 5.97$\pm$0.29 &  260 \\
 1162635 & 1.000 & 3774$\pm$ 58 & PHOT & -0.17$\pm$0.14 & PHOT & 0.50$\pm$0.04 & 0.52$\pm$0.05 & 5.64$\pm$0.27 &  255 \\
 1292688 & 1.000 & 3735$\pm$ 57 & PHOT & -0.19$\pm$0.14 & PHOT & 0.48$\pm$0.04 & 0.50$\pm$0.05 & 5.76$\pm$0.27 &  245 \\
 1293177 & 1.000 & 3420$\pm$ 57 & PHOT & -0.49$\pm$0.14 & PHOT & 0.29$\pm$0.04 & 0.29$\pm$0.04 & 7.22$\pm$0.33 &  136 \\
 1296779 & 0.999 & 3898$\pm$212 & PHOT &  0.28$\pm$0.09 & PHOT & 0.66$\pm$0.11 & 0.68$\pm$0.10 & 4.76$\pm$0.60 &   94 \\
 1431845 & 1.000 & 3472$\pm$ 59 & PHOT &  0.12$\pm$0.12 & PHOT & 0.42$\pm$0.04 & 0.44$\pm$0.04 & 6.17$\pm$0.27 &  168 \\
 1432978 & 1.000 & 3382$\pm$ 58 & PHOT & -0.27$\pm$0.13 & PHOT & 0.30$\pm$0.03 & 0.31$\pm$0.04 & 7.05$\pm$0.30 &  116 \\
 1433760 & 1.000 & 3381$\pm$ 61 & PHOT & -0.40$\pm$0.15 & PHOT & 0.28$\pm$0.04 & 0.28$\pm$0.04 & 7.24$\pm$0.35 &  150 \\
 1569863 & 0.996 & 3479$\pm$ 16 & PHOT & -0.34$\pm$0.12 & PHOT & 0.34$\pm$0.02 & 0.35$\pm$0.03 & 6.77$\pm$0.18 &  141 \\
\hline
\end{tabular}
\end{minipage}
$^{a}$Only the first few lines are shown to indicate content and form.  The full table is available in electronic form.
\end{table}

%% file: binary_table.tex
\begin{table*}
\centering
\begin{minipage}{140mm}
\caption{Candidate Stellar Companions of KOIs \label{tab.binaries}}
\begin{tabular}{@{}rrllrlrrl@{}}
\hline
\multicolumn{1}{c}{KIC ID} & \multicolumn{1}{c}{KOI} & \multicolumn{1}{c}{$\Delta K'$} & \multicolumn{1}{c}{$\theta$ ($\arcsec$)} & \multicolumn{1}{c}{P.A. $(\deg)$} & \multicolumn{1}{c}{$\Delta K'$} & \multicolumn{1}{c}{\teff} & \multicolumn{1}{c}{$\sigma$\teff{}} & \multicolumn{1}{c}{$p_{\rm BACK}$}\\
\hline
 9390653 &  249 &  0.85 & 4.322 &  29 &  0.85 & 3359 &  26 & 0.0018 \\
10489206 &  251 &  4.17 & 3.564 & 124 &  4.17 & 2791 &  28 & 0.0362 \\
11752906 &  253 &  1.58 & 4.604 & 164 &  1.58 & 3399 &  57 & 0.0073 \\
11752906 &  253 &  5.53 & 4.746 & 172 &  5.53 & 2693 &  18 & 0.0868 \\
 5794240 &  254 &  6.11 & 6.084 & 311 &  6.11 & 2589 &  17 & 0.5191 \\
 5794240 &  254 &  6.18 & 3.295 &  69 &  6.18 & 2585 &  19 & 0.1992 \\
 5794240 &  254 &  6.75 & 2.716 &   8 &  6.75 & 2535 &  18 & 0.1784 \\
 7021681 &  255 &  3.68 & 3.378 & 358 &  3.68 & 3022 &  38 & 0.0149 \\
 8120608 &  571 &  7.09 & 3.856 & 234 &  7.09 &   0. &  0. & 0.2774 \\
 6435936 &  854 &  0.30 & 0.016 & 209 &  0.30 & 3527 &  73 & 0.0000 \\
 6435936 &  854 &  3.59 & 0.154 & 182 &  3.59 & 2832 &  25 & 0.0000 \\
 7455287 &  886 &  4.56 & 3.682 & 356 &  4.56 & 2533 &  18 & 0.0977 \\
 7907423 &  899 &  6.91 & 3.000 & 218 &  6.91 &   0. &  0. & 0.1766 \\
 9710326 &  947 &  7.30 & 5.122 & 219 &  7.30 &   0. &  0. & 0.3134 \\
 9710326 &  947 &  7.80 & 3.876 & 128 &  7.80 &   0. &  0. & 0.2329 \\
 8351704 & 1146 &  7.94 & 3.856 &  36 &  7.94 & 2513 &   1 & 0.2124 \\
 4061149 & 1201 &  4.31 & 2.784 & 239 &  4.31 & 2533 &  43 & 0.0531 \\
 4061149 & 1201 &  5.80 & 3.814 & 267 &  5.80 & 2527 &   1 & 0.2177 \\
 4061149 & 1201 &  6.29 & 5.229 & 106 &  6.29 & 2531 &   1 & 0.4488 \\
 9202151 & 1393 &  5.76 & 5.532 & 197 &  5.76 & 2552 &  22 & 0.1677 \\
11497958 & 1422 &  1.16 & 0.214 & 218 &  1.16 & 3352 &  25 & 0.0000 \\
 5531953 & 1681 &  0.07 & 0.149 & 142 &  0.07 & 3591 & 111 & 0.0000 \\
 5531953 & 1681 &  5.05 & 4.091 & 197 &  5.05 & 2539 &  34 & 0.1284 \\
 5531953 & 1681 &  6.67 & 4.986 & 251 &  6.67 & 2620 &   4 & 0.3776 \\
 5080636 & 1843 &  7.18 & 3.113 & 267 &  7.18 & 2500 &   5 & 0.0685 \\
 8367644 & 1879 &  7.68 & 5.193 & 253 &  7.68 & 2522 &   1 & 0.5119 \\
10332883 & 1880 &  4.26 & 1.690 & 102 &  4.26 & 2813 &  24 & 0.0043 \\
10332883 & 1880 &  4.28 & 1.714 & 100 &  4.28 & 2797 &  33 & 0.0045 \\
10332883 & 1880 &  7.28 & 4.940 & 188 &  7.28 &   0. &  0. & 0.1843 \\
 2556650 & 2156 &  3.96 & 3.311 & 306 &  3.96 & 2783 & 127 & 0.0417 \\
10670119 & 2179 &  0.46 & 0.133 & 357 &  0.46 & 3538 &  58 & 0.0000 \\
 5601258 & 2191 &  5.41 & 1.741 & 234 &  5.41 & 2591 &  34 & 0.0121 \\
 5601258 & 2191 &  8.64 & 5.248 & 312 &  8.64 &   0. &  0. & 0.3970 \\
10027247 & 2418 &  2.51 & 0.106 &   3 &  2.51 & 3006 &  50 & 0.0000 \\
10027247 & 2418 &  3.72 & 4.159 &  30 &  3.72 & 2794 &  18 & 0.0575 \\
10027247 & 2418 &  6.84 & 3.928 & 328 &  6.84 & 2562 &  16 & 0.2685 \\
10027247 & 2418 &  7.80 & 2.390 & 104 &  7.80 & 2543 &   1 & 0.1605 \\
 8631751 & 2453 &  7.66 & 2.123 & 353 &  7.66 & 2534 &   0 & 0.1153 \\
 6183511 & 2542 &  0.93 & 0.765 &  29 &  0.93 & 3284 &  29 & 0.0001 \\
11768142 & 2626 &  0.48 & 0.205 & 213 &  0.48 & 3487 &  22 & 0.0000 \\
11768142 & 2626 &  1.04 & 0.163 & 185 &  1.04 & 3362 &  31 & 0.0000 \\
 3426367 & 2662 &  6.08 & 0.565 &  94 &  6.08 &   0. &  0. & 0.0014 \\
11453592 & 2705 &  2.65 & 1.893 & 305 &  2.65 & 2991 &  47 & 0.0011 \\
 6679295 & 2862 & -0.00 & 0.573 &  24 & -0.00 & 3763 &  24 & 0.0001 \\
 6679295 & 2862 &  4.71 & 4.226 & 346 &  4.71 & 2702 &  16 & 0.0694 \\
 6679295 & 2862 &  6.47 & 3.405 &  56 &  6.47 & 2508 &   9 & 0.1087 \\
 5551672 & 3119 &  3.11 & 6.418 & 219 &  3.11 & 2573 &  56 & 0.2651 \\
 5551672 & 3119 &  3.67 & 5.870 & 222 &  3.67 & 2535 &  40 & 0.3075 \\
 5551672 & 3119 &  3.93 & 1.085 & 306 &  3.93 & 2546 &  25 & 0.0146 \\
 5551672 & 3119 &  6.59 & 3.942 & 286 &  6.59 & 2528 &  43 & 0.5289 \\
11853130 & 3263 &  2.28 & 0.821 & 276 &  2.28 & 3146 &  23 & 0.0005 \\
 6497146 & 3284 &  2.04 & 0.439 & 193 &  2.04 & 3144 &  39 & 0.0000 \\
 6497146 & 3284 &  2.91 & 3.956 &   4 &  2.91 & 3015 &  45 & 0.0067 \\
 5384713 & 3444 &  2.46 & 1.083 &  10 &  2.46 & 3111 &  47 & 0.0006 \\
 5384713 & 3444 &  5.09 & 3.578 & 265 &  5.09 & 2544 &  16 & 0.0432 \\
 5384713 & 3444 &  8.09 & 3.873 & 354 &  8.09 &   0. &  0. & 0.2950 \\
10525049 & 4252 &  0.47 & 0.043 & 349 &  0.47 & 3533 &  22 & 0.0000 \\
 4172805 & 4427 &  3.21 & 5.270 & 273 &  3.21 & 2998 &  34 & 0.0994 \\
 4172805 & 4427 &  7.03 & 3.203 & 253 &  7.03 & 2500 &   0 & 0.3119 \\
 4172805 & 4427 &  7.83 & 3.377 & 147 &  7.83 & 2532 &   0 & 0.4457 \\
 2986833 & 4875 &  4.22 & 1.739 &  22 &  4.22 & 2736 &  39 & 0.0186 \\
 2986833 & 4875 &  5.77 & 5.213 & 247 &  5.77 & 2533 &  24 & 0.3408 \\
 7731281 & 5416 &  0.03 & 0.138 & 231 &  0.03 & 3668 &  19 & 0.0000 \\
\hline
\end{tabular}
\end{minipage}
\end{table*}

%% file: table_tap.tex
\begin{table}
\centering
\begin{minipage}{140mm}
\caption{Light Curve Fit Parameters}
\label{tab.tap}
\begin{tabular}{l*{6}{c}r}
\hline
KOI & KIC & $R_P/R_*$ & $b$  & $\tau$ & Period & Probability\footnote{For candidate multi-star systems, the probability of the planet orbiting the given star (ABC) is based on a comparison of stellar densities derived from transits vs. photometry or spectroscopy.}\\
 &  &  &  & days & days &  \\
\hline
\multicolumn{3}{l}{Single Star Systems}\\
 3090.01 &    6609270 & $0.0167^{+0.0009}_{-0.0008}$ & $0.2558^{+0.1703}_{-0.1703}$ & $0.0784^{+0.0033}_{-0.0046}$ & 3.8222 &  & \\
 3090.02 &    6609270 & $0.0213^{+0.0012}_{-0.0012}$ & $0.6671^{+0.0562}_{-0.0726}$ & $0.0960^{+0.0069}_{-0.0066}$ & 15.0368 &  & \\
 4463.01 &    6197344 & $0.0144^{+0.0015}_{-0.0017}$ & $0.1519^{+0.1315}_{-0.1036}$ & $0.1246^{+0.0018}_{-0.0035}$ & 36.7604 &  & \\
 4777.01 &    6592335 & $0.0103^{+0.0355}_{-0.0022}$ & $0.7865^{+0.1876}_{-0.3656}$ & $0.0219^{+0.0084}_{-0.0132}$ & 0.82400 &  & \\
 4928.01 &    1873513 & $0.1069^{+0.0008}_{-0.0008}$ & $0.3767^{+0.0202}_{-0.0233}$ & $0.0658^{+0.0005}_{-0.0004}$ & 3.2904 &  & \\
 4957.02 &    2861126 & $0.0169^{+0.0017}_{-0.0015}$ & $0.5897^{+0.1413}_{-0.2766}$ & $0.0635^{+0.0080}_{-0.0088}$ & 4.9879 &  & \\
 5327.01 &    6776555 & $0.0370^{+0.0021}_{-0.0019}$ & $0.6889^{+0.0620}_{-0.0772}$ & $0.0399^{+0.0030}_{-0.0029}$ & 5.4337 &  & \\
 5831.01 &   10813078 & $0.0276^{+0.0022}_{-0.0015}$ & $0.2746^{+0.3071}_{-0.1867}$ & $0.1163^{+0.0039}_{-0.0173}$ & 10.5524 &  & \\
 6149.01 &    6674908 & $0.0227^{+0.0014}_{-0.0014}$ & $0.6018^{+0.0960}_{-0.1423}$ & $0.0343^{+0.0026}_{-0.0029}$ & 0.5032 &  & \\
 6276.01 &    2557350 & $0.0148^{+0.0015}_{-0.0014}$ & $0.2662^{+0.2436}_{-0.1832}$ & $0.0658^{+0.0054}_{-0.0064}$ & 3.0987 &  & \\
 6475.01 &    4939265 & $0.0192^{+0.0014}_{-0.0013}$ & $0.4640^{+0.1495}_{-0.2384}$ & $0.0779^{+0.0062}_{-0.0064}$ & 4.8451 &  & \\
 6635.01 &    5951140 & $0.0150^{+0.0014}_{-0.0013}$ & $0.2120^{+0.2340}_{-0.1484}$ & $0.0425^{+0.0040}_{-0.0041}$ & 0.5274 &  & \\
 6705.01 &    6423922 & $0.0057^{+0.0027}_{-0.0031}$ & $0.4347^{+0.4482}_{-0.3029}$ & $0.0355^{+0.0046}_{-0.0169}$ & 0.9951 &  & \\
 6839.01 &    7185487 & $0.0250^{+0.0017}_{-0.0015}$ & $0.2488^{+0.2198}_{-0.1714}$ & $0.0318^{+0.0024}_{-0.0028}$ & 0.5348 &  & \\
 6863.01 &    7350067 & $0.0606^{+0.0171}_{-0.0141}$ & $0.9520^{+0.0279}_{-0.0487}$ & $0.0184^{+0.0071}_{-0.0064}$ & 4.4856 &  & \\
 7182.01 &    9513168 & $0.0239^{+0.0020}_{-0.0019}$ & $0.4763^{+0.1064}_{-0.3982}$ & $0.0752^{+0.0098}_{-0.0058}$ & 6.8538 &  & \\
 7319.01 &   10388451 & $0.0138^{+0.0013}_{-0.0013}$ & $0.2608^{+0.2107}_{-0.1785}$ & $0.1313^{+0.0090}_{-0.0102}$ & 17.1675 &  & \\
 7408.01 &   11092783 & $0.0176^{+0.0014}_{-0.0013}$ & $0.3652^{+0.2075}_{-0.2325}$ & $0.0850^{+0.0062}_{-0.0082}$ & 5.1600 &  & \\
 7592.01 &    8423344 & $0.0283^{+0.0015}_{-0.0014}$ & $0.1955^{+0.2299}_{-0.1377}$ & $0.6094^{+0.0360}_{-0.0383}$ & 381.980 &  & \\
 7617.01 &   11559304 & $0.0157^{+0.0013}_{-0.0011}$ & $0.3754^{+0.2081}_{-0.2435}$ & $0.1005^{+0.0091}_{-0.0083}$ & 12.9285 &  & \\
\hline
\multicolumn{3}{l}{Multiple Star Systems}\\
  854.01 &    6435936A & $0.0496^{+0.0027}_{-0.0021}$ & $0.5256^{+0.2314}_{-0.3259}$ & $0.1814^{+0.0049}_{-0.0045}$ & 56.0562 &  0.7455\\
  854.01 &    6435936B & $0.0654^{+0.0044}_{-0.0034}$ & $0.5355^{+0.2406}_{-0.3857}$ & $0.1844^{+0.0047}_{-0.0045}$ & 56.0562 &  0.2545\\
 1422.01 &   11497958A & $0.0445^{+0.0046}_{-0.0063}$ & $0.8293^{+0.0671}_{-0.6314}$ & $0.0770^{+0.0036}_{-0.0046}$ & 5.8416 &  0.9285\\
 1422.01 &   11497958B & $0.0912^{+0.0364}_{-0.0128}$ & $0.8204^{+0.0751}_{-0.7257}$ & $0.0667^{+0.0105}_{-0.0106}$ & 5.8416 &  0.0715\\
 1422.02 &   11497958A & $0.0496^{+0.0058}_{-0.0076}$ & $0.8293^{+0.0671}_{-0.6314}$ & $0.0770^{+0.0036}_{-0.0046}$ & 19.8503 &  0.9285\\
 1422.02 &   11497958B & $0.1315^{+0.0101}_{-0.0437}$ & $0.8204^{+0.0751}_{-0.7257}$ & $0.0667^{+0.0105}_{-0.0106}$ & 19.8503 &  0.0715\\
 1422.03 &   11497958A & $0.0373^{+0.0105}_{-0.0121}$ & $0.8293^{+0.0671}_{-0.6314}$ & $0.0770^{+0.0036}_{-0.0046}$ & 10.8644 &  0.9285\\
 1422.03 &   11497958B & $0.0706^{+0.0000}_{-0.0132}$ & $0.8204^{+0.0751}_{-0.7257}$ & $0.0667^{+0.0105}_{-0.0106}$ & 10.8644 &  0.0715\\
 1422.04 &   11497958A & $0.0378^{+0.0044}_{-0.0061}$ & $0.8293^{+0.0671}_{-0.6314}$ & $0.0770^{+0.0036}_{-0.0046}$ & 63.3363 &  0.9285\\
 1422.04 &   11497958B & $0.0503^{+0.0240}_{-0.0038}$ & $0.8204^{+0.0751}_{-0.7257}$ & $0.0667^{+0.0105}_{-0.0106}$ & 63.3363 &  0.0715\\
 1422.05 &   11497958A & $0.0371^{+0.0042}_{-0.0056}$ & $0.8293^{+0.0671}_{-0.6314}$ & $0.0770^{+0.0036}_{-0.0046}$ & 34.1420 &  0.9285\\
 1422.05 &   11497958B & $0.0664^{+0.0566}_{-0.0036}$ & $0.8204^{+0.0751}_{-0.7257}$ & $0.0667^{+0.0105}_{-0.0106}$ & 34.1420 &  0.0715\\
 1681.01 &    5531953A & $0.0289^{+0.0009}_{-0.0008}$ & $0.2311^{+0.1644}_{-0.1603}$ & $0.0519^{+0.0019}_{-0.0028}$ & 6.9391 &  1.0000\\
 1681.01 &    5531953B & $0.0394^{+0.0014}_{-0.0013}$ & $0.4036^{+0.2862}_{-0.2519}$ & $0.0312^{+0.0024}_{-0.0045}$ & 6.9391 &  0.0000\\
 1681.02 &    5531953A & $0.0214^{+0.0014}_{-0.0013}$ & $0.2311^{+0.1644}_{-0.1603}$ & $0.0519^{+0.0019}_{-0.0028}$ & 1.9928 &  1.0000\\
 1681.02 &    5531953B & $0.0307^{+0.0025}_{-0.0023}$ & $0.4036^{+0.2862}_{-0.2519}$ & $0.0312^{+0.0024}_{-0.0045}$ & 1.9928 &  0.0000\\
 1681.03 &    5531953A & $0.0219^{+0.0016}_{-0.0016}$ & $0.2311^{+0.1644}_{-0.1603}$ & $0.0519^{+0.0019}_{-0.0028}$ & 3.5311 &  1.0000\\
 1681.03 &    5531953B & $0.0313^{+0.0027}_{-0.0026}$ & $0.4036^{+0.2862}_{-0.2519}$ & $0.0312^{+0.0024}_{-0.0045}$ & 3.5311 &  0.0000\\
 1681.04 &    5531953A & $0.0275^{+0.0016}_{-0.0016}$ & $0.2311^{+0.1644}_{-0.1603}$ & $0.0519^{+0.0019}_{-0.0028}$ & 21.9140 &  1.0000\\
 1681.04 &    5531953B & $0.0326^{+0.0015}_{-0.0015}$ & $0.4036^{+0.2862}_{-0.2519}$ & $0.0312^{+0.0024}_{-0.0045}$ & 21.9140 &  0.0000\\
 1880.01 &   10332883A & $0.0227^{+0.0016}_{-0.0012}$ & $0.5701^{+0.2130}_{-0.3435}$ & $0.0433^{+0.0024}_{-0.0022}$ & 1.1512 &  1.0000\\
 1880.01 &   10332883B & $0.3527^{+0.1638}_{-0.0469}$ & $0.6060^{+0.2937}_{-0.3995}$ & $0.0420^{+0.0040}_{-0.0184}$ & 1.1512 &  0.0000\\
 2179.01 &   10670119A & $0.0351^{+0.0023}_{-0.0014}$ & $0.3344^{+0.3288}_{-0.2192}$ & $0.0945^{+0.0035}_{-0.0038}$ & 14.8716 &  0.4460\\
 2179.01 &   10670119B & $0.0445^{+0.0032}_{-0.0019}$ & $0.3538^{+0.3660}_{-0.2615}$ & $0.0931^{+0.0032}_{-0.0030}$ & 14.8716 &  0.5540\\
 2179.02 &   10670119A & $0.0342^{+0.0024}_{-0.0019}$ & $0.3344^{+0.3288}_{-0.2192}$ & $0.0945^{+0.0035}_{-0.0038}$ & 2.7328 &  0.4460\\
 2179.02 &   10670119B & $0.0418^{+0.0048}_{-0.0025}$ & $0.3538^{+0.3660}_{-0.2615}$ & $0.0931^{+0.0032}_{-0.0030}$ & 2.7328 &  0.5540\\
 2418.01 &   10027247A & $0.0238^{+0.0029}_{-0.0016}$ & $0.4552^{+0.3714}_{-0.3171}$ & $0.2289^{+0.0146}_{-0.0236}$ & 86.8291 &  1.0000\\
 2418.01 &   10027247B & $0.1190^{+0.0172}_{-0.0108}$ & $0.5787^{+0.1943}_{-0.3699}$ & $0.2380^{+0.0172}_{-0.0191}$ & 86.8291 &  0.0000\\
 2542.01 &    6183511A & $0.0241^{+0.0374}_{-0.0029}$ & $0.7293^{+0.2260}_{-0.5106}$ & $0.0438^{+0.0070}_{-0.0373}$ & 0.7273 &  0.9998\\
 2542.01 &    6183511B & $0.0391^{+0.0055}_{-0.0031}$ & $0.4489^{+0.3903}_{-0.3248}$ & $0.0460^{+0.0051}_{-0.0051}$ & 0.7273 &  0.0002\\
 2626.01 &   11768142A & $0.0372^{+0.0079}_{-0.0028}$ & $0.6025^{+0.3114}_{-0.4150}$ & $0.1248^{+0.0082}_{-0.0112}$ & 38.0973 &  0.6201\\
 2626.01 &   11768142B & $0.0564^{+0.0120}_{-0.0045}$ & $0.6147^{+0.2925}_{-0.4059}$ & $0.1236^{+0.0090}_{-0.0124}$ & 38.0973 &  0.2487\\
 2626.01 &   11768142C & $0.0721^{+0.0159}_{-0.0059}$ & $0.6285^{+0.2699}_{-0.4198}$ & $0.1236^{+0.0090}_{-0.0133}$ & 38.0973 &  0.1312\\
 2705.01 &   11453592A & $0.0240^{+0.0006}_{-0.0007}$ & $0.1686^{+0.3679}_{-0.1168}$ & $0.0339^{+0.0007}_{-0.0009}$ & 2.8868 &  0.0000\\
 2705.01 &   11453592B & $0.1266^{+0.0042}_{-0.0040}$ & $0.1534^{+0.2573}_{-0.1050}$ & $0.0333^{+0.0010}_{-0.0012}$ & 2.8868 &  1.0000\\
\hline
\end{tabular}
$^a$Table is truncated.  The full table is available in electronic form.\\
\end{minipage}
\end{table}

%% file: planet_table.tex
\begin{table}
\centering
\begin{minipage}{140mm}
\caption{\kep{} M Dwarf Candidate Planets$^a$ \label{tab.planets}}
\begin{tabular}{@{}rrr|rr|rr@{}}
\hline
\multicolumn{1}{c}{KOI} & \multicolumn{1}{c}{KIC ID} & \multicolumn{1}{c}{period (d)} & \multicolumn{1}{c}{$R_p/R_{\oplus}$} & \multicolumn{1}{c}{$S_*/S_{\oplus}$}\\
\hline
247.01  & 11852982 &  13.815 &  1.90$\pm$0.19 &    7.63$\pm$  1.12 \\
 249.01 &  9390653 &   9.549 &  1.67$\pm$0.17 &    4.90$\pm$  0.99 \\
 251.01 & 10489206 &   4.164 &  2.73$\pm$0.21 &   29.68$\pm$  4.57 \\
 251.02 & 10489206 &   5.774 &  0.90$\pm$0.09 &   19.19$\pm$  2.96 \\
 253.01 & 11752906 &   6.383 &  3.22$\pm$0.27 &   27.57$\pm$  3.49 \\
 253.02 & 11752906 &  20.618 &  1.80$\pm$0.19 &    5.77$\pm$  0.73 \\
 254.01 &  5794240 &   2.455 & 13.02$\pm$0.83 &   79.20$\pm$ 11.10 \\
 255.01 &  7021681 &  27.522 &  2.60$\pm$0.22 &    2.65$\pm$  0.44 \\
 255.02 &  7021681 &  13.603 &  0.78$\pm$0.09 &    6.78$\pm$  1.12 \\
 463.01 &  8845205 &  18.478 &  2.11$\pm$0.20 &    2.27$\pm$  0.45 \\
\hline
\end{tabular}
\end{minipage}
$^{a}$Only the first few lines are shown to indicate content and form.  The full table is available in electronic form.
\end{table}

%% file: appendix_a.tex
\appendix
\section{Photometry of M Dwarf Calibrators \label{sec.photometry}}

The nearby calibrator M dwarfs cataloged in \citet{Mann2015b} are comparatively bright and images in all-sky surveys such as Two Micron All Sky Survey (2MASS), Sloan Digital Sky Survey (SDSS), and the Amateur Photometry All Sky Surevey (APASS) are often saturated.  While some photometry in the Johnson and Cron-Cousins systems exists for most of these stars, no Sloan or psuedo-Sloan photometry comparable to that of the Kepler Input Catalog \citep[KIC,][]{Brown2011} are available for comparisons.  

Between 15 and 27 October 2013 we obtained Sloan $gri$ photometry of 30 calibrator M dwarfs with the 1.3~m McGraw-Hill Telescope at MDM Observatory on Kitt Peak in Arizona.   Not all passbands were obtained for all stars, and data on three stars were rejected because the stellar images were merged or all images were saturated.  We surmounted the problem of saturation and lack of nearby, suitably bright photometric reference stars by obtaining multiple images with varying integration times that allowed us to tie the signal from the \citet{Mann2015b} stars to much fainter stars from the SDSS Data Release 9 \citep{Ahn2012}.  Integration times ranged from 5 to 300 sec, with the longest integration times achieving adequate signal for the faint SDSS targets and the shortest integrations usually accompanied by severe defocusing of the telescope to avoid saturation of the target star.  Images were corrected using sky flats and/or dome flats obtained for each filter, usually on the same night.  

A set of between two and six integrations were used for each target and each passband, with the zero-point offsets between pairs of images calculated using sufficiently bright but linear/unsaturated images of stars present in both image pairs.  SDSS stars in the field of view were retrieved using a query of the VIZIER database (V/139).  These were matched with sources in the deepest images  with iterative fitting of a second-order, two-dimensional polynomial with decreasing tolerance with each iteration down to one arc-sec.  Zero-point offsets and $g-r$ color terms were then calculated by weighted linear regressions with weights set to the inverse square of the formal photometric errors.  As expected, $g$-band photometric calibrations exhibit the largest color-terms.  Formal photometric precision is between 0.4 and 1.1\%, but the scatter among multiple observations of a few stars is generally $\gtrsim 1$\%.  

%% file: appendix_b.tex
\section{Dwarf-Giant Classification \label{sec.classification}}

We formulated a posterior probability that an M star in the KIC catalog is a main-sequence dwarf as the product of a prior $\tilde{p}$ based on $J$-band apparent magnitude, and likelihoods $p$ based on colors and proper motion (if available).  We calculated the prior probability that a star with $r-J > 2.2$ has a given $J$ magnitude using the TRILEGAL model of Galactic stellar populations.  \footnote{Since we were able to simulate large numbers of stars in a uniform manner, the Gaussian process analysis employed for colors offers no advantage and training is very computationally expensive.}  We calculated a probability 
\begin{equation}
\tilde{p}(J;r-J) = \sum\limits_n^{\rm dwarfs} W_n  / \sum\limits_n^{\rm all} W_n,
\end{equation}
where the weight $W_n$ of the $n$th star is
\begin{equation}
W_n = \exp \left[-\frac{1}{2}\left(\frac{\theta^2}{\theta_0^2} + \frac{(J-J_n)^2}{\sigma_J^2 + \rho_J^2} + \frac{\left(r -J - (r_n-J_n)\right)^2}{\sigma_{r-J}^2 + \rho_{r-J}^2} \right)\right],
\end{equation}
$\theta$ is the angular arc between a KIC star and a model star, and $\theta_0$, $\sigma_J$, and $\sigma_{r-J}$ are smoothing parameters.  We used TRILEGAL model version 1.6 \citep{Girardi2012}, simulating the stellar population in square degree fields centered at each of the 84 \kep{} CCD pointings.  \citet{Gao2013} showed that the default settings of this model were best at reproducing observed color-magnitude distributions.  All standard values for TRILEGAL model parameters were used.  The model predicts that nearly all red ($r-J > 2.2$) stars with $J > 11.5$ are dwarfs, while nearly all of their brighter ($J < 11.5$) counterparts are giants.  This is consistent with the spectroscopy- and photometry-based findings of \citet{Mann2012}.  The optimal values for the smoothing parameters were determined by maximizing the logarithmic likelihood of a dateset comprised of candidate planet hosts (which are all presumably dwarfs) and asteroseimically-detected giants.  This yielded $\theta_0 = 1.85$~deg, $\rho_J = 0.817$, and $\rho_{r-J} = 41.384$.  TRILEGAL predicts about 0.11 M dwarf per sq. deg. with $J< 10$, somewhat less than the 0.14 found by \citet{Lepine2011}, and too few to calculate priors for the brightest stars.  This leads to artificially miniscule priors that the brightest stars are dwarfs, and thus classification of these objects as giants.  Proper motions surveys of M dwarfs in the northern sky are nearly complete to this magnitude limit \citep{Lepine2011,Lepine2013,Gaidos2014} and the choice of prior for brighter stars is less critical as long as it is reasonably finite, thus we fix the mean value at its $J = 10$ value ($\tilde{p} = 0.21$) for this range.  

We calculated likelihoods based on colors using binary Gaussian process (GP) classification and training sets of dwarf and giant stars established independently of colors.  For practical reasons, we assumed independent contributions of each color $c_n$ to the probability, i.e. we ignore covariant scatter in the distribution of colors of the stars.  Stellar colors tend to be highly correlated, and the assumption that the colors of stars in different passbands are independent is strictly false.  However we removed most of this correlation by fitting and removing a polynomial fit of color vs. $r-J$, our proxy for \teff, and used the deviate colors $\tilde{c}$ from the best-fit locus plus $r-J$ in our analysis instead.

We adopted the standard parametric relation for the correlation matrix elements $K_{mn}$
\begin{equation}
\label{eqn.corr}
K_{mn} = \exp \left[-\frac{1}{2}\sum_q \frac{\left(\tilde{c}_{qm}-\tilde{c}_{qn}\right)^2}{w_q^2 + \sigma_{qm}^2 + \sigma_{qn}^2}\right],
\end{equation}
where $\tilde{c}_{qm}$ is the $q$th deviatory color of the $m$th star (plus $r-J$), $\sigma_{qm}$ is the associated standard error, and $w_q$ is the correlation parameter for the $q$th color.  We adopted the ``probit" (cumulative Gaussian) function $\phi(f)$ as the ``quashing" function to convert the latent function $f$ to a probability of belonging to a class (dwarf or giant).  The maximum likelihood values of $f = \hat{f}$ were calculated using the Laplace approximation in an iterative fashion \citep{Rasmussen2005}.

We used training sets of giant and dwarf stars to find optimal values for the hyperparameters.  These consist of stars that were classified independently of colors, although some color information was used to select the sample.  For giants, we started with a sample of 12,592 giant stars with $\log g$ established by asteroseismology \citep{Stello2014,Huber2014}.  To this we added 195 K and M giants that were selected as \kep{} targets based on variability; \kep{} observations of all of these stars confirmed the presence of giant-like radial and non-radial pulsations \citep{Stello2013}.  We also added 345 giants that were confirmed with moderate-resolution spectra by \citet{Mann2012}.  For the dwarf training set we used the 178 calibrator stars in \citet{Mann2015b}.  We extracted the KIC+KIS photometry and initially retained all stars with $r-J > 1.8$.

\begin{figure}
\includegraphics[width=84mm]{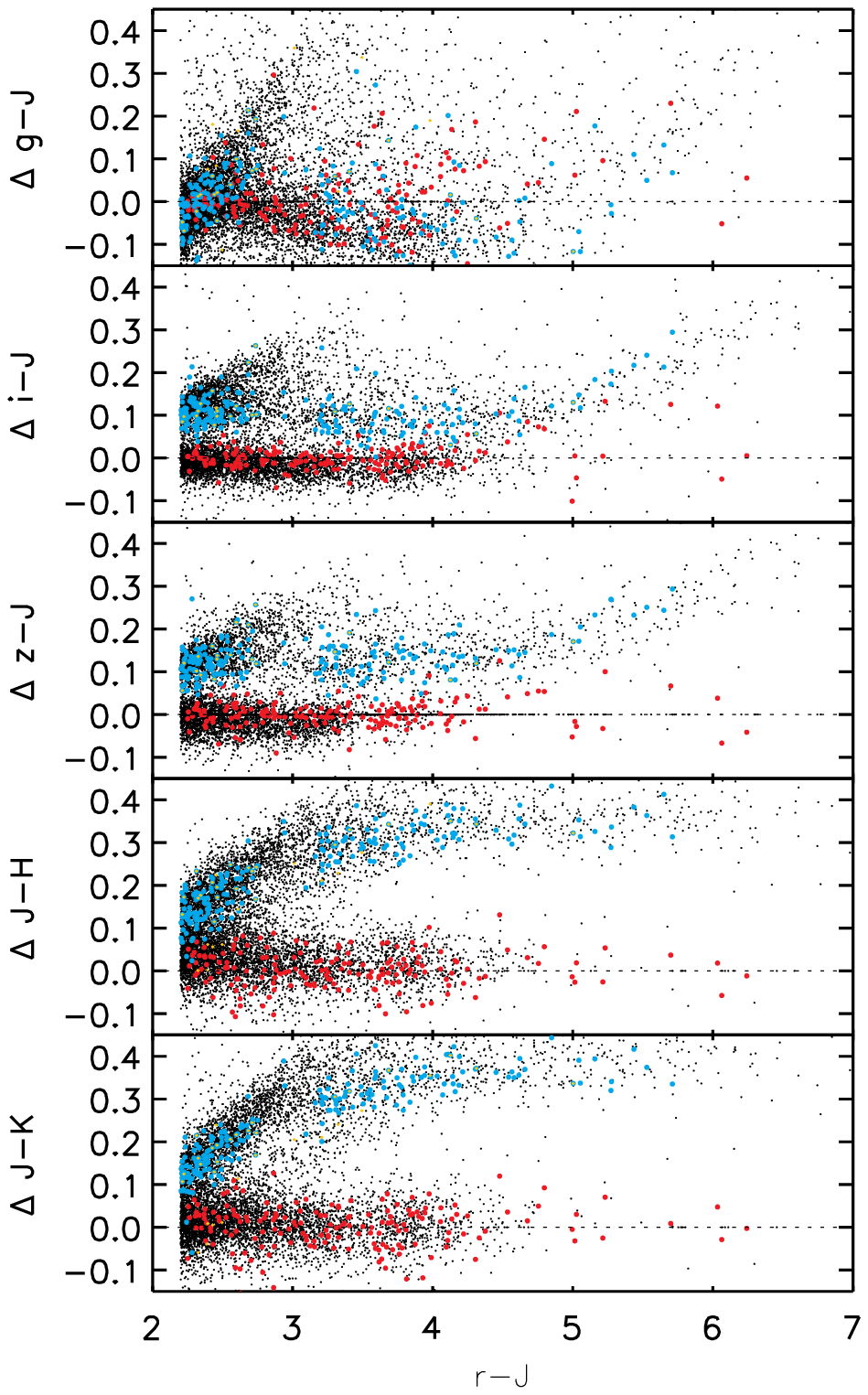}
\caption{Deviatory colors (w.r.t. a quadratic fit to the dwarf locus) of stars vs. $r-J$ color, a proxy of \teff{}.  Black points are KIC M stars, red points are the M dwarf training set, and blue points are the the M giant training set.}
 \label{fig.colors}
\end{figure}

The colors of distant giant stars can be profoundly affected by reddening; inspection of color-color diagrams shows dispersion in the direction predicted by reddening calculations.  It is thus important to thoroughly sample the distribution of reddening, however, there are only 28 stars with complete $rizJHK_S$ photometry and $r-J > 2.2$ in our catalog.  Our approach was to estimate the distribution of reddening $E_{r-J}$ for the {\it entire} giant star catalog by comparing the observed $r-J$ color with that synthesized by the PHOENIX stellar atmosphere model \citep{Husser2013} for the known \teff{} and asteroseimically determined $\log g$.  We then de-reddened the colors of each giant, and created 100 simulated giants reddened by values from the overall distribution.  We then retained all simulated stars with $r-J>2.2$.  Plots for all the five deviatory colors vs. $r-J$ (Fig. \ref{fig.colors}) illustrate that separation of giants and dwarfs based on colors alone can be achieved for most stars with $r-J>2.2$.

We determined optimal values of the six hyperparameters by maximizing the quantity
\begin{equation}
\log L = \sum\limits_n^{\rm dwarfs} \log \phi (f_n) + \sum\limits_m^{\rm giants} \log \left(1 - \phi(f_m)\right), 
\end{equation}
using the {\it tnmin} routine from \citet{Markwardt2009}.  We found that the $g-J$ colors of M giants and dwarfs heavily overlap and that $g-J$ is not useful for discriminating between the two luminosity classes.  The signal-to-noise in $g$-band for these red stars is also often low, therefore we excluded this color from further consideration here.  The optimal hyperparameter values for the other colors are reported in Table \ref{tab.hypers}.  The $z-J$ color has the smallest value of $w$ relative to the dynamic range of colors indicating that, as found by \citet{Gaidos2013}, $z$-band is the most useful Sloan passband for discriminating between dwarfs and giants.       

\begin{table}
\label{tab.hypers}
\caption{Hyperparameters for Gaussian Process Classification}
\centering
\begin{tabular}{l|r}
Color & $w$ \\
\hline
$r-J$           &      4.7292\\
deviatory $i-J$ &      27.7956 \\
deviatory $z-J$ &      0.2793 \\
deviatory $J-H$ &      0.6338 \\
deviatory $J-K$ &      0.4851 \\
\hline
\end{tabular}
\end{table}

\begin{figure}
\includegraphics[width=84mm]{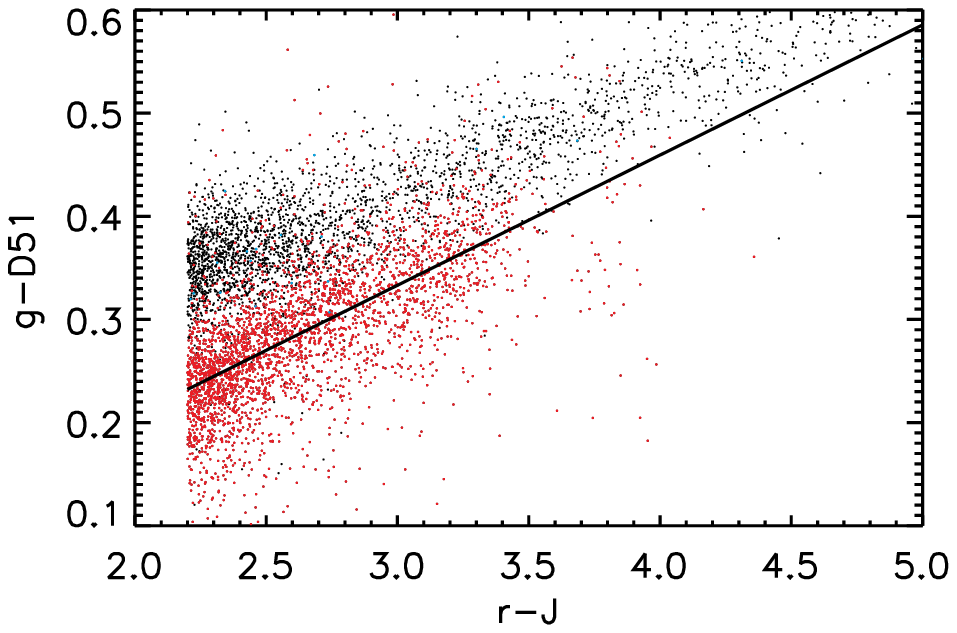}
\caption{$g-D51$ vs. $r-J$ colors for KIC stars.  Red stars are classified as dwarfs based on other colors and $J$-magnitudes; blue stars are asteroseimically-identified giants.  The solid curve is a best-fit line to the dwarfs.}
 \label{fig.d51_a}
\end{figure}

\begin{figure}
\includegraphics[width=84mm]{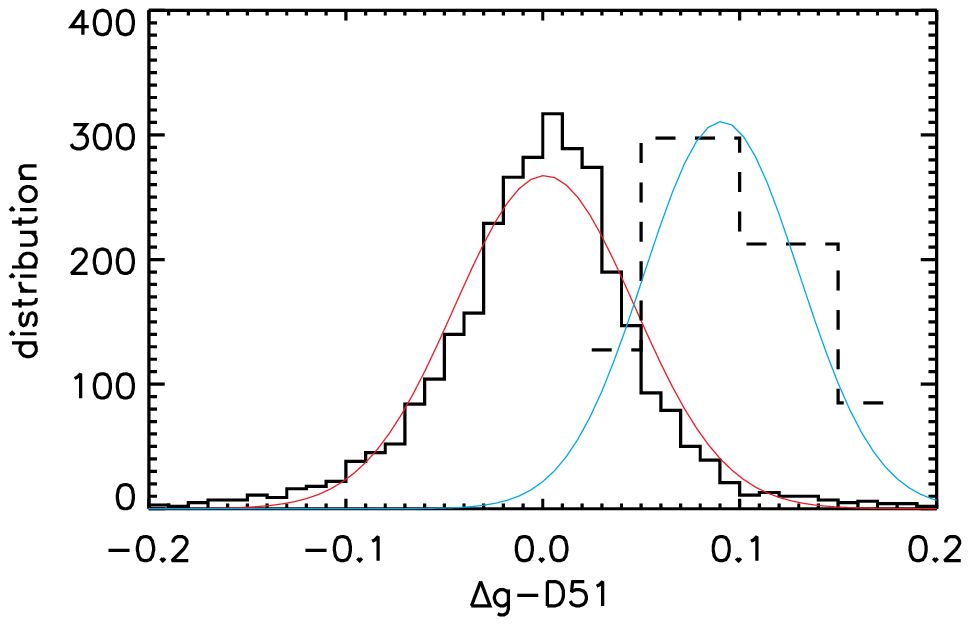}
\caption{Distribution of  $\Delta g-D51$ for dwarfs (solid histogram) and asteroseimically confirmed giants (dashed histogram) along with best-fit Gaussian functions (red and blue, respectively) centered on the median of the distribution and with robust standard deviation.  The giant distribution has been scaled to the size of the dwarf distribution.}
 \label{fig.d51_b}
\end{figure}

For some stars, the KIC catalog also includes measurements of flux in a narrow-band ($\sim 150$\AA{}-wide) D51 filter, which is centered at 5100\AA{} and includes the metallicity- and gravity-sensitive Mg~Ib resonant line \citep{Brown2011}.  As described in \citet{Brown2011}, $g-D51$ colors are useful for discriminating between giants and dwarfs with $g-r \approx 0.8-1.3$, i.e K- and early M-type stars.  The D51 bandpass falls near the boundary between the two spectral channels of SNIFS spectrograph and we cannot generate accurate synthetic magnitudes as was the case for the Sloan filters.  Thus we use an internal comparison approach in including this information.  Figure \ref{fig.d51_a} plots $g-D51$ vs. $r-J$ colors for \kep{} targets, with stars identified as dwarfs indicated as red and stars identified as giants based on asteroseismic signals indicated by blue.  The solid line is a linear robust fit to the dwarf locus.  The separation of dwarfs (bluer) and giants (redder) with $g-D51$ color is apparent, but only for stars with $r-J < 3.5$.  Figure \ref{fig.d51_b} shows the distributions for $r-J > 3.5$ dwarfs and giants, respectively.  The solid lines are Gaussians with centered on $9\times 10^{-4}$ and 0.089, respectively, and standard deviations of 0.0466 and 0.0407, respectively.  These were determined from filtered medians and robust standard deviations of the data.  For stars with $r-J < 3.5$, we calculated classification likelihood factors for dwarfs and giants using normalized versions of these curves.

GP classification can be ineffective where the training sets do not span the possible range of parameter space.  Our training set of giants, while large, is not exhaustive; the apparent colors of many giant stars are affected by reddening and variability over the interval between observations in different filters.  Extremely reddened giants falling outside both the loci of the giant and dwarf training sets will be assigned ambiguous classifications with probabilities near 0.5.  We addressed this problem by multiplying the dwarf classification probability by the product of one-sided Gaussian functions of each of the available deviatory colors relative to the reddest dwarf in the training set.  The width of the Gaussian was set to the standard error of the measurements.  

\begin{figure}
\includegraphics[width=84mm]{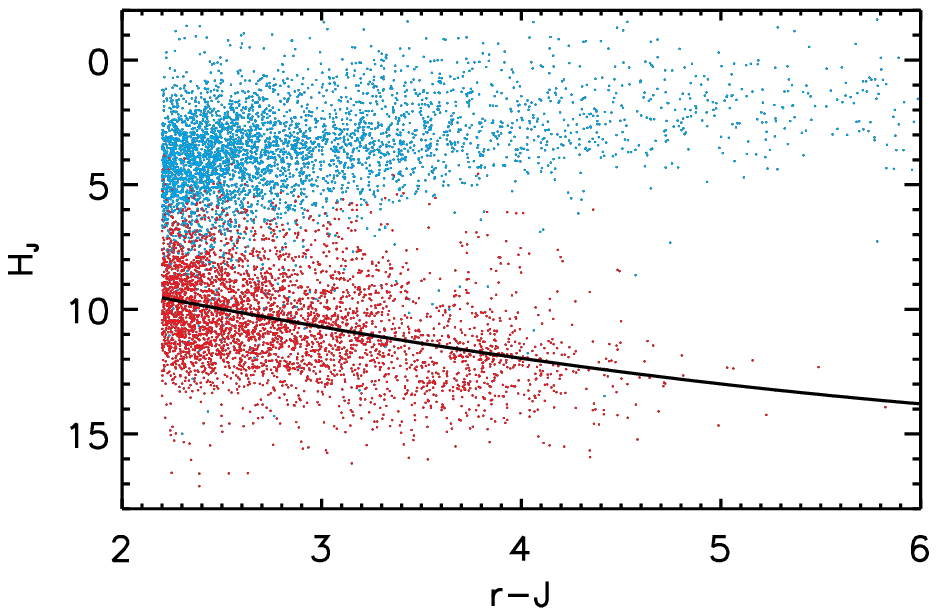}
\caption{Reduced proper motion $H_J$ vs. $V-J$ color for dwarf stars (red) and giants classified according to color and apparent $J$-band magnitude.  The solid line is a polynomial fit to the dwarf locus using a robust-fitting algorithm.}
 \label{fig.rpm_a}
\end{figure}

\begin{figure}
\includegraphics[width=84mm]{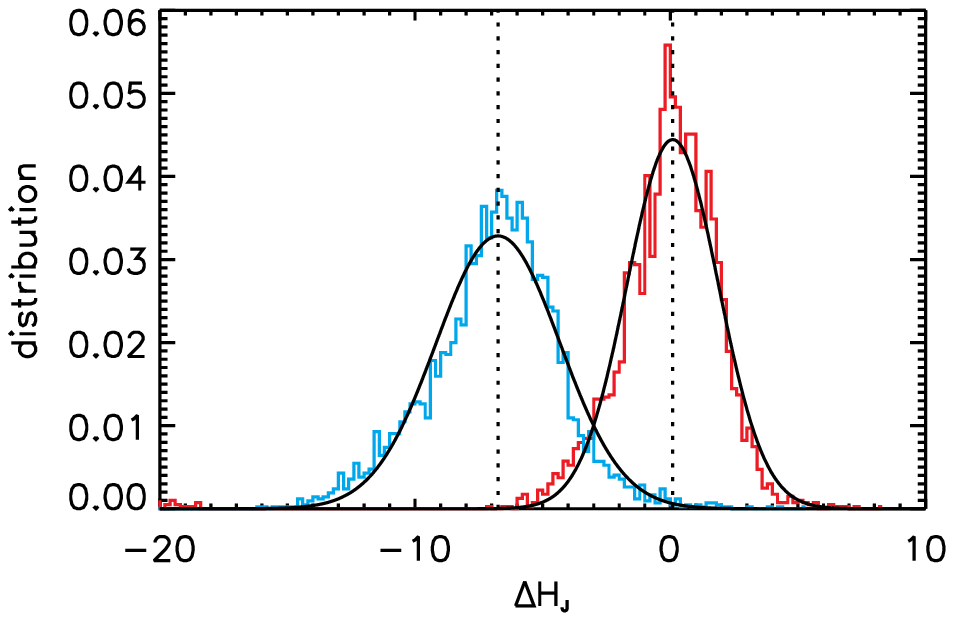}
\caption{Distributions of deviations of reduced proper motion $\Delta H_J$ from the best-fit locus vs. $r-J$ for dwarfs (red) and giants (blue).  Black curves are Gaussian function centered at the medians with widths determined by a robust estimate of standard deviation.}
 \label{fig.rpm_b}
\end{figure}

Relatively bright stars that are significantly bluer than either the giant or dwarf locus cannot be classified based on colors and our GP scheme; they are classified as giants based on magnitude alone, i.e. $J \lesssim 11.5$.  We found that placing a limit on how blue a giant star could be only led to mis-classification of many Mira stars and long-period variables as dwarfs.  Rather, we identified as many dwarf star as possible based on the reduced proper motion, which is a proxy for absolute magnitude for stars lacking a measured parallax:
\begin{equation}
\label{eqn.rpm}
H_J = J + 5 \log \mu + 5
\end{equation}
Proper motions were extracted from the Position and Proper Motion Catalog \citep[PPMXL,][]{Roeser2010} using the NASA/IPAC Infrared Science Archive, identifying the closest entry, if any, within 10 arc-sec to each \kep{} target and within 0.1 magnitudes of $J$-magnitude.  In addition, entries from the SuperBlink catalog of \citet[][S. L\'{e}pine, private communication,]{Lepine2005,Lepine2011} were matched with \kep{} targets using the same criteria.  Figure \ref{fig.rpm_a} plots reduced proper motions for stars with independent classifications based on color and apparent magnitude (red = dwarfs, blue = giants).  The luminosity classes are distinct but partly overlap for the bluest colors.  The solid curve is a robust best-fit quadratic to the dwarf locus:  $H_J = 9.535 + 1.566(r-J-2.2) -0.115 \left(r-J-2.2\right)^2$. This locus is relatively immune to the effect of interstellar reddening since the reddening vector is mostly parallel to the curve.  Fig. \ref{fig.rpm_b} plots the dwarf and giant distribution of deviations $\Delta H_J$ about this locus and Gaussian distributions centered on the medians (0.095 and -6.748, respectively) and width found by robust standard deviations (1.796 and 2.430, respectively).  Stars with significantly non-zero PMs ($>3\sigma$) to with $\Delta H_J$ fainter (to the right) of the dashed line (which presents three standard deviations are classified as dwarfs.  Following this procedure, 19 previously classified as giants stars were re-classified as dwarfs.

Seven stars with asteroseismic signals were classified as dwarfs by our procedure: KIC~4732678, 5471005, 5599911, 8428756, , 8654244, 9112438, and 9913261.  The classifications of KIC~4732678, 5471005, and 8654244 are marginal (P(dwarf)$<0.8$).  The interpretation of the variability of 5599911 as giant oscillations is doubtful (D. Huber, private communication).  The three other stars have valid asteroseismic signals (D. Huber, private communication) but could be blends of dwarfs and background giants in the 4" pixels of \kep{}. 

%% file: appendix_c.tex
\section{Estimation of Dwarf Properties \label{sec.parameters}}

In contrast to our approach for classification, we used the colors themselves, rather than differentials, to estimate the properties of the dwarf stars.  We also retained $g-J$, which is strongly correlated and thus predictive of \teff{}.  We ignored the effect of interstellar reddening and extinction on the \kep{} M dwarfs due to the low column densities of dust in the ``Local Bubble".  \citet{Reis2011} found that $E_{b-y} < 0.07$ to 500~pc in the direction of the \kep{} field; this distance includes nearly all of our M dwarfs.  Adopting reddening coefficients of 4.3 for $b-y$ and 2.29 for $r-J$ the corresponding $E_{r-J}$ should be $\ll 0.037$, corresponding to a change of \teff{} of $\ll 17$K.  We carried out a Gaussian process (GP) regression of \teff{} and \feh{} with all six possible colors.  A GP regression produces interpolated or extrapolated values for independent coordinate values using a parametric model of the covariance based on a set of dependent-independent coordinate values.  

As in the case of classification (Appendix B), we adopted a ``stationary squared exponential" function for the covariance of the stellar property (\teff{} or \feh{}).  The elements of the covariance matrix $k_{mn}$ between the $m$th and $n$th observations are prescribed by:
\begin{equation}
\label{eqn.covariance}
k_{mn} = \sigma_m^2\delta_{mn} + \Delta^2 \exp \left[- \frac{1}{2} \sum\limits_{c=0}^{N} - \frac{\left(x_m(c)-x_n(c)\right)^2}{w(c)^2 + \sigma_m(c)^2 + \sigma_n(c)^2}\right],
\end{equation}
where $x_m(c)$ is the $c$th color $(1...N)$ of the $m$th star, and $\Delta$ and $w(c)$ are parameters that describe the intrinsic width of the covariance with color and the magnitude of the covariance, respectively.  The first term on the right-hand side accounts for measurement error $\sigma_m$.  The larger the value of $\Delta$, the greater the overall covariance between colors and the stellar property, and the smaller the value of $w$, the more sensitive the covariance is to a particular color.  In the case of zero dependence, $w(c) \rightarrow \infty$ and observations are only related if they are indistinguishable within errors.  Equation \ref{eqn.covariance} assumes that the stellar property-color relation is a smooth function with a single scale that is of order the range of color itself, i.e. no smaller-scale structure.

\begin{figure}
\includegraphics[width=84mm]{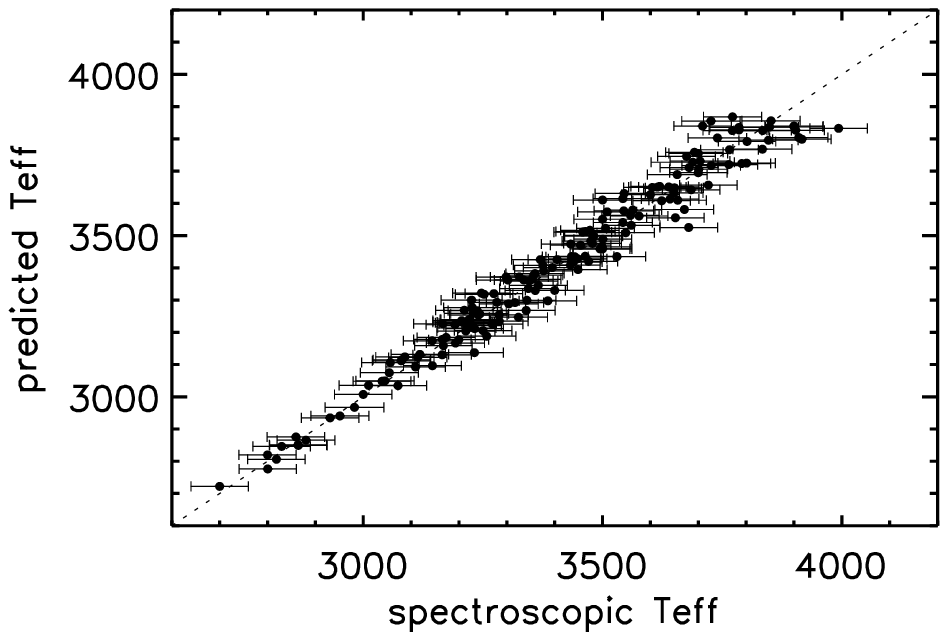}
\caption{Photometrically-derived values of \teff{} vs. spectroscopic-based values for the calibrator stars.  The dotted line is equality.}
 \label{fig.teffpred}
\end{figure}

\begin{figure}
\includegraphics[width=84mm]{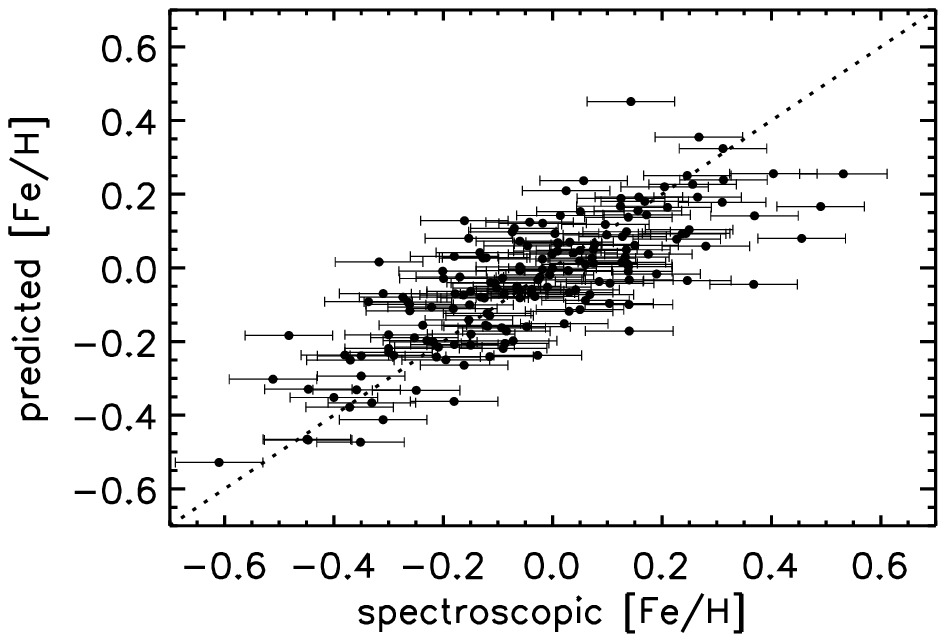}
\caption{Photometrically-derived values of \feh{} vs. spectroscopic-based values for the calibrator stars.  The dotted line is equality.}
 \label{fig.fehpred}
\end{figure}

The key to successful GP analysis is the determination of optimal values for the parameters $\Delta$ and $w(c)$.  Our criterion for success was reproduction of the known properties of each calibrator star using a GP constructed from the properties of all other calibrator stars, i.e. minimization of 
\begin{equation}
\label{eqn.chi2}
\chi^2 = \sum_n \left(\frac{Y_{\rm pred}- Y_{\rm obs}}{\sigma_n}\right)^2,
\end{equation}
where $Y$ represents either \teff{} or [Fe/H].  We used the MPFIT routine of \citep{Markwardt2009} to find the minimum $\chi^2$ parameter values.  The best-performance GP parameters and $\chi^2$ values ($\nu = 158$) are reported in Table \ref{tab.gp_params} and the predicted vs. actual (spectroscopic) values are plotted in Figs. \ref{fig.teffpred} and \ref{fig.fehpred}.  The mean offsets between predicted and actual values are negligible (0.6~K and 0.003 dex).  We found that $r-J$ color is the best predictor of \teff{} while the best predictor of \feh{} is $J-H$ color.  The reduced $\chi^2$ for \teff{} is $< 1$, suggesting that the spectroscopic errors are overestimated, while that for \feh{} indicates that the available colors cannot be used to estimate the [Fe/H] of M dwarfs with the precision attainable by spectroscopy ($\sim 0.05$ dex). 

\begin{table}
\label{tab.gp_params}
\caption{GP Hyperparameters for Characterization}
\centering
\begin{tabular}{l|r|r}
Parameter & \teff{} & [Fe/H]\\
\hline
$\chi^2 (\nu = 158)$ &   107.4 & 467.2\\
$\Delta$ &   8382.590 &   0.422 \\
$w_{g-J}$ &     32.592 &      1.785 \\
$w_{r-J}$ &      0.783 &      1.498 \\
$w_{i-J}$ &     10.529 &  37445.283 \\
$w_{z-J}$ &      6.163 &      1.059 \\
$w_{J-H}$ &     16.459 &      0.071 \\
$w_{J-K}$ &   2277.978 & 665255.974 \\
\hline
\end{tabular}
\end{table}

We identified 101 dwarfs in our catalog (esssentially all KOIs) with spectroscopic properties (\teff{} or \feh{}) tabulated in \cite{Muirhead2012}, \citet{Mann2012}, \citet{Mann2013}, \citet{Muirhead2014}, or \citet{Newton2015}.  We identified offsets between the photometric and spectroscopic values of \teff{} and [Fe/H] that were small (weighted means of +61~K and +0.19 dex) but significant ($p =$ 0.0.026 and 0.015).  We investigated and excluded the following explanations for this offset: (i) an error in the GP interpolation algorithm; (ii) interstellar reddening, which would make photometric \teff{} cooler, not hotter; (iii) systematic errors in the synthetic magnitudes generated from spectra of the calibrator stars; and (iv) a systematic error in spectroscopic \teff{} (the mean offset of revised \teff{} for a few stars incorporating parallaxes is within 25~K).

\teff{} is determined primarily by $r-J$ color and [Fe/H] by $H-J$.  These colors have the $J$-band in common, and we found that the entire \teff{} offset and nearly all of the [Fe/H] offset could be removed by adjusting the $J$-band values by 0.037 magnitudes, making either the KIC stars brighter or the calibrators fainter.  The 2MASS Point Source Catalog is uniform to the $\sim$4\% level in unconfused fields, so the required bias is marginally plausible and could arise from multiple sources of systematic error.  The 2MASS photometry in the KIC is based on profile fitting and under-sampling by the 2" pixels can produce biases.  We compared the aperture and profile-fit photometry of 1243 M dwarfs with $J< 13$, but found the median offset is only 0.002 magnitudes.   $J$-band is also the most affected by telluric water vapor absorption \citet{Cohen2003} and this could be another source of systematic error.   

\begin{figure}
\includegraphics[width=84mm]{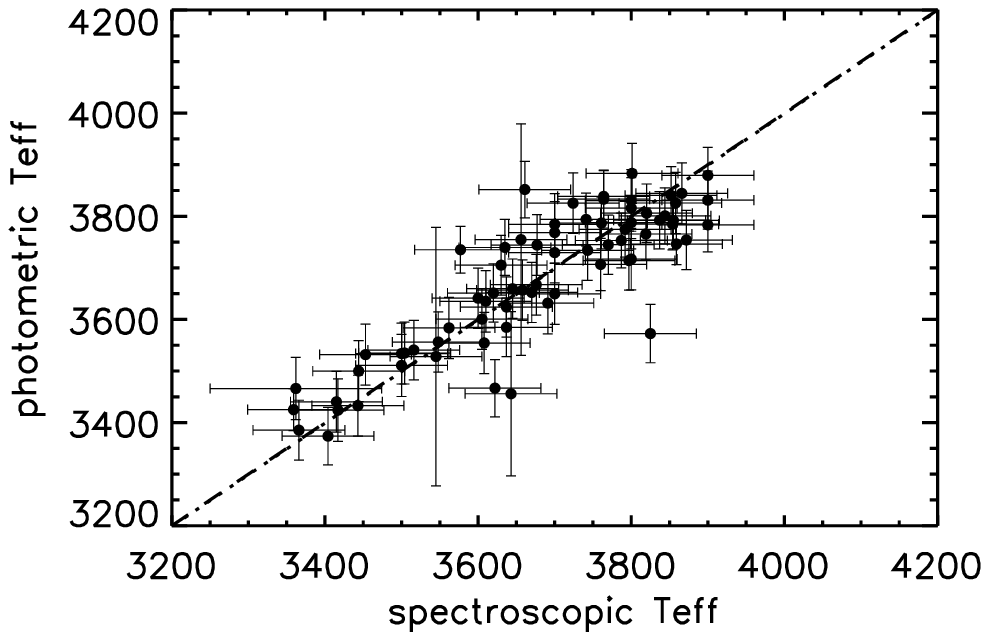}
\caption{Photometrically-derived values of \teff{} (after $J$-band correction) vs. spectroscopic-based values.  The dashed line is equality.}
 \label{fig.teffcompare}
\end{figure}

\begin{figure}
\includegraphics[width=84mm]{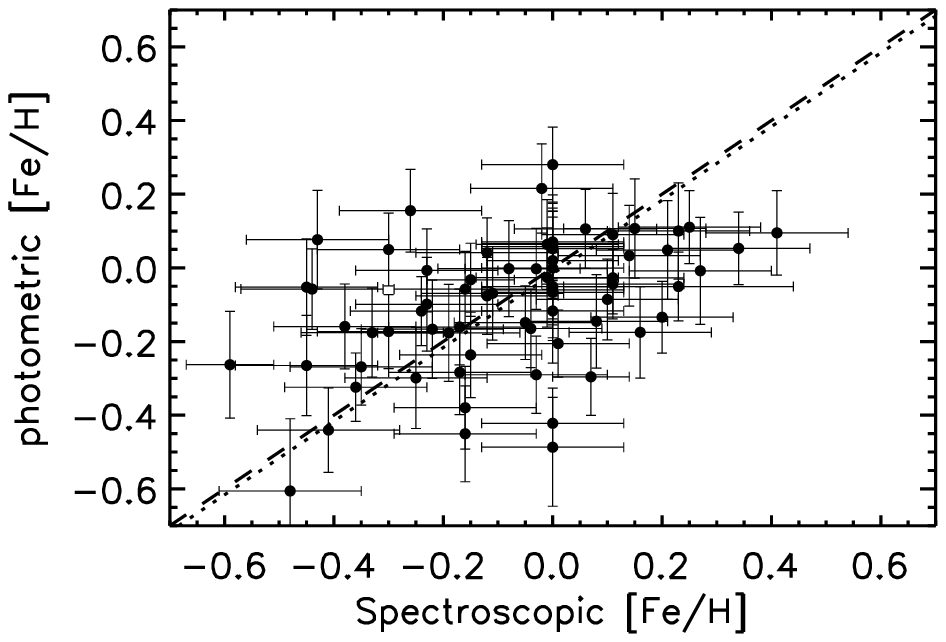}
\caption{Photometrically-derived values of \feh{} (after $J$-band correction) vs. spectroscopic-based values.  The dashed line is equality.}
 \label{fig.fehcompare}
\end{figure}

Figures \ref{fig.teffcompare} and \ref{fig.fehcompare} compares \teff{} and [Fe/H]  values based on the corrected $J$-band photometry with those based on spectroscopy.  The $\chi^2$ values are 61 ($\nu = 70$) and 109.4 ($\nu = 56$) for \teff{} and [Fe/H], respectively and this performance is comparable to that on the calibrator stars.  We replaced \teff{}, [Fe/H] (if available), and the dependent parameters with spectroscopic values, where available.  In addition, the regression failed for two dwarfs and these were excluded from subsequent analysis.

\label{LastPage}

%% file: manuscript.bbl
\clearpage

\begin{thebibliography}{115}
\expandafter\ifx\csname natexlab\endcsname\relax\def\natexlab#1{#1}\fi

\bibitem[{{Ahn} {et~al}\mbox{.}(2012){Ahn}, {Alexandroff}, {Allende Prieto},
  {Anderson}, {Anderton}, {Andrews}, {Aubourg}, {Bailey}, {Balbinot}, {Barnes},
  \& et~al.}]{Ahn2012}
{Ahn} C.~P. {et~al.}, 2012, \apjs, 203, 21

\bibitem[{{Artigau} {et~al}\mbox{.}(2014){Artigau}, {Kouach}, {Donati},
  {Doyon}, {Delfosse}, {Baratchart}, {Lacombe}, {Moutou}, {Rabou}, {Par{\`e}s},
  {Micheau}, {Thibault}, {Reshetov}, {Dubois}, {Hernandez}, {Vall{\'e}e},
  {Wang}, {Dolon}, {Pepe}, {Bouchy}, {Striebig}, {H{\'e}nault}, {Loop},
  {Saddlemyer}, {Barrick}, {Vermeulen}, {Dupieux}, {H{\'e}brard}, {Boisse},
  {Martioli}, {Alencar}, {do Nascimento}, \& {Figueira}}]{Artigau2014}
{Artigau} {\'E}. {et~al.}, 2014, in Society of Photo-Optical Instrumentation
  Engineers (SPIE) Conference Series, Vol. 9147, Society of Photo-Optical
  Instrumentation Engineers (SPIE) Conference Series, p.~15

\bibitem[{{Ballard} \& {Johnson}(2016)}]{Ballard2015}
{Ballard} S., {Johnson} J.~A., 2016, \apj, 816, 66 

\bibitem[{{Barclay} {et~al}\mbox{.}(2015){Barclay}, {Quintana}, {Adams},
  {Ciardi}, {Huber}, {Foreman-Mackey}, {Montet}, \& {Caldwell}}]{Barclay2015}
{Barclay} T., {Quintana} E.~V., {Adams} F.~C., {Ciardi} D.~R., {Huber} D.,
  {Foreman-Mackey} D., {Montet} B.~T., {Caldwell} D., 2015, \apj, 809, 7

\bibitem[{{Barclay} {et~al}\mbox{.}(2013){Barclay}, {Rowe}, {Lissauer},
  {Huber}, {Fressin}, {Howell}, {Bryson}, {Chaplin}, {D{\'e}sert}, {Lopez},
  {Marcy}, {Mullally}, {Ragozzine}, {Torres}, {Adams}, {Agol}, {Barrado},
  {Basu}, {Bedding}, {Buchhave}, {Charbonneau}, {Christiansen},
  {Christensen-Dalsgaard}, {Ciardi}, {Cochran}, {Dupree}, {Elsworth},
  {Everett}, {Fischer}, {Ford}, {Fortney}, {Geary}, {Haas}, {Handberg},
  {Hekker}, {Henze}, {Horch}, {Howard}, {Hunter}, {Isaacson}, {Jenkins},
  {Karoff}, {Kawaler}, {Kjeldsen}, {Klaus}, {Latham}, {Li}, {Lillo-Box},
  {Lund}, {Lundkvist}, {Metcalfe}, {Miglio}, {Morris}, {Quintana}, {Stello},
  {Smith}, {Still}, \& {Thompson}}]{Barclay2013}
{Barclay} T. {et~al.}, 2013, \nat, 494, 452

\bibitem[{{Benz} {et~al}\mbox{.}(2014){Benz}, {Ida}, {Alibert}, {Lin}, \&
  {Mordasini}}]{Benz2014}
{Benz} W., {Ida} S., {Alibert} Y., {Lin} D., {Mordasini} C., 2014, Protostars
  and Planets VI, 691

\bibitem[{{Borucki} {et~al}\mbox{.}(2010){Borucki}, {Koch}, {Basri}, {Batalha},
  {Brown}, {Caldwell}, {Caldwell}, {Christensen-Dalsgaard}, {Cochran},
  {DeVore}, {Dunham}, {Dupree}, {Gautier}, {Geary}, {Gilliland}, {Gould},
  {Howell}, {Jenkins}, {Kondo}, {Latham}, {Marcy}, {Meibom}, {Kjeldsen},
  {Lissauer}, {Monet}, {Morrison}, {Sasselov}, {Tarter}, {Boss}, {Brownlee},
  {Owen}, {Buzasi}, {Charbonneau}, {Doyle}, {Fortney}, {Ford}, {Holman},
  {Seager}, {Steffen}, {Welsh}, {Rowe}, {Anderson}, {Buchhave}, {Ciardi},
  {Walkowicz}, {Sherry}, {Horch}, {Isaacson}, {Everett}, {Fischer}, {Torres},
  {Johnson}, {Endl}, {MacQueen}, {Bryson}, {Dotson}, {Haas}, {Kolodziejczak},
  {Van Cleve}, {Chandrasekaran}, {Twicken}, {Quintana}, {Clarke}, {Allen},
  {Li}, {Wu}, {Tenenbaum}, {Verner}, {Bruhweiler}, {Barnes}, \&
  {Prsa}}]{Borucki2010}
{Borucki} W.~J. {et~al.}, 2010, Science, 327, 977

\bibitem[{{Boyajian} {et~al}\mbox{.}(2012){Boyajian}, {von Braun}, {van Belle},
  {McAlister}, {ten Brummelaar}, {Kane}, {Muirhead}, {Jones}, {White},
  {Schaefer}, {Ciardi}, {Henry}, {L{\'o}pez-Morales}, {Ridgway}, {Gies}, {Jao},
  {Rojas-Ayala}, {Parks}, {Sturmann}, {Sturmann}, {Turner}, {Farrington},
  {Goldfinger}, \& {Berger}}]{Boyajian2012}
{Boyajian} T.~S. {et~al.}, 2012, \apj, 757, 112

\bibitem[{{Brown} {et~al}\mbox{.}(2011){Brown}, {Latham}, {Everett}, \&
  {Esquerdo}}]{Brown2011}
{Brown} T.~M., {Latham} D.~W., {Everett} M.~E., {Esquerdo} G.~A., 2011, \aj,
  142, 112

\bibitem[{{Bryson} {et~al}\mbox{.}(2013){Bryson}, {Jenkins}, {Gilliland},
  {Twicken}, {Clarke}, {Rowe}, {Caldwell}, {Batalha}, {Mullally}, {Haas}, \&
  {Tenenbaum}}]{Bryson2013}
{Bryson} S.~T. {et~al.}, 2013, \pasp, 125, 889

\bibitem[{{Buchhave} {et~al}\mbox{.}(2014){Buchhave}, {Bizzarro}, {Latham},
  {Sasselov}, {Cochran}, {Endl}, {Isaacson}, {Juncher}, \&
  {Marcy}}]{Buchhave2014}
{Buchhave} L.~A. {et~al.}, 2014, \nat, 509, 593

\bibitem[{{Burke} {et~al}\mbox{.}(2015){Burke}, {Christiansen}, {Mullally},
  {Seader}, {Huber}, {Rowe}, {Coughlin}, {Thompson}, {Catanzarite}, {Clarke},
  {Morton}, {Caldwell}, {Bryson}, {Haas}, {Batalha}, {Jenkins}, {Tenenbaum},
  {Twicken}, {Li}, {Quintana}, {Barclay}, {Henze}, {Borucki}, {Howell}, \&
  {Still}}]{Burke2015}
{Burke} C.~J. {et~al.}, 2015, \apj, 809, 8

\bibitem[{Cappe, Godsill \& Moulines(2007)Cappe, Godsill, \&
  Moulines}]{Cappe2007}
Cappe O., Godsill S., Moulines E., 2007, Proceedings of the IEEE, 95, 899

\bibitem[{{Cartier} {et~al}\mbox{.}(2015){Cartier}, {Gilliland}, {Wright}, \&
  {Ciardi}}]{Cartier2015}
{Cartier} K.~M.~S., {Gilliland} R.~L., {Wright} J.~T., {Ciardi} D.~R., 2015,
  \apj, 804, 97

\bibitem[{{Chatterjee} \& {Tan}(2014)}]{Chatterjee2014}
{Chatterjee} S., {Tan} J.~C., 2014, \apj, 780, 53

\bibitem[{{Christiansen} {et~al}\mbox{.}(2015){Christiansen}, {Clarke},
  {Burke}, {Seader}, {Jenkins}, {Twicken}, {Catanzarite}, {Smith}, {Batalha},
  {Haas}, {Thompson}, {Campbell}, {Sabale}, \& {Kamal
  Uddin}}]{Christiansen2015}
{Christiansen} J.~L. {et~al.}, 2015, \apj, 810, 95

\bibitem[{{Claret} \& {Bloemen}(2011)}]{Claret2011}
{Claret} A., {Bloemen} S., 2011, \aap, 529, A75

\bibitem[{{Cohen}, {Wheaton} \& {Megeath}(2003){Cohen}, {Wheaton}, \&
  {Megeath}}]{Cohen2003}
{Cohen} M., {Wheaton} W.~A., {Megeath} S.~T., 2003, \aj, 126, 1090

\bibitem[{{Crossfield} {et~al}\mbox{.}(2015){Crossfield}, {Petigura},
  {Schlieder}, {Howard}, {Fulton}, {Aller}, {Ciardi}, {L{\'e}pine}, {Barclay},
  {de Pater}, {de Kleer}, {Quintana}, {Christiansen}, {Schlafly},
  {Kaltenegger}, {Crepp}, {Henning}, {Obermeier}, {Deacon}, {Weiss},
  {Isaacson}, {Hansen}, {Liu}, {Greene}, {Howell}, {Barman}, \&
  {Mordasini}}]{Crossfield2015}
{Crossfield} I.~J.~M. {et~al.}, 2015, \apj, 804, 10

\bibitem[{{Dawson} \& {Johnson}(2012)}]{Dawson:2012fk}
{Dawson} R.~I., {Johnson} J.~A., 2012, \apj, 756, 122

\bibitem[{{de Bruijne}(2012)}]{deBruijne2012}
{de Bruijne} J.~H.~J., 2012, \apss, 341, 31

\bibitem[{{De Cat} {et~al}\mbox{.}(2015){De Cat}, {Fu}, {Ren}, {Yang}, {Shi},
  {Luo}, {Yang}, {Wang}, {Zhang}, {Shi}, {Zhang}, {Dong}, {Catanzaro},
  {Corbally}, {Frasca}, {Gray}, {Molenda-{\.Z}akowicz}, {Uytterhoeven},
  {Briquet}, {Bruntt}, {Frandsen}, {Kiss}, {Kurtz}, {Marconi}, {Niemczura},
  {{\O}stensen}, {Ripepi}, {Smalley}, {Southworth}, {Szab{\'o}}, {Telting},
  {Karoff}, {Silva Aguirre}, {Wu}, {Hou}, {Jin}, \& {Zhou}}]{DeCat2015}
{De Cat} P. {et~al.}, 2015, \apjs, 220, 19

\bibitem[{{Deacon} {et~al}\mbox{.}(2016){Deacon}, {Kraus}, {Mann}, {Magnier},
  {Chambers}, {Wainscoat}, {Tonry}, {Kaiser}, {Waters}, {Flewelling}, {Hodapp},
  \& {Burgett}}]{Deacon2015}
{Deacon} N.~R. {et~al.}, 2016, \mnras, 455, 4212

\bibitem[{{Dotter} {et~al}\mbox{.}(2008){Dotter}, {Chaboyer}, {Jevremovi{\'c}},
  {Kostov}, {Baron}, \& {Ferguson}}]{Dotter2008}
{Dotter} A., {Chaboyer} B., {Jevremovi{\'c}} D., {Kostov} V., {Baron} E.,
  {Ferguson} J.~W., 2008, \apjs, 178, 89

\bibitem[{{Dressing} \& {Charbonneau}(2013)}]{Dressing2013}
{Dressing} C.~D., {Charbonneau} D., 2013, \apj, 767, 95

\bibitem[{{Dressing} \& {Charbonneau}(2015)}]{Dressing2015}
{Dressing} C.~D., {Charbonneau} D., 2015, \apj, 807, 45

\bibitem[{{Fang} \& {Margot}(2012)}]{Fang2012}
{Fang} J., {Margot} J.-L., 2012, \apj, 761, 92

\bibitem[{{Foreman-Mackey}, {Hogg} \& {Morton}(2014){Foreman-Mackey}, {Hogg},
  \& {Morton}}]{Foreman-Mackey2014}
{Foreman-Mackey} D., {Hogg} D.~W., {Morton} T.~D., 2014, \apj, 795, 64

\bibitem[{{Gaidos}(2013)}]{Gaidos2013}
{Gaidos} E., 2013, \apj, 770, 90

\bibitem[{{Gaidos} {et~al}\mbox{.}(2013){Gaidos}, {Fischer}, {Mann}, \&
  {Howard}}]{Gaidos2013b}
{Gaidos} E., {Fischer} D.~A., {Mann} A.~W., {Howard} A.~W., 2013, \apj, 771, 18

\bibitem[{{Gaidos} {et~al}\mbox{.}(2012){Gaidos}, {Fischer}, {Mann}, \&
  {L{\'e}pine}}]{Gaidos2012}
{Gaidos} E., {Fischer} D.~A., {Mann} A.~W., {L{\'e}pine} S., 2012, \apj, 746,
  36

\bibitem[{{Gaidos} \& {Mann}(2014)}]{Gaidos2014b}
{Gaidos} E., {Mann} A.~W., 2014, \apj, 791, 54

\bibitem[{{Gaidos}, {Mann} \& {Ansdell}(2015){Gaidos}, {Mann}, \&
  {Ansdell}}]{Gaidos2015b}
{Gaidos} E., {Mann} A.~W., {Ansdell} M., 2016, \apj, 817, 50 

\bibitem[{{Gaidos} {et~al}\mbox{.}(2014){Gaidos}, {Mann}, {L{\'e}pine},
  {Buccino}, {James}, {Ansdell}, {Petrucci}, {Mauas}, \& {Hilton}}]{Gaidos2014}
{Gaidos} E. {et~al.}, 2014, \mnras, 443, 2561

\bibitem[{{Gao}, {Just} \& {Grebel}(2013){Gao}, {Just}, \& {Grebel}}]{Gao2013}
{Gao} S., {Just} A., {Grebel} E.~K., 2013, \aap, 549, A20

\bibitem[{{Gazak} {et~al}\mbox{.}(2012){Gazak}, {Johnson}, {Tonry}, {Dragomir},
  {Eastman}, {Mann}, \& {Agol}}]{Gazak2012}
{Gazak} J.~Z., {Johnson} J.~A., {Tonry} J., {Dragomir} D., {Eastman} J., {Mann}
  A.~W., {Agol} E., 2012, Advances in Astronomy, 2012, 30

\bibitem[{{Girardi} {et~al}\mbox{.}(2012){Girardi}, {Barbieri}, {Groenewegen},
  {Marigo}, {Bressan}, {Rocha-Pinto}, {Santiago}, {Camargo}, \& {da
  Costa}}]{Girardi2012}
{Girardi} L. {et~al.}, 2012, {TRILEGAL, a TRIdimensional modeL of thE GALaxy:
  Status and Future}, {Miglio} A., {Montalb{\'a}n} J., {Noels} A., eds., p. 165

\bibitem[{{Greiss} {et~al}\mbox{.}(2012{\natexlab{a}}){Greiss}, {Steeghs},
  {G{\"a}nsicke}, {Mart{\'{\i}}n}, {Groot}, {Irwin}, {Gonz{\'a}lez-Solares},
  {Greimel}, {Knigge}, {{\O}stensen}, {Verbeek}, {Drew}, {Drake}, {Jonker},
  {Ripepi}, {Scaringi}, {Southworth}, {Still}, {Wright}, {Farnhill}, {van
  Haaften}, \& {Shah}}]{Greiss2012a}
{Greiss} S. {et~al.}, 2012{\natexlab{a}}, \aj, 144, 24

\bibitem[{{Greiss} {et~al}\mbox{.}(2012{\natexlab{b}}){Greiss}, {Steeghs},
  {G{\"a}nsicke}, {Mart{\'{\i}}n}, {Groot}, {Irwin}, {Gonz{\'a}lez-Solares},
  {Greimel}, {Gentile Fusillo}, {Still}, \& {the KIS
  collaboration}}]{Greiss2012b}
{Greiss} S. {et~al.}, 2012{\natexlab{b}}, arXiv:1212.3613

\bibitem[{{Hansen} \& {Murray}(2013)}]{Hansen2013}
{Hansen} B.~M.~S., {Murray} N., 2013, \apj, 775, 53

\bibitem[{{Hirano} {et~al}\mbox{.}(2015){Hirano}, {Masuda}, {Sato}, {Benomar},
  {Takeda}, {Omiya}, {Harakawa}, \& {Kobayashi}}]{Hirano2015}
{Hirano} T., {Masuda} K., {Sato} B., {Benomar} O., {Takeda} Y., {Omiya} M.,
  {Harakawa} H., {Kobayashi} A., 2015, \apj, 799, 9

\bibitem[{{Howard} {et~al}\mbox{.}(2012){Howard}, {Marcy}, {Bryson}, {Jenkins},
  {Rowe}, {Batalha}, {Borucki}, {Koch}, {Dunham}, {Gautier}, {Van Cleve},
  {Cochran}, {Latham}, {Lissauer}, {Torres}, {Brown}, {Gilliland}, {Buchhave},
  {Caldwell}, {Christensen-Dalsgaard}, {Ciardi}, {Fressin}, {Haas}, {Howell},
  {Kjeldsen}, {Seager}, {Rogers}, {Sasselov}, {Steffen}, {Basri},
  {Charbonneau}, {Christiansen}, {Clarke}, {Dupree}, {Fabrycky}, {Fischer},
  {Ford}, {Fortney}, {Tarter}, {Girouard}, {Holman}, {Johnson}, {Klaus},
  {Machalek}, {Moorhead}, {Morehead}, {Ragozzine}, {Tenenbaum}, {Twicken},
  {Quinn}, {Isaacson}, {Shporer}, {Lucas}, {Walkowicz}, {Welsh}, {Boss},
  {Devore}, {Gould}, {Smith}, {Morris}, {Prsa}, {Morton}, {Still}, {Thompson},
  {Mullally}, {Endl}, \& {MacQueen}}]{Howard2012}
{Howard} A.~W. {et~al.}, 2012, \apjs, 201, 15

\bibitem[{{Howell} {et~al}\mbox{.}(2014){Howell}, {Sobeck}, {Haas}, {Still},
  {Barclay}, {Mullally}, {Troeltzsch}, {Aigrain}, {Bryson}, {Caldwell},
  {Chaplin}, {Cochran}, {Huber}, {Marcy}, {Miglio}, {Najita}, {Smith},
  {Twicken}, \& {Fortney}}]{Howell2014}
{Howell} S.~B. {et~al.}, 2014, \pasp, 126, 398

\bibitem[{{Huber} {et~al}\mbox{.}(2014){Huber}, {Silva Aguirre}, {Matthews},
  {Pinsonneault}, {Gaidos}, {Garc{\'{\i}}a}, {Hekker}, {Mathur}, {Mosser},
  {Torres}, {Bastien}, {Basu}, {Bedding}, {Chaplin}, {Demory}, {Fleming},
  {Guo}, {Mann}, {Rowe}, {Serenelli}, {Smith}, \& {Stello}}]{Huber2014}
{Huber} D. {et~al.}, 2014, \apjs, 211, 2

\bibitem[{{Husser} {et~al}\mbox{.}(2013){Husser}, {Wende-von Berg}, {Dreizler},
  {Homeier}, {Reiners}, {Barman}, \& {Hauschildt}}]{Husser2013}
{Husser} T.-O., {Wende-von Berg} S., {Dreizler} S., {Homeier} D., {Reiners} A.,
  {Barman} T., {Hauschildt} P.~H., 2013, \aap, 553, A6

\bibitem[{{Jin} {et~al}\mbox{.}(2014){Jin}, {Mordasini}, {Parmentier}, {van
  Boekel}, {Henning}, \& {Ji}}]{Sheng2014}
{Jin} S., {Mordasini} C., {Parmentier} V., {van Boekel} R., {Henning} T., {Ji}
  J., 2014, \apj, 795, 65

\bibitem[{{Johnson} {et~al}\mbox{.}(2010){Johnson}, {Aller}, {Howard}, \&
  {Crepp}}]{Johnson2010}
{Johnson} J.~A., {Aller} K.~M., {Howard} A.~W., {Crepp} J.~R., 2010, \pasp,
  122, 905

\bibitem[{{Johnson} {et~al}\mbox{.}(2011){Johnson}, {Apps}, {Gazak}, {Crepp},
  {Crossfield}, {Howard}, {Marcy}, {Morton}, {Chubak}, \&
  {Isaacson}}]{Johnson2011}
{Johnson} J.~A. {et~al.}, 2011, \apj, 730, 79

\bibitem[{{Johnson} {et~al}\mbox{.}(2012){Johnson}, {Gazak}, {Apps},
  {Muirhead}, {Crepp}, {Crossfield}, {Boyajian}, {von Braun}, {Rojas-Ayala},
  {Howard}, {Covey}, {Schlawin}, {Hamren}, {Morton}, {Marcy}, \&
  {Lloyd}}]{Johnson2012}
{Johnson} J.~A. {et~al.}, 2012, \aj, 143, 111

\bibitem[{{Kopparapu} {et~al}\mbox{.}(2013){Kopparapu}, {Ramirez}, {Kasting},
  {Eymet}, {Robinson}, {Mahadevan}, {Terrien}, {Domagal-Goldman}, {Meadows}, \&
  {Deshpande}}]{Kopparapu2013}
{Kopparapu} R.~K. {et~al.}, 2013, \apj, 765, 131

\bibitem[{{Kotani} {et~al}\mbox{.}(2014){Kotani}, {Tamura}, {Suto},
  {Nishikawa}, {Sato}, {Aoki}, {Usuda}, {Kurokawa}, {Kashiwagi}, {Nishiyama},
  {Ikeda}, {Hall}, {Hodapp}, {Hashimoto}, {Morino}, {Okuyama}, {Tanaka},
  {Suzuki}, {Inoue}, {Kwon}, {Suenaga}, {Oh}, {Baba}, {Narita}, {Kokubo},
  {Hayano}, {Izumiura}, {Kambe}, {Kudo}, {Kusakabe}, {Ikoma}, {Hori}, {Omiya},
  {Genda}, {Fukui}, {Fujii}, {Guyon}, {Harakawa}, {Hayashi}, {Hidai}, {Hirano},
  {Kuzuhara}, {Machida}, {Matsuo}, {Nagata}, {Onuki}, {Ogihara}, {Takami},
  {Takato}, {Takahashi}, {Tachinami}, {Terada}, {Kawahara}, \&
  {Yamamuro}}]{Kotani2014}
{Kotani} T. {et~al.}, 2014, in Society of Photo-Optical Instrumentation
  Engineers (SPIE) Conference Series, Vol. 9147, Society of Photo-Optical
  Instrumentation Engineers (SPIE) Conference Series, p.~14

\bibitem[{{Kraus} {et~al}\mbox{.}(2015){Kraus}, {Ireland}, {Huber}, {Mann}, \&
  {Dupuy}}]{Kraus2015}
{Kraus} A.~L., {Ireland} M.~J., {Huber} D., {Mann} A.~W., {Dupuy} T.~J., 2015,
  \aj, submitted

\bibitem[{{Kraus} {et~al}\mbox{.}(2011){Kraus}, {Ireland}, {Martinache}, \&
  {Hillenbrand}}]{Kraus2011b}
{Kraus} A.~L., {Ireland} M.~J., {Martinache} F., {Hillenbrand} L.~A., 2011,
  \apj, 731, 8

\bibitem[{{Kraus} {et~al}\mbox{.}(2008){Kraus}, {Ireland}, {Martinache}, \&
  {Lloyd}}]{Kraus2008}
{Kraus} A.~L., {Ireland} M.~J., {Martinache} F., {Lloyd} J.~P., 2008, \apj,
  679, 762

\bibitem[{{Laughlin}, {Bodenheimer} \& {Adams}(2004){Laughlin}, {Bodenheimer},
  \& {Adams}}]{Adams2004}
{Laughlin} G., {Bodenheimer} P., {Adams} F.~C., 2004, \apjl, 612, L73

\bibitem[{{L{\'e}pine} \& {Gaidos}(2011)}]{Lepine2011}
{L{\'e}pine} S., {Gaidos} E., 2011, \aj, 142, 138

\bibitem[{{L{\'e}pine} {et~al}\mbox{.}(2013){L{\'e}pine}, {Hilton}, {Mann},
  {Wilde}, {Rojas-Ayala}, {Cruz}, \& {Gaidos}}]{Lepine2013}
{L{\'e}pine} S., {Hilton} E.~J., {Mann} A.~W., {Wilde} M., {Rojas-Ayala} B.,
  {Cruz} K.~L., {Gaidos} E., 2013, \aj, 145, 102

\bibitem[{{L{\'e}pine} \& {Shara}(2005)}]{Lepine2005}
{L{\'e}pine} S., {Shara} M.~M., 2005, \aj, 129, 1483

\bibitem[{{Lillo-Box}, {Barrado} \& {Bouy}(2014){Lillo-Box}, {Barrado}, \&
  {Bouy}}]{LilloBox2014}
{Lillo-Box} J., {Barrado} D., {Bouy} H., 2014, \aap, 566, A103

\bibitem[{{Lissauer} {et~al}\mbox{.}(2011){Lissauer}, {Ragozzine}, {Fabrycky},
  {Steffen}, {Ford}, {Jenkins}, {Shporer}, {Holman}, {Rowe}, {Quintana},
  {Batalha}, {Borucki}, {Bryson}, {Caldwell}, {Carter}, {Ciardi}, {Dunham},
  {Fortney}, {Gautier}, {Howell}, {Koch}, {Latham}, {Marcy}, {Morehead}, \&
  {Sasselov}}]{Lissauer2011}
{Lissauer} J.~J. {et~al.}, 2011, \apjs, 197, 8

\bibitem[{{Luger} {et~al}\mbox{.}(2015){Luger}, {Barnes}, {Lopez}, {Fortney},
  {Jackson}, \& {Meadows}}]{Lopez2015}
{Luger} R., {Barnes} R., {Lopez} E., {Fortney} J., {Jackson} B., {Meadows} V.,
  2015, Astrobiology, 15, 57

\bibitem[{{Mahadevan} {et~al}\mbox{.}(2014){Mahadevan}, {Ramsey}, {Terrien},
  {Halverson}, {Roy}, {Hearty}, {Levi}, {Stefansson}, {Robertson}, {Bender},
  {Schwab}, \& {Nelson}}]{Mahadevan2014}
{Mahadevan} S. {et~al.}, 2014, in Society of Photo-Optical Instrumentation
  Engineers (SPIE) Conference Series, Vol. 9147, Society of Photo-Optical
  Instrumentation Engineers (SPIE) Conference Series, p.~1

\bibitem[{{Mandel} \& {Agol}(2002)}]{MandelAgol2002}
{Mandel} K., {Agol} E., 2002, \apjl, 580, L171

\bibitem[{{Mann} {et~al}\mbox{.}(2015{\natexlab{a}}){Mann}, {Feiden}, {Gaidos},
  {Boyajian}, \& {von Braun}}]{Mann2015b}
{Mann} A.~W., {Feiden} G.~A., {Gaidos} E., {Boyajian} T., {von Braun} K.,
  2015{\natexlab{a}}, \apj, 804, 64

\bibitem[{{Mann}, {Gaidos} \& {Ansdell}(2013){Mann}, {Gaidos}, \&
  {Ansdell}}]{Mann2013}
{Mann} A.~W., {Gaidos} E., {Ansdell} M., 2013, \apj, 779, 188

\bibitem[{{Mann}, {Gaidos} \& {Gaudi}(2010){Mann}, {Gaidos}, \&
  {Gaudi}}]{Mann2010}
{Mann} A.~W., {Gaidos} E., {Gaudi} B.~S., 2010, \apj, 719, 1454

\bibitem[{{Mann} {et~al}\mbox{.}(2013){Mann}, {Gaidos}, {Kraus}, \&
  {Hilton}}]{Mann2013b}
{Mann} A.~W., {Gaidos} E., {Kraus} A., {Hilton} E.~J., 2013, \apj, 770, 43

\bibitem[{{Mann} {et~al}\mbox{.}(2012){Mann}, {Gaidos}, {L{\'e}pine}, \&
  {Hilton}}]{Mann2012}
{Mann} A.~W., {Gaidos} E., {L{\'e}pine} S., {Hilton} E.~J., 2012, \apj, 753, 90

\bibitem[{{Mann} {et~al}\mbox{.}(2015{\natexlab{b}}){Mann}, {Gaidos}, {Mace},
  {Johnson}, {Bowler}, {LaCourse}, {Jacobs}, {Vanderburg}, {Kraus}, {Kaplan},
  \& {Jaffe}}]{Mann2015c}
{Mann} A.~W. {et~al.}, 2015{\natexlab{b}}, arXiv:1512.00483M

\bibitem[{{Mann} \& {von Braun}(2015)}]{Mann2015a}
{Mann} A.~W., {von Braun} K., 2015, \pasp, 127, 102

\bibitem[{{Marcy} {et~al}\mbox{.}(2008){Marcy}, {Butler}, {Vogt}, {Fischer},
  {Wright}, {Johnson}, {Tinney}, {Jones}, {Carter}, {Bailey}, {O'Toole}, \&
  {Upadhyay}}]{Marcy2008}
{Marcy} G.~W. {et~al.}, 2008, Physica Scripta Volume T, 130, 014001

\bibitem[{{Markwardt}(2009)}]{Markwardt2009}
{Markwardt} C.~B., 2009, in Astronomical Society of the Pacific Conference
  Series, Vol. 411, Astronomical Data Analysis Software and Systems XVIII,
  {Bohlender} D.~A., {Durand} D., {Dowler} P., eds., p. 251

\bibitem[{{Mayor} \& {Queloz}(1995)}]{Mayor1995}
{Mayor} M., {Queloz} D., 1995, \nat, 378, 355

\bibitem[{{Montet} {et~al}\mbox{.}(2015){Montet}, {Morton}, {Foreman-Mackey},
  {Johnson}, {Hogg}, {Bowler}, {Latham}, {Bieryla}, \& {Mann}}]{Montet2015}
{Montet} B.~T. {et~al.}, 2015, \apj, 809, 25

\bibitem[{{Morton}(2012)}]{Morton2012}
{Morton} T.~D., 2012, \apj, 761, 6

\bibitem[{{Morton} \& {Swift}(2014)}]{Morton2014}
{Morton} T.~D., {Swift} J., 2014, \apj, 791, 10

\bibitem[{{Muirhead} {et~al}\mbox{.}(2014){Muirhead}, {Becker}, {Feiden},
  {Rojas-Ayala}, {Vanderburg}, {Price}, {Thorp}, {Law}, {Riddle}, {Baranec},
  {Hamren}, {Schlawin}, {Covey}, {Johnson}, \& {Lloyd}}]{Muirhead2014}
{Muirhead} P.~S. {et~al.}, 2014, \apjs, 213, 5

\bibitem[{{Muirhead} {et~al}\mbox{.}(2012{\natexlab{a}}){Muirhead}, {Hamren},
  {Schlawin}, {Rojas-Ayala}, {Covey}, \& {Lloyd}}]{Muirhead2012}
{Muirhead} P.~S., {Hamren} K., {Schlawin} E., {Rojas-Ayala} B., {Covey} K.~R.,
  {Lloyd} J.~P., 2012{\natexlab{a}}, \apjl, 750, L37

\bibitem[{{Muirhead} {et~al}\mbox{.}(2012{\natexlab{b}}){Muirhead}, {Johnson},
  {Apps}, {Carter}, {Morton}, {Fabrycky}, {Pineda}, {Bottom}, {Rojas-Ayala},
  {Schlawin}, {Hamren}, {Covey}, {Crepp}, {Stassun}, {Pepper}, {Hebb}, {Kirby},
  {Howard}, {Isaacson}, {Marcy}, {Levitan}, {Diaz-Santos}, {Armus}, \&
  {Lloyd}}]{Muirhead2012b}
{Muirhead} P.~S. {et~al.}, 2012{\natexlab{b}}, \apj, 747, 144

\bibitem[{{Muirhead} {et~al}\mbox{.}(2015){Muirhead}, {Mann}, {Vanderburg},
  {Morton}, {Kraus}, {Ireland}, {Swift}, {Feiden}, {Gaidos}, \&
  {Gazak}}]{Muirhead2015}
{Muirhead} P.~S. {et~al.}, 2015, \apj, 801, 18

\bibitem[{{Muirhead} {et~al}\mbox{.}(2013){Muirhead}, {Vanderburg}, {Shporer},
  {Becker}, {Swift}, {Lloyd}, {Fuller}, {Zhao}, {Hinkley}, {Pineda}, {Bottom},
  {Howard}, {von Braun}, {Boyajian}, {Law}, {Baranec}, {Riddle}, {Ramaprakash},
  {Tendulkar}, {Bui}, {Burse}, {Chordia}, {Das}, {Dekany}, {Punnadi}, \&
  {Johnson}}]{Muirhead2013}
{Muirhead} P.~S. {et~al.}, 2013, \apj, 767, 111

\bibitem[{{Mulders}, {Pascucci} \& {Apai}(2015){Mulders}, {Pascucci}, \&
  {Apai}}]{Mulders2015}
{Mulders} G.~D., {Pascucci} I., {Apai} D., 2015, \apj, 798, 112

\bibitem[{{Mullally} {et~al}\mbox{.}(2015){Mullally}, {Coughlin}, {Thompson},
  {Rowe}, {Burke}, {Latham}, {Batalha}, {Bryson}, {Christiansen}, {Henze},
  {Ofir}, {Quarles}, {Shporer}, {Van Eylen}, {Van Laerhoven}, {Shah},
  {Wolfgang}, {Chaplin}, {Xie}, {Akeson}, {Argabright}, {Bachtell}, {Barclay},
  {Borucki}, {Caldwell}, {Campbell}, {Catanzarite}, {Cochran}, {Duren},
  {Fleming}, {Fraquelli}, {Girouard}, {Haas}, {He{\l}miniak}, {Howell},
  {Huber}, {Larson}, {Gautier}, {Jenkins}, {Li}, {Lissauer}, {McArthur},
  {Miller}, {Morris}, {Patil-Sabale}, {Plavchan}, {Putnam}, {Quintana},
  {Ramirez}, {Silva Aguirre}, {Seader}, {Smith}, {Steffen}, {Stewart},
  {Stober}, {Still}, {Tenenbaum}, {Troeltzsch}, {Twicken}, \&
  {Zamudio}}]{Mullaly2015}
{Mullally} F. {et~al.}, 2015, \apjs, 217, 31

\bibitem[{{Newton} {et~al}\mbox{.}(2015){Newton}, {Charbonneau}, {Irwin}, \&
  {Mann}}]{Newton2015}
{Newton} E.~R., {Charbonneau} D., {Irwin} J., {Mann} A.~W., 2015, \apj, 800, 85

\bibitem[{{Owen} \& {Wu}(2013)}]{Owen2013}
{Owen} J.~E., {Wu} Y., 2013, \apj, 775, 105

\bibitem[{{Petigura}, {Howard} \& {Marcy}(2013){Petigura}, {Howard}, \&
  {Marcy}}]{Petigura2013}
{Petigura} E.~A., {Howard} A.~W., {Marcy} G.~W., 2013, Proceedings of the
  National Academy of Science, 110, 19273

\bibitem[{{Pinsonneault} {et~al}\mbox{.}(2012){Pinsonneault}, {An},
  {Molenda-{\.Z}akowicz}, {Chaplin}, {Metcalfe}, \&
  {Bruntt}}]{Pinsonneault2012}
{Pinsonneault} M.~H., {An} D., {Molenda-{\.Z}akowicz} J., {Chaplin} W.~J.,
  {Metcalfe} T.~S., {Bruntt} H., 2012, \apjs, 199, 30

\bibitem[{{Quintana} {et~al}\mbox{.}(2014){Quintana}, {Barclay}, {Raymond},
  {Rowe}, {Bolmont}, {Caldwell}, {Howell}, {Kane}, {Huber}, {Crepp},
  {Lissauer}, {Ciardi}, {Coughlin}, {Everett}, {Henze}, {Horch}, {Isaacson},
  {Ford}, {Adams}, {Still}, {Hunter}, {Quarles}, \& {Selsis}}]{Quintana2014}
{Quintana} E.~V. {et~al.}, 2014, Science, 344, 277

\bibitem[{{Quirrenbach} {et~al}\mbox{.}(2014){Quirrenbach}, {Amado},
  {Caballero}, {Mundt}, {Reiners}, {Ribas}, {Seifert}, {Abril}, {Aceituno},
  {Alonso-Floriano}, {Ammler-von Eiff}, {Antona Jim{\'e}nez},
  {Anwand-Heerwart}, {Azzaro}, {Bauer}, {Barrado}, {Becerril}, {B{\'e}jar},
  {Ben{\'{\i}}tez}, {Berdi{\~n}as}, {C{\'a}rdenas}, {Casal}, {Claret},
  {Colom{\'e}}, {Cort{\'e}s-Contreras}, {Czesla}, {Doellinger}, {Dreizler},
  {Feiz}, {Fern{\'a}ndez}, {Galad{\'{\i}}}, {G{\'a}lvez-Ortiz},
  {Garc{\'{\i}}a-Piquer}, {Garc{\'{\i}}a-Vargas}, {Garrido}, {Gesa}, {G{\'o}mez
  Galera}, {Gonz{\'a}lez {\'A}lvarez}, {Gonz{\'a}lez Hern{\'a}ndez},
  {Gr{\"o}zinger}, {Gu{\`a}rdia}, {Guenther}, {de Guindos},
  {Guti{\'e}rrez-Soto}, {Hagen}, {Hatzes}, {Hauschildt}, {Helmling}, {Henning},
  {Hermann}, {Hern{\'a}ndez Casta{\~n}o}, {Herrero}, {Hidalgo}, {Holgado},
  {Huber}, {Huber}, {Jeffers}, {Joergens}, {de Juan}, {Kehr}, {Klein},
  {K{\"u}rster}, {Lamert}, {Lalitha}, {Laun}, {Lemke}, {Lenzen}, {L{\'o}pez del
  Fresno}, {L{\'o}pez Mart{\'{\i}}}, {L{\'o}pez-Santiago}, {Mall}, {Mandel},
  {Mart{\'{\i}}n}, {Mart{\'{\i}}n-Ruiz}, {Mart{\'{\i}}nez-Rodr{\'{\i}}guez},
  {Marvin}, {Mathar}, {Mirabet}, {Montes}, {Morales Mu{\~n}oz}, {Moya},
  {Naranjo}, {Ofir}, {Oreiro}, {Pall{\'e}}, {Panduro}, {Passegger},
  {P{\'e}rez-Calpena}, {P{\'e}rez Medialdea}, {Perger}, {Pluto}, {Ram{\'o}n},
  {Rebolo}, {Redondo}, {Reffert}, {Reinhardt}, {Rhode}, {Rix}, {Rodler},
  {Rodr{\'{\i}}guez}, {Rodr{\'{\i}}guez-L{\'o}pez},
  {Rodr{\'{\i}}guez-P{\'e}rez}, {Rohloff}, {Rosich}, {S{\'a}nchez-Blanco},
  {S{\'a}nchez Carrasco}, {Sanz-Forcada}, {Sarmiento}, {Sch{\"a}fer},
  {Schiller}, {Schmidt}, {Schmitt}, {Solano}, {Stahl}, {Storz}, {St{\"u}rmer},
  {Su{\'a}rez}, {Ulbrich}, {Veredas}, {Wagner}, {Winkler}, {Zapatero Osorio},
  {Zechmeister}, {Abell{\'a}n de Paco}, {Anglada-Escud{\'e}}, {del Burgo},
  {Klutsch}, {Lizon}, {L{\'o}pez-Morales}, {Morales}, {Perryman}, {Tulloch}, \&
  {Xu}}]{Quirrenbach2014}
{Quirrenbach} A. {et~al.}, 2014, in Society of Photo-Optical Instrumentation
  Engineers (SPIE) Conference Series, Vol. 9147, Society of Photo-Optical
  Instrumentation Engineers (SPIE) Conference Series, p.~1

\bibitem[{Rasmussen \& Williams(2005)}]{Rasmussen2005}
Rasmussen C.~E., Williams C. K.~I., 2005, Gaussian Processes for Machine
  Learning (Adaptive Computation and Machine Learning). The MIT Press

\bibitem[{{Reid} {et~al}\mbox{.}(2004){Reid}, {Cruz}, {Allen}, {Mungall},
  {Kilkenny}, {Liebert}, {Hawley}, {Fraser}, {Covey}, {Lowrance},
  {Kirkpatrick}, \& {Burgasser}}]{Reid2004}
{Reid} I.~N. {et~al.}, 2004, \aj, 128, 463

\bibitem[{{Reid}, {Cruz} \& {Allen}(2007){Reid}, {Cruz}, \& {Allen}}]{Reid2007}
{Reid} I.~N., {Cruz} K.~L., {Allen} P.~R., 2007, \aj, 133, 2825

\bibitem[{{Reinhold}, {Reiners} \& {Basri}(2013){Reinhold}, {Reiners}, \&
  {Basri}}]{Reinhold2013}
{Reinhold} T., {Reiners} A., {Basri} G., 2013, \aap, 560, A4

\bibitem[{{Reis} {et~al}\mbox{.}(2011){Reis}, {Corradi}, {de Avillez}, \&
  {Santos}}]{Reis2011}
{Reis} W., {Corradi} W., {de Avillez} M.~A., {Santos} F.~P., 2011, \apj, 734, 8

\bibitem[{{Riaz}, {Gizis} \& {Harvin}(2006){Riaz}, {Gizis}, \&
  {Harvin}}]{Riaz2006}
{Riaz} B., {Gizis} J.~E., {Harvin} J., 2006, \aj, 132, 866

\bibitem[{{Roeser}, {Demleitner} \& {Schilbach}(2010){Roeser}, {Demleitner}, \&
  {Schilbach}}]{Roeser2010}
{Roeser} S., {Demleitner} M., {Schilbach} E., 2010, \aj, 139, 2440

\bibitem[{{Rowe} {et~al}\mbox{.}(2014){Rowe}, {Bryson}, {Marcy}, {Lissauer},
  {Jontof-Hutter}, {Mullally}, {Gilliland}, {Issacson}, {Ford}, {Howell},
  {Borucki}, {Haas}, {Huber}, {Steffen}, {Thompson}, {Quintana}, {Barclay},
  {Still}, {Fortney}, {Gautier}, {Hunter}, {Caldwell}, {Ciardi}, {Devore},
  {Cochran}, {Jenkins}, {Agol}, {Carter}, \& {Geary}}]{Rowe2014}
{Rowe} J.~F. {et~al.}, 2014, \apj, 784, 45

\bibitem[{{Rowe} {et~al}\mbox{.}(2015){Rowe}, {Coughlin}, {Antoci}, {Barclay},
  {Batalha}, {Borucki}, {Burke}, {Bryson}, {Caldwell}, {Campbell},
  {Catanzarite}, {Christiansen}, {Cochran}, {Gilliland}, {Girouard}, {Haas},
  {He{\l}miniak}, {Henze}, {Hoffman}, {Howell}, {Huber}, {Hunter},
  {Jang-Condell}, {Jenkins}, {Klaus}, {Latham}, {Li}, {Lissauer}, {McCauliff},
  {Morris}, {Mullally}, {Ofir}, {Quarles}, {Quintana}, {Sabale}, {Seader},
  {Shporer}, {Smith}, {Steffen}, {Still}, {Tenenbaum}, {Thompson}, {Twicken},
  {Van Laerhoven}, {Wolfgang}, \& {Zamudio}}]{Rowe2015}
{Rowe} J.~F. {et~al.}, 2015, \apjs, 217, 16

\bibitem[{{Schlaufman}(2015)}]{Schlaufman2015}
{Schlaufman} K.~C., 2015, \apjl, 799, L26

\bibitem[{{Seager} \& {Mall{\'e}n-Ornelas}(2003)}]{Seager:2003lr}
{Seager} S., {Mall{\'e}n-Ornelas} G., 2003, \apj, 585, 1038

\bibitem[{{Silburt}, {Gaidos} \& {Wu}(2015){Silburt}, {Gaidos}, \&
  {Wu}}]{Silburt2015}
{Silburt} A., {Gaidos} E., {Wu} Y., 2015, \apj, 799, 180

\bibitem[{{Skrutskie} {et~al}\mbox{.}(2006){Skrutskie}, {Cutri}, {Stiening},
  {Weinberg}, {Schneider}, {Carpenter}, {Beichman}, {Capps}, {Chester},
  {Elias}, {Huchra}, {Liebert}, {Lonsdale}, {Monet}, {Price}, {Seitzer},
  {Jarrett}, {Kirkpatrick}, {Gizis}, {Howard}, {Evans}, {Fowler}, {Fullmer},
  {Hurt}, {Light}, {Kopan}, {Marsh}, {McCallon}, {Tam}, {Van Dyk}, \&
  {Wheelock}}]{Skrutskie:2006lr}
{Skrutskie} M.~F. {et~al.}, 2006, \aj, 131, 1163

\bibitem[{{Slawson} {et~al}\mbox{.}(2011){Slawson}, {Pr{\v s}a}, {Welsh},
  {Orosz}, {Rucker}, {Batalha}, {Doyle}, {Engle}, {Conroy}, {Coughlin},
  {Gregg}, {Fetherolf}, {Short}, {Windmiller}, {Fabrycky}, {Howell}, {Jenkins},
  {Uddin}, {Mullally}, {Seader}, {Thompson}, {Sanderfer}, {Borucki}, \&
  {Koch}}]{Slawson2011}
{Slawson} R.~W. {et~al.}, 2011, \aj, 142, 160

\bibitem[{{Sliski} \& {Kipping}(2014)}]{Sliski2014}
{Sliski} D.~H., {Kipping} D.~M., 2014, \apj, 788, 148

\bibitem[{{Sousa} {et~al}\mbox{.}(2008){Sousa}, {Santos}, {Mayor}, {Udry},
  {Casagrande}, {Israelian}, {Pepe}, {Queloz}, \& {Monteiro}}]{Sousa2008}
{Sousa} S.~G. {et~al.}, 2008, \aap, 487, 373

\bibitem[{{Stello} {et~al}\mbox{.}(2014){Stello}, {Compton}, {Bedding},
  {Christensen-Dalsgaard}, {Kiss}, {Kjeldsen}, {Bellamy}, {Garc{\'{\i}}a}, \&
  {Mathur}}]{Stello2014}
{Stello} D. {et~al.}, 2014, \apjl, 788, L10

\bibitem[{{Stello} {et~al}\mbox{.}(2013){Stello}, {Huber}, {Bedding},
  {Benomar}, {Bildsten}, {Elsworth}, {Gilliland}, {Mosser}, {Paxton}, \&
  {White}}]{Stello2013}
{Stello} D. {et~al.}, 2013, \apjl, 765, L41

\bibitem[{{Stumpe} {et~al}\mbox{.}(2012){Stumpe}, {Smith}, {Van Cleve},
  {Twicken}, {Barclay}, {Fanelli}, {Girouard}, {Jenkins}, {Kolodziejczak},
  {McCauliff}, \& {Morris}}]{Stumpe2012}
{Stumpe} M.~C. {et~al.}, 2012, \pasp, 124, 985

\bibitem[{{Swift} {et~al}\mbox{.}(2013){Swift}, {Johnson}, {Morton}, {Crepp},
  {Montet}, {Fabrycky}, \& {Muirhead}}]{Swift2013}
{Swift} J.~J., {Johnson} J.~A., {Morton} T.~D., {Crepp} J.~R., {Montet} B.~T.,
  {Fabrycky} D.~C., {Muirhead} P.~S., 2013, \apj, 764, 105

\bibitem[{{Swift} {et~al}\mbox{.}(2015){Swift}, {Montet}, {Vanderburg},
  {Morton}, {Muirhead}, \& {Johnson}}]{Swift2015}
{Swift} J.~J., {Montet} B.~T., {Vanderburg} A., {Morton} T., {Muirhead} P.~S.,
  {Johnson} J.~A., 2015, \apjs, 218, 26

\bibitem[{{Udry} {et~al}\mbox{.}(2006){Udry}, {Mayor}, {Benz}, {Bertaux},
  {Bouchy}, {Lovis}, {Mordasini}, {Pepe}, {Queloz}, \& {Sivan}}]{Udry2006}
{Udry} S. {et~al.}, 2006, \aap, 447, 361

\bibitem[{{Van Eylen} \& {Albrecht}(2015)}]{Van-Eylen2015}
{Van Eylen} V., {Albrecht} S., 2015, \apj, 808, 126

\bibitem[{{Wang} \& {Fischer}(2015)}]{Wang2015}
{Wang} J., {Fischer} D.~A., 2015, \aj, 149, 14

\bibitem[{{Yelda} {et~al}\mbox{.}(2010){Yelda}, {Lu}, {Ghez}, {Clarkson},
  {Anderson}, {Do}, \& {Matthews}}]{Yelda2010}
{Yelda} S., {Lu} J.~R., {Ghez} A.~M., {Clarkson} W., {Anderson} J., {Do} T.,
  {Matthews} K., 2010, \apj, 725, 331

\bibitem[{{Youdin}(2011)}]{Youdin2011}
{Youdin} A.~N., 2011, \apj, 742, 38

\end{thebibliography}
